\documentclass[11pt, oneside]{article}   	
\usepackage{geometry}                		
\usepackage[parfill]{parskip}    		
\usepackage{amsmath}
\usepackage{amssymb}
\usepackage{amsfonts}
\usepackage{float}
\usepackage{hyperref}
\usepackage{multirow}
\usepackage{algorithmic,algorithm}
\usepackage{amsthm}
\usepackage{latexsym}
\usepackage{graphicx}				
\usepackage{natbib}
\usepackage{color}

\newcommand{\argmin}{arg\,min}

\newtheorem{theorem}{Theorem}
\newtheorem{lemma}{Lemma}
\newtheorem{remark}{Remark}

\begin{document}

\title{Sparse Group Fused Lasso for Model Segmentation\\ \smallskip  \Large{A Hybrid Approach}}

\author{David Degras}

\maketitle

\begin{abstract}
\noindent This article introduces the sparse group fused lasso (SGFL) as a statistical framework for segmenting sparse regression models with multivariate time series.  To compute solutions of the SGFL, a nonsmooth and nonseparable convex program, we develop a hybrid optimization method that is fast, requires no tuning parameter selection, and is guaranteed to converge to a global minimizer. In numerical experiments, the hybrid method compares favorably to state-of-the-art techniques with respect to computation time and numerical accuracy; benefits are particularly substantial in high dimension. The method's statistical performance is satisfactory in recovering nonzero regression coefficients  and excellent in change point detection. An application to air quality data is presented. 
The hybrid method is implemented in the R package \texttt{sparseGFL} available on the author's Github page. 
\end{abstract}

\section{Introduction}

In the  analysis of complex signals,  using a single statistical model 
with a fixed set of parameters is rarely enough to track data variations 
over their entire range. 
In long and/or high-dimensional time series for example, the presence of nonstationarity,
either in the form of slowly drifting dynamics or of abrupt regime changes,
requires that statistical models  flexibly account for temporal variations in signal characteristics.
To overcome the intrinsic limitations of approaches based on a single model  
vis-\`a-vis heterogeneous and nonstationary signals, 
\emph{model segmentation} techniques have been successfully employed 
in various fields including  
image processing (\cite{Alaiz2013,Friedman2007}) 
genetics (\cite{Bleakley2011,Tibshirani2007}),
brain imaging \cite{Beer2019,Cao2018,Ombao2015,Xu2015,Zhou2013},  \break
finance \cite{Hallac2019,Nystrup2017}, 
industrial monitoring \cite{Saxen2016}, 
oceanography \cite{Ranalli2018}, 
seismology \cite{Ohlsson2010},
and ecology \cite{Alewijnse2018}. 
Model segmentation consists in 
partitioning the domain of the signal 
(e.g. the temporal range of a time series 
or the lattice of a digital image) 
into a small number of  segments 
or regions such that for each segment, 
the data are suitably represented with a single model. 
The models used to segment the data 
are typically of the same type (e.g. linear model) 
but differ by their parameters.  
The task of model segmentation is closely related 
to \emph{change point detection}   
and is commonly referred to as (hybrid or time-varying) 
system identification in the engineering literature.

This work considers model segmentation in the following setup:  
\begin{itemize}
\item \emph{Structured multivariate data.} The observed data are multivariate predictor and response variables 
measured over a time grid. 
\item \emph{Regression.} 
Predictor and response variables are related through a regression model, e.g. a
 linear model, generalized linear model, or vector autoregressive model. 
 \item \emph{High dimension.}  
There are far more predictors than response variables. 
However, at each measurement point, the responses only depend on a small number of predictors.

\end{itemize}

For simplicity, we  present our methods and results in the context of  linear regression, 
keeping in mind that our work readily extends to other regression models. 
Let $(X_t)_{ 1\le t \le T}$ and  $(y_t)_{ 1\le t \le T}$ be multivariate time series 
where $y_{t}\in \mathbb{R}^d$ is a response vector and  $X_{t}\in \mathbb{R}^{d\times p}$ a predictor matrix. 
We consider the time-varying multivariate linear model 
 \begin{equation}\label{model yxb} 
 y_t = X_t \beta_t + \varepsilon_t 
 \end{equation}
where $\beta_t \in \mathbb{R}^p$ is an unknown regression vector 
and $ \varepsilon_t$ a  random vector with mean zero. 
As noted above, we assume that $p \gg d$,   
that the  $\beta_t$ are sparse, 
and that $\beta_{t}=\beta_{t+1}$ 
for most values of $t$, that is, $\beta = (\beta_t)_{1\le t \le T}$ 
is a piecewise constant function of $t$ with few  change points. 
Our goal is to develop  efficient computational methods for 
estimating $\beta $ and its change points 
$ t: \beta_{t-1} \ne \beta_{t} $.

Before introducing the optimization problem at the core of this study, 
namely the  \emph{sparse group fused lasso} (SGFL), 
 we review relevant work on model segmentation, change point detection, and structured sparse regression.

\subsection*{Related work}

We first introduce some notations. Throughout the paper,  $\| \cdot \|_q$ denotes the standard $\ell_q$ norm:  
$\| x \|_{q}  = (\sum_{i=1}^n |x_i|^q )^{1/q} $ if $0 < q < \infty$ and $\| x \|_{\infty} = \max_{1\le i \le n}( |x_i |) $ if $q=\infty$ 
for $x\in \mathbb{R}^n$.  
For convenience, we use the same notation  $\beta = (\beta_1, \ldots, \beta_T)$ to refer to regression coefficients either as a single vector in $\mathbb{R}^{pT}$ or 
as a sequence of $T$ vectors in $\mathbb{R}^p$. 

\paragraph{Combinatorial approaches to change point detection.} 
There is an extensive literature on change point detection spanning multiple fields and decades, 
which we only very partially describe here. For estimating changes in linear regression models, 
if the number $K\ge 2$ of segments (or equivalently the number $K-1$ of change points) is fixed, 
the segmentation problem can be expressed as    
\begin{equation}\label{linreg l0 penalty}
\min_{\substack{(\beta_1,\ldots,\beta_K)  \in \mathbb{R}^{p  K} \\  1 < T_{1} < \cdots < T_{K-1} \le T }} 
\frac{1}{2} \sum_{k=1}^{K} \sum_{t=T_{k-1}}^{T_{k}-1}  \big\|  y_t - X_t\beta_{k}    \big\|_2^2 
\end{equation}
with $T_0 = 1$ and $T_K=T+1$. For a given set of change points $(T_1,\ldots,T_{K-1})$, 
the minimizing argument $\beta = (\beta_1,\ldots,\beta_K)$ and associated objective value 
are obtained by ordinary least squares regression. Accordingly the optimization reduces 
to a combinatorial problem solvable by dynamic programming \cite{Bai2003}. 
This technique is computationally demanding as it requires performing $O(T^2)$ linear regressions 
before carrying out the dynamic program per se;  
the time spent in linear regression can however be reduced through recursive calculations. 
 A fundamental instance of model segmentation  in \eqref{linreg l0 penalty} occurs 
 when the design matrix $X_t$  is the identity matrix. 
In this case the problem is to approximate the signal $(y_t)$ itself 
with  a piecewise constant function.

If $K$ is not prespecified, one may add a penalty function to \eqref{linreg l0 penalty} 
so as to strike a compromise between fitting the data and keeping the model complexity low.  
Examples of penalty functions on $K$ include the Akaike Information Criterion (AIC), Bayesian Information Criterion (BIC), 
as well as more recent variants for high-dimensional data \cite{Chen2008,Yao1988}. 
Another way to select $K$ is to add/remove change points based on statistical tests or other criteria in top/down or bottom/up approaches. See  \cite{Basseville1993} for a classical book on statistical change point detection and \cite{Truong2018} for a more recent survey. Readers interested in  the popular method of binary segmentation may also consult 
\cite{Bai1997,Fryzlewicz2014,Leonardi2016}.

\paragraph{Total variation penalty methods.} 
Studying the piecewise constant  approximation of 1-D signals, 
\cite{Friedman2007} utilize a convex relaxation of  \eqref{linreg l0 penalty} 
called the \emph{fused lasso signal approximation} (FLSA): 
 \begin{equation}\label{FLSA}
\min_{ \beta \in \mathbb{R}^T} 
\frac{1}{2} \sum_{t=1}^T
 (  y_t - \beta_t  )^2 + \lambda_1 \sum_{t=1}^T  | \beta_t  |
 + \lambda_2 \sum_{t=1}^{T-1}   | \beta_{t+1} - \beta_t  | \, .
\end{equation}
Here, hard constraints or penalties on the number $K$ of segments 
are replaced by a  penalty on the increments $\beta_{t+1} - \beta_{t}$. 
This total variation penalty promotes flatness in the profile of $\beta$, 
that is, a small number of change points. 
The $\ell_1$ penalty on $\beta$, 
called a \emph{lasso penalty} in the statistical literature,
favors sparsity in $\beta$. 
The regularization parameters $\lambda_1, \lambda_2 > 0$ determine a balance between fidelity to the data, 
sparsity of $\beta $, and number of change points. 
They can be specified by the user or selected from the data, for example by cross-validation.   
\cite{Friedman2007} derive an efficient coordinate descent method to calculate the solution $\hat{\beta} = \hat{\beta}(\lambda_1,\lambda_2)$ to \eqref{FLSA} along a path of values of $(\lambda_1,\lambda_2)$. 
 Their method can also be applied to the more general problem of \emph{fused lasso regression}   
  \begin{equation}\label{fused lasso}
\min_{ \beta \in \mathbb{R}^T} 
\frac{1}{2} \sum_{i=1}^n
 (  y_i - x_i' \beta  )^2 + \lambda_1 \sum_{t=1}^T  | \beta_t  |
 + \lambda_2 \sum_{t=1}^{T-1}   | \beta_{t+1} - \beta_t  | 
\end{equation}
where $x_i\in \mathbb{R}^T$ is a vector of predictors, 
 although it is not guaranteed to yield a global minimizer in this case. 
 One may recover the FLSA \eqref{FLSA} 
  by setting $n=T$ and taking the $x_i $ as the canonical basis of $\mathbb{R}^T$ in  \eqref{fused lasso}. 
More recent approaches to fused lasso regression include \cite{Hoefling2010b,Liu2010,Wang2015}.

The FLSA and fused lasso can easily be adapted to the multivariate setup as follows:  
\begin{equation}\label{FL multivariate}
 \min_{ \beta \in \mathbb{R}^{p  T}} 
\frac{1}{2} \sum_{t=1}^T
 \| y_t - X_t  \beta_t  \|_2^2 + \lambda_1 \sum_{t=1}^T  \| \beta_t  \|_1
 + \lambda_2 \sum_{t=1}^{T-1}   \| \beta_{t+1} - \beta_t  \|_1 
\end{equation}
where $X_t \in \mathbb{R}^{d\times p}$, $y_t \in \mathbb{R}^d$, and $\beta = (\beta_1, \ldots, \beta_T)$.
These approaches are however not suitable for segmenting multivariate signals/models as they typically produce change points that are only shared by few predictor variables. This is because the $\ell_1$ norm in the total variation penalty affects each of the $p$ predictors separately. 
A simple way to induce change points common to all predictors is to replace this $\ell_1$ norm by an $\ell_q$ norm with $q>1$. Indeed for $q>1$, the $\ell_q$ norm of $\mathbb{R}^p$ is differentiable everywhere except at the origin, which promotes $\| \beta_{t+1} - \beta_{t} \|_q =0$. Typically, for the model estimate to have a change point at time $t+1$, a jump of at least modest size must occur in a significant fraction of the $p$ time-varying regression coefficients between $t$ and $t+1$. Due to its computational simplicity, the $\ell_2$ norm is often used in practice. 
For example, a common approach to denoising multivariate signals is to solve 
 \begin{equation}\label{TV-L2 denoising}
\min_{\beta  \in \mathbb{R}^{d  T}} 
\frac{1}{2} \sum_{t=1}^T
 \|  y_t - \beta_t \|_2^2 + \lambda_2 \sum_{t=1}^{T-1} w_t  \| \beta_{t+1} - \beta_t  \|_2   
\end{equation}
where the $w_t $ are positive weights. 
\cite{Bleakley2011} reformulate this problem as a group lasso regression and apply the group LARS algorithm (see \cite{Yuan2006}) to efficiently find solution paths $\hat{\beta} = \hat{\beta}(\lambda_2)$ as $\lambda_2$ varies. 
\cite{Wytock2014} propose Newton-type methods for \eqref{TV-L2 denoising} that extend to multichannel images.  
These two papers refer to problem \eqref{TV-L2 denoising} as the  \emph{group fused lasso} (GFL). 

Related to \eqref{TV-L2 denoising} is the \emph{convex clustering problem}
\begin{equation}\label{convex clustering} 
\min_{\beta \in \mathbb{R}^{p  T}} 
\frac{1}{2} \sum_{t=1}^T
 \|  y_t- \beta_t   \|_2^2 + \lambda_2 \sum_{1\le s<t \le T} w_{st}  \| \beta_{s} - \beta_t  \|_2 
\end{equation}
where the $\beta_{t}$ are cluster centroids  (see \cite{Chi2015,Hocking2011,Wang2018,Zhu2014} and the reference therein). 
This problem is arguably more difficult because there, the penalty pertains to all pairs of data points, as opposed to all pairs of  successive data points.  
Also  related to  \eqref{TV-L2 denoising}-\eqref{convex clustering} is the problem
 of  \emph{penalized multinomial logistic regression} (\cite{Price2019}) where the squared loss function is replaced by the negative log-likelihood, $w_{st}=1$,  and the $\beta_t$ represent regression vectors for $T$ response categories. 
In this context the total variation penalty facilitates combining  similar  categories.

To segment multivariate regression models with group sparsity structure,  
\cite{Alaiz2013} consider a generalization of \eqref{TV-L2 denoising} 
 that they also call group fused lasso:
\begin{equation}\label{GFL}
\min_{\beta \in \mathbb{R}^{p  T}} 
\frac{1}{2} \sum_{t=1}^T
 \|  y_t- X_t \beta_t   \|_2^2 + \lambda_1 \sum_{t=1}^T  \| \beta_t  \|_2 + 
\lambda_2 \sum_{t=1}^{T-1} w_t  \| \beta_{t+1} - \beta_t  \|_2 .
\end{equation}
They handle the optimization with a proximal splitting method similar to Dykstra's projection algorithm. 
\cite{Songsiri2015} studies \eqref{GFL} in the context of vector autoregressive models, 
using the well-known alternative direction method of multipliers (ADMM). 
See e.g. \cite{Combettes2011} for an overview of proximal methods and  ADMM.

\subsection*{Sparse Group Fused Lasso}

Under our assumptions, the set of regression coefficients 
$\beta = (\beta_1,\ldots, \beta_T)$ 
in the time-varying model $y_t = X_t \beta_t + \varepsilon_t$ 
is sparse and piecewise constant with few change points. 
To enforce these assumptions in fitting the model to data, we propose to solve 
 \begin{equation}\label{objective}
\min_{\beta \in \mathbb{R}^{pT}} F(\beta)  : = 
\frac{1}{2} \sum_{t=1}^{T}  \|  y_{t} -  X_{t} \beta_{t}    \|_2^2 
+ \lambda_1\sum_{t=1}^{T}  \| \beta_{t}  \|_1 
+ \lambda_2 \sum_{t=1}^{T-1} w_{t}   \, \| \beta_{t+1} - \beta_{t}  \|_2 \, .
\end{equation}
Problem \eqref{objective} has common elements
with the fused lasso \eqref{FL multivariate} and the group fused lasso \eqref{GFL} 
but the three problems are distinct and not reducible to one another.  
For example,  \eqref{FL multivariate} uses an $\ell_1$ TV penalty 
that favors equality between successive regression vectors at the coordinate level, i.e. $(\beta_t)_k = (\beta_{t+1})_k$, 
whereas \eqref{objective} uses an $\ell_2$ TV penalty to promote blockwise equality  $\beta_t = \beta_{t+1}$. 
This distinction is essential because the former tends to produce \emph{local} change points, which may be difficult to interpret and unsuitable for model segmentation, whereas the latter promotes \emph{global} change points at which most or all regression coordinates $(\beta_t)_k$ $(1\le k \le p$) change value; see \cite{Bleakley2011} for a numerical illustration. 
 Also, unlike \eqref{GFL} which exploits an $\ell_2$ 
 penalty to induce group sparsity in $\beta$, \eqref{objective} 
 features a standard lasso penalty. 
To distinguish \eqref{objective} from the group fused lasso problems \eqref{TV-L2 denoising}-\eqref{GFL}, we call it \emph{sparse group fused lasso} (SGFL). (Problem \eqref{objective} is referred to as $\ell_2$ variable fusion in \cite{Barbero2011} but we have not found this terminology elsewhere in the literature.) The GFL \eqref{TV-L2 denoising} is a special case of \eqref{objective} where $X_t = I_{d}$ (identity matrix) for all $t$ and $\lambda_1 = 0$.

 \medskip

\begin{remark}[Intercept]
\label{rmk: intercept}
A time-varying intercept vector $\delta_t $ can be added  to the regression model, 
yielding  $y_t = X_t \beta_t +\delta_t +\varepsilon_t$. While intercepts are typically not penalized in lasso regression, 
 one must assume some sparsity in the increments $\delta_{t+1}-\delta_{t}$ for the extended model to be meaningful. Accordingly, the extended SGFL expresses as
 \begin{equation}\label{SGFEN with intercept}
 \begin{split}
 \min_{\beta ,\, \delta }  
&\ \frac{1}{2} \sum_{t=1}^{T}  \|  y_{t} -  X_{t} \beta_{t}   -\delta_{t}  \|_2^2 
+ \lambda_1 \sum_{t=1}^{T}   \| \beta_{t}  \|_1  \\ 
& \qquad + \lambda_2 \sum_{t=1}^{T-1} w_{t}  \sqrt{ \| \beta_{t+1} - \beta_{t}  \|_2^2 
+   \| \delta_{t+1} - \delta_{t}  \|_2^2 } \, .
\end{split}
\end{equation}
For simplicity of exposition, we only consider problem \eqref{objective} in this paper, 
noting that all methods and results easily extend to \eqref{SGFEN with intercept}. 
\end{remark}

 \medskip

The objective function $F$ in \eqref{objective} has three components: 
a smooth function (squared loss), 
a nonsmooth but separable function (elastic net penalty), 
and a nonsmooth, nonseparable function (total variation penalty). 
We recall that a function $ f(\beta_1, \ldots , \beta_T)$ is said to be (block-)separable 
if it can be expressed as a sum of functions $\sum_{t=1}^T f_t(\beta_t)$. 
All three functions are convex. Accordingly, the SGFL \eqref{objective}
is a \emph{nonsmooth, nonseparable convex program}.  
Several off-the-shelf methods can be found in the convex optimization literature for this type of problem, 
among which primal-dual algorithms take a preeminent place \cite{Condat2013,Yan2018}.  
One could also utilize general-purpose convex optimization tools such as proximal methods (for instance, the Dykstra-like approach of \cite{Alaiz2013} can easily be adapted to \eqref{objective}),  ADMM and its variants, or even subgradient methods.  
However, these approaches do not take full advantage of the structure of  \eqref{objective}, 
which may  cause computational inefficiencies. 
In addition, these approaches aim at function minimization and not model segmentation or change point detection. 
As a result, they typically produce solutions for which every time $t$ is a change point and where 
the task of recovering the ``true" underlying change points (or segments) may be nontrivial. 
By devising customized methods for SGFL, 
one may expect substantial gains in computational speed 
while at the same time producing well-defined  model segmentations.

\pagebreak

\subsection*{Contributions and organization the paper}

We make the following contributions with this paper. 

\begin{enumerate}
\item We introduce the sparse group fused lasso (SGFL) for model segmentation in high dimension and develop a hybrid algorithm that efficiently solves the SGFL (Algorithm \ref{alg:main}). 
The algorithm produces a sequence of solutions that monotonically decrease the objective function and converge to a global minimizer. It yields exact model segmentations, as opposed to general optimization methods that only provide approximate segmentations. 
Importantly, the hybrid algorithm does not require any complicated selection of tuning parameters from the user.

\item A key component of the hybrid algorithm is an iterative soft-thresholding scheme for computing the proximal operator of sums of $\ell_1$ and $\ell_2$ norms. This scheme, which is shown to converge linearly, is of independent interest and can serve as a building block in other optimization problems.

\item 
Regarding the implementation of  the SGFL, we compare our hybrid approach to state-of-the-art optimization methods (e.g. 
ADMM, linearized ADMM, primal-dual) in terms of computational speed and numerical accuracy via numerical experiments.
We also examine the statistical accuracy of the SGFL method, comparing it to recent methods for change point detection in high dimensional regression. In this second set of experiments, we pay particular attention to recovery of nonzero coefficients, change point detection, sparsity of the solution,  and goodness of fit.  
We also illustrate the SGFL method with an application to air quality monitoring.

\item We implement  the hybrid algorithm in the R package \texttt{sparseGFL} available at \url{https://github.com/ddegras/sparseGFL}.

\end{enumerate}

The paper is organized as follows. 
Section \ref{sec:overview} gives an overview of the hybrid algorithm. 
Section \ref{sec:detailed computations}  details the calculations involved in each part of the algorithm. 
The full algorithm is summarized in Section \ref{sec: main algorithm} and its global convergence is stated.   
Section \ref{sec:experiments} presents numerical experiments comparing the proposed algorithm to state-of-the-art approaches both in terms of optimization performance and of statistical accuracy. An illustration of SGFL with air quality data is also given. 
Section \ref{sec:discussion} summarizes our results and outlines directions for future research. 
Appendix \ref{sec: linear convergence} contains a proof of linear convergence for the iterative soft-thresholding scheme used in the algorithm. Appendix \ref{Appendix B} contains a proof of global convergence for the hybrid algorithm. 
Supplementary Materials include additional results related to the numerical experiments of Section \ref{sec:experiments}. 


\section{Algorithm overview} 
\label{sec:overview}

\subsection*{Optimization strategy}

The proposed algorithm operates at different levels across iterations or cycles. 
By increasing order of complexity and generality, 
the optimization of $F$ in \eqref{objective} 
may be conducted with respect to: 
1) a single block $\beta_t$,
2) a chain of blocks $(\beta_t, \ldots, \beta_{t+k} )$ 
such that    $\beta_t = \cdots =  \beta_{t+k}$ (\emph{fusion chain}),
3) all fusion chains,
4)  all blocks.
The rationale for this hybrid optimization is to exploit problem structure for fast calculations 
while guaranteeing convergence to a global solution. 
By problem structure, we refer both to the block structure of the regression coefficients $\beta = (\beta_1,\ldots,\beta_T)$ 
and to the piecewise nature of the regression model \eqref{model yxb} over the time range $\{ 1,\ldots, T\}$. 
The first two levels of optimization (single block and single chain) involve block coordinate descent methods that can be implemented very quickly in a serial or parallel fashion. Because the objective \eqref{objective} is  \emph{not separable} in $\beta_t$, these methods do not guarantee convergence to a global solution. They only establish that the current solution $\hat{\beta} $ is \emph{blockwise optimal} (\cite{Tseng2001}), i.e. $F$ cannot be further reduced by changing a single block or chain of fused blocks in $\hat{\beta} $. 
 The next level  (all fusion chains) involves an \emph{active set} approach: assuming to have identified the optimal model segmentation, the associated fusion chains are fixed and $F$ is minimized with respect to these chains. Denoting by $K$ the number of chains, the dimension of the search space decreases from $pT$ variables to $pK$ where typically $K \ll T$.  
This third level of optimization is also not sufficient for global convergence. It ensures however that the minimum of $F$ over the current model segmentation has been attained. 
The fourth level consists in a single iteration of the subgradient method, which is known to converge (albeit very slowly) to a global minimizer of $F$ (e.g. \cite[Chap. 3]{Bertsekas2015}). Of the four optimization levels, this is the most general and most computationally intensive one.  Once combined to form the proposed  hybrid algorithm (Algorithm \ref{alg:main} in Section \ref{sec: main algorithm}),  these four optimization components guarantee convergence to a (global) minimizer of $F$ (Theorem \ref{thm: convergence} in Section \ref{sec: main algorithm}).

The general strategy of the hybrid algorithm is to optimize at the lowest (and fastest) level whenever possible. 
More precisely, optimization is realized at a higher level only when it is impossible to reduce the objective $F$
at the current level or a lower level. The flow of these operations is outlined below.
\medskip

\hspace*{-6mm}\fbox{
\addtolength{\linewidth}{-2\fboxsep}%
\addtolength{\linewidth}{-2\fboxrule}%
\begin{minipage}{\linewidth}
\begin{description}
\item [\textbf{Step 1.}] For each $t$, optimize the objective function $F$ in \eqref{objective} with respect to $\beta_t$ while keeping all other blocks fixed to their current values.
Cycle through blocks until $F$ cannot be further reduced, then go to Step 2. 

\item[\textbf{Step 2.}] Let $T_2 < \dots < T_K$ be the change points associated to the current solution $\hat{\beta}$ (if any) 
and $C_k = \{ t: T_k \le t <  T_{k+1} \}$  the corresponding fusion chains ($1\le k \le K$) with   
 $T_1=1 $ and $T_{K+1}=T+1$. For each $k$,  
 optimize $F$ with respect to $(\beta_t)_{ t\in C_k}$ under the constraint $\beta_{T_k} = \cdots = \beta_{T_{k+1}-1} $ while keeping all blocks outside $C_k$ fixed to their current values. 
Temporarily fuse $C_k$ with $C_{k+1}$ and repeat the optimization under the constraint 
$ \beta_{T_{K}} = \ldots , =\beta_{T_{K+2}-1}$. If $F$ has been reduced, update the blocks $(\beta_t)_{t\in C_k \cup C_{k+1}}$ 
to their common value and fuse $C_k$ with $C_{k+1}$. Cycle through fusion chains until $F$ cannot be further reduced. If $F$ has been reduced at any point during Step 2, go back to Step 1. Otherwise, go to Step 3. 

\item[\textbf{Step 3.}] Fix the current fusion chains and optimize $F$ with respect to $(\beta_{T_1},\ldots, \beta_{T_K})$ 
under the constraints $\beta_{T_k} = \ldots =\beta_{T_{k+1} - 1}$  ($1\le k \le K$). Go to Step 4.

\item[\textbf{Step 4.}] Check the optimality of the current solution $\hat{\beta}$ by calculating a subgradient $g \in \partial F(\hat{\beta})$ of minimal norm. If $g$ is (approximately) zero, terminate the algorithm and output $\hat{\beta}$. 
Otherwise, take a subgradient step in the direction $-g$ and go back to Step 1.  

\end{description}
\end{minipage}
}

\bigskip

We now give an overview of  each of the four algorithm steps above.

\subsection{Block coordinate descent}

The principle of block coordinate descent is to partition the 
optimization variables into blocks and to optimize the objective function at each iteration 
with respect to a given block while keeping the other blocks fixed.  
In the optimization \eqref{objective}, time provides a natural blocking structure. 
 Given a current solution $\hat{\beta} = (\hat{\beta}_1,\ldots, \hat{\beta}_T )$ and a time index $t \in \{ 1,\ldots,T\}$, 
 the problem formulates as
\begin{equation*}
\min_{\beta_t \in \mathbb{R}^p} F( \hat{\beta}_1,\ldots, \hat{\beta}_{t-1}, \beta_t , \hat{\beta}_{t+1},\ldots, \hat{\beta}_T).
\end{equation*}
Eliminating terms in $F$ that do not depend on $\beta_t$, this amounts to 
\begin{equation}\label{block descent}
 \min_{\beta_t \in \mathbb{R}^p } \frac{1}{2} \left\|   y_t - X_t \beta_t    \right\|_2^2 
+ \lambda_1   \big\| \beta_t \big\|_1  + 
\lambda_2 \left( w_{t-1} \big\| \beta_t - \hat{\beta}_{t-1} \big\|_2 +   w_{t} \big\|\hat{\beta}_{t+1} - \beta_t \big\|_2 \right) . 
\end{equation}
To accommodate the cases $t=1$ and $t=T$, we set $w_0 = w_T =0$ and $\hat{\beta}_0 =\hat{\beta}_{T+1} = 0_p$.

Problem \eqref{block descent} cannot be solved in closed form. Instead, we solve it using the  fast iterative soft-thresholding algorithm (FISTA) of \cite{Beck2009}, a proximal gradient method that enjoys the accelerated convergence rate $O(1/n^2)$, with $n$  the number of iterations. This algorithm is described in Section \ref{sec:FISTA}. 
The application of FISTA to \eqref{block descent}  entails calculating the proximal operator 
of the sum of the lasso and total variation penalties.
As a reminder, the proximal operator of a convex function $g: \mathbb{R}^p \to \mathbb{R}$ 
is defined by $\mathrm{prox}_{g}(x) = \mathrm{argmin}_{y\in  \mathbb{R}^p } g(y) + (1/2) \| y -x \|_2^2 $. 
Although the proximal operator of each penalty  easily obtains in closed form, 
determining the proximal operator of their sum is highly nontrivial. 
For this purpose, we develop an iterative soft-thresholding algorithm described in  Section \ref{sec:IST}.

The optimization \eqref{block descent} is repeated over a sequence of blocks 
and the solution $\hat{\beta}$ is updated each time   
until the objective function $F$ in \eqref{objective} cannot be further reduced. 
The order in which the blocks are selected for optimization is called the \emph{sweep pattern}. 
Common examples of sweep patterns include cyclic  (e.g. \cite{Tseng2001}), 
cyclic permutation, (e.g. \cite{Nesterov2012}), and greedy selection (e.g. \cite{Li2009}). 
The block coordinate descent is summarized in  Algorithm \ref{alg:BCD}.

\begin{algorithm}
\caption{Block Coordinate Descent}
\label{alg:BCD} 
\begin{algorithmic}
\REQUIRE $\beta^{n-1} \in \mathbb{R}^{pT}$, sweeping pattern $(t_1,\ldots,t_T)$
\ENSURE $\beta^{n}  \in \mathbb{R}^{pT}$
\STATE $\hat{\beta} \leftarrow \beta^{n-1}$
\FOR{$t=t_1,t_2,\ldots,t_T$}
\STATE Check  \eqref{local simple solution 1}-\eqref{local simple solution 2}-\eqref{local simple solution 3}
for a simple solution to \eqref{block descent} 
\IF{simple solution}
\STATE  $\hat{\beta}_t \leftarrow \hat{\beta}_{t-1}$ or 
 $\hat{\beta}_t \leftarrow \hat{\beta}_{t+1}$ as required
\ELSE[FISTA]
\STATE Set $f(\beta_t) = \frac{1}{2} \left\|   y_t - X_t \beta_t   \right\|_2^2  ,  
\ g(\beta_t)= \lambda_1   \| \beta_t  \|_1 + 
\lambda_2 \big( w_{t-1}  \| \beta_t - \hat{\beta}_{t-1}  \|_2 
+   w_{t}  \|\hat{\beta}_{t+1} - \beta_t  \|_2 \big) $
\STATE Apply  Algorithm \ref{alg:FISTA constant step size} to $f+g$ 
with starting point $\hat{\beta}_t$, 
Lipschitz constant $L = \| X_t'X_t \|_2 $, 
and $\mathrm{prox}_{g/L}$ given by \eqref{fp operator}-\eqref{fp iteration}. 
Output $\beta_t^{+}$
\STATE $\hat{\beta}_t \leftarrow \beta_t^{+}$
\ENDIF
\STATE $\beta_t^{n} \leftarrow \hat{\beta}_t$
\ENDFOR
\end{algorithmic}
\end{algorithm}

\subsection{Fusion cycle:  single chain}

Because the total variation penalty in $F$ is nonsmooth and nonseparable, 
the block coordinate descent can get stuck 
in points that are blockwise optimal but not globally optimal; 
see  \cite{Tseng2001} for a theoretical justification 
and  \cite{Friedman2007} for an example. 
To overcome this difficulty, one may 
constrain two or more consecutive blocks $\beta_t , \beta_{t+1}, \ldots$ to be equal 
and optimize $F $   
with respect to their common value while keeping other blocks fixed. 
This fusion strategy is well suited to  segmentation  because
 it either preserves segments or merges them into larger ones. 
Given a current solution $\hat{\beta} = (\hat{\beta}_1,\ldots, \hat{\beta}_T)$, 
 the time range $\{1,\ldots,T\}$ is partitioned  into segments or fusion chains 
$C_k = \{ t: T_{k} \le t <  T_{k+1} \}$  ($1 \le  k \le K$) 
 such that the $\hat{\beta}_{T_k} = \cdots = \hat{\beta}_{T_{k+1}-1}$
 and that $\hat{\beta}_{T_{k}} \ne \hat{\beta}_{T_{k+1}} $.  
By convention  we set $T_{1}=1$ and $T_{K+1} = T+1$. 
If $K>1$, $T_2,\ldots,T_{K}$  are the estimated change points 
 of the regression model $y_t = X_t\beta_t +\varepsilon_t$. 
The algorithm successively optimizes \eqref{objective} over each fusion chain $C_k$ 
while enforcing the equality constraint $\beta_{T_k} = \cdots = \beta_{T_{k+1}-1}$: 
\begin{equation*}
 \min_{\beta_t \in \mathbb{R}^p} F(\hat{\beta}_1,\ldots,\hat{\beta}_{T_k-1}, 
\beta_t,\ldots,\beta_t,
  \hat{\beta}_{T_{k+1}},\ldots,\hat{\beta}_T) 
\end{equation*}
where $\beta_t$ is repeated $n_k= T_{k+1}-T_{k}$ times.  
This works out as 
\begin{equation}\label{optimization single fusion chain}
\begin{split}
& \min_{\beta_{t} \in \mathbb{R}^p} \bigg\{ \frac{1}{2} \sum_{s=T_k}^{T_{k+1}-1} \left\| y_s - X_s \beta_{t} \right\|_2^2 + 
\lambda_1 n_k    \| \beta_{t}  \|_1   \\
& \qquad \qquad + 
\lambda_2 \left( w_{T_k-1} \big\|  \beta_{t} - \hat{\beta}_{T_k-1} \big\|_2 + w_{T_{k+1}} \big\|  \beta_{t} - \hat{\beta}_{T_{k+1}} \big\|_2 \right) \bigg\}\, .
\end{split}
\end{equation}

The algorithm may also try to merge two consecutive fusion chains to form a larger chain. 
To be precise, as $t$ follows a given sweeping pattern   $ t_1, \ldots,t_T$,  
 the algorithm either: 
(i) solves  \eqref{optimization single fusion chain} if $t=T_k$ 
and $T_{k+1}-T_{k}>1$ (start of a non-singleton chain), 
(ii) solves  \eqref{optimization single fusion chain} with 
each $T_{k+1}$ replaced by $T_{k+2}$ and $n_k$ by $n_k+n_{k+1}$ 
 if $t=T_{k+1}-1$ and $t<T$ (end of a chain),  
 or (iii) skips to the next value of $t$ in other cases. 
The optimization \eqref{optimization single fusion chain} is performed 
in the same way as the block coordinate descent 
\eqref{block descent} (FISTA + iterative soft-thresholding). 
The fusion cycle for single chains is summarized in Algorithm \ref{alg:BCD fusion}.  

\begin{algorithm}
\caption{Fusion Cycle: Single Chain}
\label{alg:BCD fusion} 
\begin{algorithmic}
\REQUIRE $\beta^{n-1} \in \mathbb{R}^{pT}$, sweeping pattern $(t_1,\ldots,t_T)$
\ENSURE $\beta^{n}$
\STATE $\hat{\beta} \leftarrow \hat{\beta}^{n-1}$
\STATE Determine fusion chains $C_1,\ldots, C_K$ and chain starts $T_1  \le \ldots \le T_K$
\FOR{$t=t_1,t_2,\ldots,t_T$}
\IF[chain start]{$t =  T_k$ for some $k$ and $T_{k+1}-T_{k} >1$}
\STATE $a\leftarrow T_k\, , \quad  b\leftarrow T_{k+1}-1$ 
\ELSIF[chain end]{$t =  T_{k+1}-1$ for some $k<K$}
\STATE $ a\leftarrow T_k\, , \quad b\leftarrow T_{k+2}-1$ 
\ELSE[chain interior]
\STATE Skip to next $t$
\ENDIF
\STATE Set $f(\beta_t) = \frac{1}{2} \sum_{s=a}^{b} \|   y_s - X_s \beta_t   \|_2^2  $
\STATE Set $g(\beta_t)= \lambda_1   (b-a+1)  \| \beta_t  \|_1 + 
\lambda_2 \big( w_{a-1}  \| \beta_t - \hat{\beta}_{a-1}  \|_2 +   w_{b}  \|\hat{\beta}_{b+1} - \beta_t  \|_2 \big) $
\STATE Check \eqref{fusion simple solution 1}-\eqref{fusion simple solution 2}-\eqref{fusion simple solution 3}
for a simple solution to $\min(f+g)$
\IF{simple solution}
\STATE 
$\beta_t^{+}\leftarrow   \hat{\beta}_{a-1}$ or $ \beta_t^{+} \leftarrow \hat{\beta}_{b+1}$ as required 
\ELSE[FISTA]
\STATE Apply Algorithm \ref{alg:FISTA constant step size} to $f+g$ 
with starting point $\hat{\beta}_t$, Lipschitz constant 
$L = \| \sum_{s=a}^{b} X_s'X_s \|_2$,  
and $\mathrm{prox}_{g/L}$ given by \eqref{fp operator chain}. Output $\beta_t^{+}$ 
\ENDIF
\STATE $\beta^{+} \leftarrow (\hat{\beta}_1,\ldots, \hat{\beta}_{a-1} , \beta_t^{+}, \ldots,\beta_t^{+}, \hat{\beta}_{b+1},\ldots , \hat{\beta}_T) \in\mathbb{R}^{pT}$
\IF[merge $C_k$ and $C_{k+1}$]{$b = T_{k+2}-1$ and $F(\beta^{+}) < F(\hat{\beta})$}
\STATE Remove $T_{k+1}$ from $\{ T_1, \ldots, T_{K}\}$, set $K \leftarrow K-1$, 
relabel chain starts as $T_1 \le \cdots \le T_K$, and set $T_{K+1}\leftarrow T+1$ 
\ENDIF
\STATE $\hat{\beta}_{s} \leftarrow \beta_t^{+}$ 
and $\beta_{s}^{n} \leftarrow \beta_t^{+}$ for $ T_k \le s < T_{k+1}$
\ENDFOR
\end{algorithmic}
\end{algorithm}

 \subsection{Fusion cycle: all chains} 
 
When no further reduction can be achieved in $F $ 
by changing a single block or single fusion chain 
in the current solution $\hat{\beta}\in\mathbb{R}^{pT}$,
a logical next step is to optimize $F$ with respect to \emph{all} fusion chains. 
Specifically, one identifies the fusion chains $C_k = \{ t: T_k \le t < T_{k+1}\}$  $(1\le k \le K)$ 
over which $\hat{\beta} = (\hat{\beta}_t)$ is constant 
and optimizes $F$ with respect to all blocks $\beta_t$ under the equality 
constraints induced by the fusion chains:  
\begin{equation*}
\min_{\beta_{T_1},\ldots,\beta_{T_{K}}\in\mathbb{R}^{p}} F(\beta_{T_1},\ldots,\beta_{T_{1}}, 
\ldots, \beta_{T_K},\ldots,\beta_{T_{K}})
\end{equation*}
with each $\beta_{T_k}$ repeated $n_k =T_{k+1}-T_k $ times. Explicitly, this amounts to  
 \begin{equation}\label{fixed chains}
 \begin{split}
& \min_{\beta_{T_1},\ldots,\beta_{T_{K}}} 
\bigg\{ \frac{1}{2} \sum_{k=1}^{K} \sum_{t = T_{k} }^{T_{k+1}-1} \left\| y_t - X_t \beta_{T_{k}}  \right\|_2^2
+ \lambda_1 \sum_{k=1}^K n_k   \| \beta_{T_k}  \|_1   \\
& \qquad \qquad + \lambda_2 \sum_{k=1}^{K-1} w_{T_{k+1}-1}  \| \beta_{T_{k+1}} - \beta_{T_{k}}  \|_2 \bigg\} \, .
 \end{split}
\end{equation}
To solve \eqref{fixed chains} we employ a version of FISTA slightly different from the one used in  
\eqref{block descent} and \eqref{optimization single fusion chain}. 
In particular this version (Algorithm \ref{alg:FISTA backtracking}) operates under the requirement that 
$\beta_{T_{k}} \ne \beta_{T_{k+1}} $ for all $k$. 
If two blocks $\beta_{T_{k}} $ and $\beta_{T_{k+1}} $ become equal during the optimization, 
the corresponding fusion chains $C_k$ and $C_{k+1}$  are merged 
and problem \eqref{fixed chains} is restarted. 
Details are given in Section \ref{sec: FISTA all chains}.

\subsection{Checking the optimality of a solution}
\label{sec: checking global optimality}

A vector $x\in \mathbb{R}^n$ ($n \ge 1$) minimizes  a convex function $f :  \mathbb{R}^n \to   \mathbb{R}$ 
if and only if $0_n$ is a \emph{subgradient} of $f$ at $x$. 
(The concept of subgradient generalizes the gradient to possibly nondifferentiable convex functions.) 
This expresses equivalently  as the membership of $0_n$ 
to the \emph{subdifferential} $\partial f (x)$, that is, the set of all subgradients of $f$ at $x$. 
Definition, basic properties, and examples of subgradients and subdifferentials can be found in textbooks on convex analysis, 
e.g. \cite{Rockafellar2015}.

In order to formulate the optimality conditions of the SGFL problem \eqref{objective}, we define the sign operator 
$$ \mathrm{sgn}(x) = \begin{cases} \{ 1\} & \textrm{if } x > 0 ,\\  \{ -1\} & \textrm{if } x< 0, \\  [-1,1] & \textrm{if } x= 0 ,\end{cases} $$  
for $x\in\mathbb{R}$ and extend it as a set-valued function from $\mathbb{R}^n $ to $\mathbb{R}^n $ in a componentwise fashion: 
 $( \mathrm{sgn}(x) )_i =  \mathrm{sgn}(x_i) $ ($1\le i \le n$). 
Now, a vector $\hat{\beta}=(\hat{\beta}_{1},\ldots,\hat{\beta}_{T}) \in \mathbb{R}^{pT}$  minimizes  $F$ 
if and only if $0_{pT}$ is a subgradient at $\hat{\beta}$,  that is, if and only if 
there exist  vectors $u_1,\ldots,u_T \in \mathbb{R}^{p}$ and 
$v_1,\ldots,v_{T-1}  \in \mathbb{R}^{p}$ satisfying
\begin{subequations}
\begin{equation}\label{global optimality}
 X_t' (X_{t} \hat{\beta}_{t} -   y_{t} ) + 
 \lambda_1  u_{t} + 
\lambda_2  ( w_{t-1} v_{t-1} - w_t v_{t}  ) = 0_{p} 
\end{equation}
 and
\begin{equation}\label{constraints u}
u_t \in \mathrm{sgn}(\hat{\beta}_t) 
\end{equation}
 for $1 \le t \le T$ as well as 
\begin{equation}\label{constraints v}
 \begin{cases}
 \displaystyle v_{t}  = \frac{ \hat{\beta}_{t+1} -  \hat{\beta}_{t} } { \|  \hat{\beta}_{t+1} -  \hat{\beta}_{t} \|_2  } & \textrm{ if } \hat{\beta}_{t} \ne \hat{\beta}_{t+1},\\
\| v_{t}  \|_2 \le 1 & \textrm{ if } \hat{\beta}_{t} = \hat{\beta}_{t+1},
 \end{cases}
 \end{equation}
for $1 \le t \le T-1$. 
 \end{subequations}
 By convention we take  $v_0 =v_T= 0_p$. 
Conditions \eqref{constraints u}-\eqref{constraints v} arise 
from the facts that the subdifferential of the $\ell_1$ norm is the sign operator
and that the subdifferential of the $\ell_2$ norm at $0_p$ 
is the $\ell_2$-unit ball of $\mathbb{R}^p$.

The optimality conditions \eqref{global optimality}-\eqref{constraints u}-\eqref{constraints v} 
can be checked by solving 
\begin{equation}\label{global optimality minimization}
\min_{U \in \mathcal{C}_1, V\in \mathcal{C}_2} \frac{1}{2} \left\| Z + \lambda_1 \alpha U + \lambda_2 VW D' \right\|_F^2 
\end{equation}
where $U=(u_1,\ldots,u_T)\in \mathbb{R}^{p\times T}$, $V=(v_1,\ldots,v_{T-1})\in \mathbb{R}^{p \times (T-1)}$, 
$Z=(z_1,\ldots,z_T)\in \mathbb{R}^{p\times T}$ with  $z_t =    X_t' (X_{t} \hat{\beta}_{t} -   y_{t} ) + \lambda_1 (1-\alpha)\hat{\beta}_{t} $, and $D\in \mathbb{R}^{T\times (T-1)}$ is the differencing matrix given by $(D)_{ij} = -1$ if $i=j$, $(D)_{ij} = 1$ if $i=j+1$, and $(D)_{ij} = 0$ otherwise. (Here we use matrix formalism to express \eqref{global optimality} more simply.) 
The sets $\mathcal{C}_1 $ and $\mathcal{C}_2$ embody the constraints \eqref{constraints u}  and 
\eqref{constraints v}, respectively. If the minimum of \eqref{global optimality minimization} is zero, then $0_{p T}$ 
is a subgradient of $F$ at $\hat{\beta}$ and $\hat{\beta}$  minimizes  $F$. In this case the optimization is over.

A closer examination of \eqref{global optimality}-\eqref{constraints u}-\eqref{constraints v}  
reveals that change points in $\hat{\beta} $ break the global problem \eqref{global optimality minimization} 
into independent subproblems. More precisely, let $T_2 < \ldots < T_K$ be the change points induced by $\hat{\beta}$ 
(assuming there is at least one) and $C_1,\ldots,C_K$ the associated segmentation of $\{ 1, \ldots, T\}$. 
The constraints  \eqref{constraints v} entirely determine the vectors $v_{T_k - 1}$  ($k \ge 2$), 
which breaks the coupling of the $v_t$ separated by change points in \eqref{global optimality}. 
On the other hand the constraints \eqref{constraints u} clearly affect each block $u_t$ separately. 
Therefore, problem \eqref{global optimality minimization} can be solved separately (and in parallel) on each fusion chain $C_k$. 
We tackle \eqref{global optimality minimization} on each $C_k$ using \emph{gradient projection}. We embed this method inside FISTA  for faster convergence. The necessary gradient calculation and projections on $\mathcal{C}_1 $ and $\mathcal{C}_2$ are 
described in Section \ref{sec: gradient projection}.

\subsection{Subgradient step}

If the attained minimum in \eqref{global optimality minimization} 
is greater than zero, then  $\hat{\beta}$ is not a minimizer of $F$. 
By design of the hybrid algorithm, 
 this implies that the segmentation $C_1,\ldots,C_K$ associated with $\hat{\beta}$ is suboptimal 
and that, starting from $\hat{\beta}$, $F$  cannot be further reduced at  
the first three levels of optimization.   
In this case, arguments $(U^{\ast},V^{\ast})$ that minimize 
\eqref{global optimality minimization} provide a subgradient 
$G = Z + \lambda_1 \alpha U^{\ast} + \lambda_2 V^{\ast}W D' $ of minimum norm. 
Denoting the vectorized version of $G$ by $g \in \mathbb{R}^{pT}$, 
the opposite of $g$ is a direction of steepest descent for $F$ at $\hat{\beta}$ (e.g. \cite{Shor1985}). 
Accordingly, at the fourth level of optimization, the algorithm takes a step in the direction $-g$ with 
step length obtained by exact line search. 
The updated solution expresses as $ \beta^{+} = \hat{\beta} - \alpha^{\ast} g  $ where $\alpha^{\ast} =\mathrm{argmin}_{\alpha > 0} F(\hat{\beta} - \alpha g)$.  
The subgradient step accomplishes two important things: first, it moves the optimization away from the suboptimal segmentation $C_1,\ldots,C_K$ and second, by reducing the objective, it ensures that this segmentation will not be visited again later in the optimization. 
These properties and their consequence, namely the global convergence of the hybrid algorithm  
will be established in Theorem \ref{thm: convergence} in Section \ref{sec: main algorithm}.


\section{Algorithm details}
\label{sec:detailed computations}

This section gives a detailed account of how optimization 
is carried out at each level (single block, single fusion chain, all fusion chains, all blocks) 
in  the main Algorithm \ref{alg:main}. We first present the fast iterative soft-thresholding algorithm 
(FISTA) of \cite{Beck2009} which we  extensively use in  Algorithm \ref{alg:main}.

\subsection{FISTA}
\label{sec:FISTA}

Beck and Teboulle \cite{Beck2009} consider the convex program  
$$ \min_{x\in \mathbb{R}^n} \left\{  f(x) + g(x) \right\} $$
where $f:\mathbb{R}^n \rightarrow \mathbb{R}$ is a smooth convex function
and $g:\mathbb{R}^n \rightarrow \mathbb{R}$ is a continuous convex function, possibly nonsmooth. 
Regarding the proposed hybrid algorithm for SGFL (Algorithm \ref{alg:main}),  
the specific functions $f,g$ used in applying  FISTA 
are given in Algorithm \ref{alg:BCD} for block-level optimization, 
Algorithm \ref{alg:BCD fusion} for optimization over a single chain, 
section \ref{sec: FISTA all chains} for optimization over all chains, 
and section \ref{sec: gradient projection} for verification of optimality.

The function $f$ is  assumed to be differentiable with Lipschitz-continuous gradient: 
$$ \left\| \nabla f(x) - \nabla f(y) \right\|_2 \le L \left\| x - y\right\|_2 $$
   for all $x,y \in\mathbb{R}^n$ and some finite Lipschitz constant $L>0$. 
The function $g$ is assumed to be proximable, 
that is, its proximal operator 
$ \mathrm{prox}_{\gamma g} (x) = \argmin_{y \in \mathbb{R}^n}  \big\{ g(x) + 1/(2\gamma) \| y - x\|^2  \big\} $
should be easy to calculate for all $\gamma > 0$. 

Their Iterative Soft Thresholding Algorithm (ISTA) is an iterative method that replaces at each iteration
the difficult optimization of the objective $f+g$ 
by the simpler optimization of a quadratic approximation $Q_L$. 
Given a suitable vector $y \in \mathbb{R}^n$, the goal is to minimize 
\begin{equation}\label{bd2}
Q_L(x ,  y ) =  f(y) + \nabla f(y) '  ( x-y  ) + g(x) + \frac{L}{2} \left\|x-y \right\|_2^2 . 
\end{equation}
with respect to $x \in \mathbb{R}^n$. 
With a few algebraic manipulations and omitting irrelevant additive constants, 
$Q_L$ can be rewritten as
\begin{equation*}
Q_L(x ,  y ) =  g(x) + \frac{L}{2}\,  \left\| x- \left(y - \frac{1}{L}\nabla f(y)\right) \right\|_2^2  
\end{equation*}
so that 
$ \mathrm{argmin}_{x} Q(x,y) = \mathrm{prox}_{g/L} \left(y - (1/L) \nabla f(y) \right)  $. 
In other words, the minimization of $Q_L$ is achieved through a gradient step with respect to $f$ 
followed by a proximal step with respect to $g$. ISTA can thus be viewed as a proximal gradient method, 
also known as forward-backward method (e.g. \cite{Combettes2011}).  
Observing that $Q_L (\cdot, y)$ majorizes $f+g$, ISTA can also be viewed as a majorization-minimization method.  
 Formally, an  iteration of ISTA expresses as $x^{n+1} = \mathrm{argmin}_{x} Q_L( x,x^{n}) = \mathrm{prox}_{g/L}  \left(x^n - (1/L) \nabla f(x^n) \right)$.

Proximal gradient methods are not new: they have been used for decades. 
The innovation of \cite{Beck2009}  is to accelerate the convergence of ISTA 
by introducing an auxiliary sequence $(y^{k})$ such that $y^{k}$ is a well-chosen   
linear combination of $x^{k-1}$ and $x^{k}$, the main solution iterates 
(see also \cite{Becker2011,Nesterov2005} for related work). With this technique, the convergence 
rate of proximal gradient improves from $O(1/k)$ to $O(1/k^2)$. 
Algorithm \ref{alg:FISTA constant step size} presents this accelerated method, called Fast 
Iterative Soft Thresholding Algorithm (FISTA), 
in the case where a Lipschitz constant $L$ is prespecified  and kept constant through iterations. 
Algorithm \ref{alg:FISTA backtracking} presents FISTA in the case where $L$ is difficult to determine ahead of time and is chosen by backtracking at each iteration. This version of FISTA requires an initial guess $L^{0}$ for the Lipschitz constant as well as a factor $\eta > 1$ by which to increase the candidate value $L$ in backtracking steps. 

\begin{algorithm}[ht]
\caption{FISTA with constant step size}
\label{alg:FISTA constant step size}
\begin{algorithmic}
\REQUIRE $x^{0} \in \mathbb{R}^{n}$, Lipschitz constant $L >0 $
\ENSURE $x^{k} $
\STATE $y^{1} \leftarrow x^{0} $, $\alpha_{1} \leftarrow 1$
\FOR{$k = 1,2,\ldots $}
\STATE $x^{k} \leftarrow   \mathrm{prox}_{g/L}\left(y^{k} - \frac{1}{L} \nabla f(y^{k})  \right)  $ \smallskip
\STATE $\alpha_{k+1} \leftarrow \frac{1 + \sqrt{1 + 4(\alpha_{k})^2}}{2} $ \smallskip
\STATE $ y^{k+1} \leftarrow x^{k} + \left( \frac{\alpha_{k} - 1}{\alpha_{k+1} }\right) \left( x^{k} - x^{k-1} \right)$
\ENDFOR
\end{algorithmic}
\end{algorithm}

\begin{algorithm}[ht]
\caption{FISTA with backtracking}
\label{alg:FISTA backtracking}
\begin{algorithmic}
\REQUIRE $x^{0} \in \mathbb{R}^{n}$, $L^{0}>0$, $\eta > 1 $
\ENSURE $x^{k}$
\STATE $y^{1} \leftarrow x^{0} $, $\alpha_{1} \leftarrow 1$
\FOR{$k = 1,2,\ldots $}
\STATE $i \leftarrow 0$
\REPEAT 
\STATE $L  \leftarrow \eta^{i} L^{k-1} $
\STATE $x^{k} \leftarrow   \mathrm{prox}_{g/L}\left( y^{k} - (1/L)\nabla f(y^{k})\right)  $ \smallskip
\STATE $i \leftarrow i+1$
\UNTIL{ $  (f+g) (x^{k}) \le Q_L (x^{k} , y^{k}) $} \smallskip
\STATE $L^{k} \leftarrow L$
\STATE $\alpha_{k+1} \leftarrow \frac{1 + \sqrt{1 + 4(\alpha_{k})^2}}{2} $ \smallskip
\STATE $ y^{k+1} \leftarrow x^{k} + \left( \frac{\alpha_{k} - 1}{\alpha_{k+1} }\right) \left( x^{k} - x^{k-1} \right)$
\ENDFOR
\end{algorithmic}
\end{algorithm}

\subsection{Iterative soft-thresholding}
\label{sec:IST}

In this section we present a novel iterative soft-thresholding algorithm 
for computing the proximal operators required in the application of FISTA 
to problems \eqref{block descent} and \eqref{optimization single fusion chain}.
We first examine the case of \eqref{block descent} (block coordinate descent) and then show how to 
adapt the algorithm to  \eqref{optimization single fusion chain} 
(optimization of $F$ with respect to a single fusion chain).   
Of crucial importance is the soft-thresholding operator 
  $$  S(x , \lambda)   =
  \begin{cases} 
x + \lambda ,  &\textrm{if } x < -\lambda,  \\
0, &\textrm{if }  | x | \le \lambda ,\\
x - \lambda , &\textrm{if } x > \lambda,
\end{cases} 
 $$   
where $x\in \mathbb{R}$ and $\lambda \ge 0$ is a threshold. 
This operator accommodates vector arguments 
$x \in \mathbb{R}^p$ in a componentwise fashion: 
$(S(x,\lambda))_ i = S(x_i , \lambda)$  ($1\le i \le p$).

\paragraph{Checking for simple solutions.}

It is advantageous to verify whether $\hat{\beta}_{t-1}$ 
or $ \hat{\beta}_{t+1}$ solves  \eqref{block descent} 
before applying FISTA, which is more computationally demanding. 
The optimality conditions for \eqref{block descent} 
are very similar to those for the global problem \eqref{objective}, 
namely  \eqref{global optimality}-\eqref{constraints u}-\eqref{constraints v},  
although of course the conditions for \eqref{block descent} pertain to a single time $t$. 
Hereafter we state these conditions in an easily computable form. 
Let $\phi : \mathbb{R}^p \times \mathbb{R}^p \times \mathbb{R}_{+} \to \mathbb{R}^p$ 
be defined in a componentwise fashion by  
$$ (\phi (x,s,\lambda) )_{i} = 
\begin{cases} 
x_i + \lambda_i &  \textrm{if } s_i > 0, \\
x_i - \lambda_i &  \textrm{if } s_i < 0 ,\\
S(x_i , \lambda_i )&  \textrm{if } s_i = 0. \\
\end{cases} 
$$
If $\hat{\beta}_{t-1} = \hat{\beta}_{t+1} $, this vector solves \eqref{block descent} if and only if 
\begin{equation}\label{local simple solution 1}
 \big\| \phi\big(X_t' (X_t \hat{\beta}_{t-1} - y_t)  , \hat{\beta}_{t-1} , \lambda_1  \big) \big\|_2 \le \lambda_2 (w_{t-1} + w_{t} ).
\end{equation}
If $\hat{\beta}_{t-1} \ne \hat{\beta}_{t+1} $, 
$\hat{\beta}_{t-1}$ solves \eqref{block descent} if and only if 
\begin{equation}\label{local simple solution 2}
\hspace*{-3mm} \bigg\| \phi\bigg(X_t' (X_t \hat{\beta}_{t-1} - y_t) 
 +  \lambda_2 w_t \frac{\hat{\beta}_{t-1} - \hat{\beta}_{t+1}} { \| \hat{\beta}_{t-1} - \hat{\beta}_{t+1} \|_2 } 
  , \hat{\beta}_{t-1} , \lambda_1  \bigg) \bigg\|_2 \le \lambda_2 w_{t-1} 
\end{equation}
and $\hat{\beta}_{t+1}$ solves \eqref{block descent} if and only if 
\begin{equation}\label{local simple solution 3}
\hspace*{-4mm} \bigg\| \phi\bigg(X_t' (X_t \hat{\beta}_{t+1} - y_t) 
 +  \lambda_2 w_{t-1} \frac{\hat{\beta}_{t+1} - \hat{\beta}_{t-1}} { \| \hat{\beta}_{t+1} - \hat{\beta}_{t-1} \|_2 }  , 
 \hat{\beta}_{t+1} , \lambda_1  \bigg) \bigg\|_2 \le \lambda_2  w_{t} .
\end{equation}

\paragraph{Fixed point iteration.} 
After verifying  that neither $\hat{\beta}_{t-1} $ nor $ \hat{\beta}_{t+1} $ is a solution of \eqref{block descent}, 
we apply Algorithm \ref{alg:FISTA constant step size} (FISTA with constant step size) 
to  \eqref{block descent} using the decomposition 
\begin{equation}\label{fg BCD}
\left\{ \begin{array}{l}
f(\beta_t) =\displaystyle  \frac{1}{2} \left\| y_t - X_t \beta_t \right\|_2^2 , \\
g(\beta_t) = \lambda_1  \| \beta_t \|_1 + \lambda_2 \big(w_{t-1}\| \beta_t - \hat{\beta}_{t-1} \|_2 + w_t \| \beta_t - \hat{\beta}_{t+1} \|_2 \big)\, .
\end{array} \right. 
\end{equation}
 The gradient of the smooth component $f$ is $\nabla f(\beta_t) = X_t' (X_t \beta_t - y_t)  $ 
with Lipschitz constant $L_t = \| X_t' X_t \|_2 $. The main task is to calculate the proximal operator of $g$. 
Given a vector $z_t\in\mathbb{R}^p$, we seek 
\begin{equation}\label{prox g}
\mathrm{prox}_{g/L_t}(z_t) = \mathrm{argmin}_{\beta_t \in \mathbb{R}^p} g(\beta_t) + \frac{L_t}{2} \left\| \beta_t - z_t \right\|_2^2 .
\end{equation}
 The optimality conditions for this problem are 
\begin{equation}\label{optimality prox}
0_p \in L_{t} \left( \beta_t - z_t \right) + 
\lambda_1  \, \mathrm{sgn}(  \beta_t ) + \frac{ \lambda_2  w_{t-1}}{ \| \beta_t - \hat{\beta}_{t-1}  \|_2}  (  \beta_t - \hat{\beta}_{t-1} ) + 
 \frac{ \lambda_2 w_t} { \| \beta_t - \hat{\beta}_{t+1} \|_2 }  ( \beta_t - \hat{\beta}_{t+1}  )  
\end{equation}
or equivalently 
\begin{equation*}
\begin{split}
 & \left(L_t  + \frac{  \lambda_2 w_{t-1} }{\big\| \beta_t - \hat{\beta}_{t-1} \big\|_2} + 
 \frac{  \lambda_2 w_{t} } {\big\| \beta_t - \hat{\beta}_{t+1} \big\|_2 }
  \right) 
\beta_t  \\
& \qquad \in  \left( L_t z_t  +  \frac{ \lambda_2 w_{t-1} \hat{\beta}_{t-1} }{ \| \beta_t - \hat{\beta}_{t-1} \|_2} + 
  \frac{\lambda_2  w_t \hat{\beta}_{t+1} } { \| \beta_t - \hat{\beta}_{t+1} \|_2 } \right)
- \lambda_1 \alpha \, \mathrm{sgn}(  \beta_t ) \,.
\end{split}
\end{equation*}
Given $\hat{\beta}_{t-1}$, $\hat{\beta}_{t+1}$ and $z_t$, 
we define the  operator 
\begin{equation}\label{fp operator}
 \mathcal{T}(\beta_t ) = \frac{\displaystyle S\left(  L_t z_t + 
 \frac{ \lambda_2 w_{t-1}  \hat{\beta}_{t-1}}{ \| \beta_t - \hat{\beta}_{t-1}  \|_2} + 
  \frac{\lambda_2  w_t \hat{\beta}_{t+1}} { \| \beta_t - \hat{\beta}_{t+1}  \|_2 } \, 
, \, \lambda_1  \right) }
{  \displaystyle L_t  + \frac{  \lambda_2 w_{t-1} }{ \| \beta_t - \hat{\beta}_{t-1}  \|_2} + 
 \frac{  \lambda_2 w_{t} } { \| \beta_t - \hat{\beta}_{t+1}  \|_2 }
} 
\end{equation}
for $\beta_t \in \mathbb{R}^{p}\setminus \{ \hat{\beta}_{t-1}, \hat{\beta}_{t+1}\}$ and extend it by continuity: 
 $\mathcal{T}(\hat{\beta}_{t-1} )=  \hat{\beta}_{t-1}$ and $\mathcal{T}(\hat{\beta}_{t+1} )=  \hat{\beta}_{t+1}$. 
The optimality conditions \eqref{optimality prox} now express as the fixed point equation 
\begin{equation*}
\mathcal{T}(\beta_t) = \beta_t   .
\end{equation*}
The operator $\mathcal{T}$ admits the fixed points $\hat{\beta}_{t-1}$, 
$\hat{\beta}_{t+1}$, and $\mathrm{prox}_{g/L_t}(z_t)$. 
It can be shown that if 
$\mathrm{prox}_{g/L_t}(z_t) \notin \{\hat{\beta}_{t-1},\hat{\beta}_{t+1}\}$,  
the fixed points $\hat{\beta}_{t-1}$ and $\hat{\beta}_{t+1}$
are \emph{repulsive} 
in the sense that there exist $\eta, \epsilon>0$ such that 
 $\| \mathcal{T}(\beta_t) - \hat{\beta}_{t- 1} \|_2 
\ge (1+\epsilon) \| \beta_t - \hat{\beta}_{t - 1} \|_2$
for $\| \beta_t - \hat{\beta}_{t- 1} \| \le \eta$ (same for $\hat{\beta}_{t+1}$). 
 This suggests calculating
 $\mathrm{prox}_{g/L_t}(z_t)$ with 
 the  iterative soft-thresholding   
 \begin{equation}\label{fp iteration}
\beta_t^{n+1} =   \mathcal{T}(\beta_t^{n} ) \, .
\end{equation}

\medskip

\begin{remark}[proximal gradient]
\label{fixed point as forward-backward}
The fixed point iteration \eqref{fp operator}-\eqref{fp iteration} can be viewed as a proximal gradient algorithm. 
Writing 
$g_1(\beta_t) = \lambda_1   \| \beta_t \|_1$ and 
$g_2(\beta_t) = \lambda_2 w_{t-1}  \| \beta_t - \hat{\beta}_{t-1}  \|_2  +\lambda_2w_t \| \beta_t- \hat{\beta}_{t+1}   \|_2  + (L_t/2) \|  \beta_t -z_t \|_2^2 $, it holds that 
\begin{equation}\label{fp prox grad}
 \mathcal{T}(\beta_t) =  \mathrm{prox}_{\gamma g_1 }\big(\beta_t - \gamma \nabla g_2(\beta_t) \big) \,  \ 
\textrm{with }  \frac{1}{\gamma } = L_t + \frac{\lambda_2 w_{t-1} }{ \| \beta_t - \hat{\beta}_{t-1} \|_2 } +
\frac{\lambda_2 w_t }{ \| \beta_{t} - \hat{\beta}_{t+1}  \|_2 } \ . 
\end{equation}
\end{remark}
\medskip
\begin{remark}[Weiszfeld's algorithm]
The fixed point iteration \eqref{fp operator}-\eqref{fp iteration} is related in spirit to Weiszfeld's algorithm \cite{Weiszfeld2009} and its generalizations (e.g. \cite{Kuhn1973}) for the Fermat-Weber location problem 
 $\argmin_{y\in \mathbb{R}^p} \sum_{i=1}^m w_i\| y - x_i\|_2$, 
 where $x_1,\ldots,x_m\in \mathbb{R}^p$ and $w_1,\ldots,w_m >0$ are weights.  
Weiszfeld's algorithm, in its generalized version, has iterates of the form 
$$y^{n+1} = \bigg( \sum_{i=1}^m \frac{w_i  x_i}{\| y^{n}-x_i\|_2} \bigg) \bigg/ 
\bigg( \sum_{i=1}^m \frac{w_i  }{\| y^{n}-x_i\|_2}\bigg) $$ 
and is derived along the same lines as \eqref{fp operator}-\eqref{fp iteration}, 
namely by equating the gradient to zero and turning 
this equation into a fixed point equation.  
\end{remark}

By exploiting a connection to proximal gradient methods (\ref{fixed point as forward-backward}) and adapting the results of \cite{Bredies2008} to a nonsmooth setting, we can establish the linear convergence of \eqref{fp operator}-\eqref{fp iteration}. We defer the proof of this result to Section \ref{sec: linear convergence}. For convenience, let us denote the proximal operator $\mathrm{prox}_{g/L_t}(z_t) $ by $\beta_t^\ast$ and the associated objective function by $ \bar{g} (\beta_t) = g(\beta_t) + (L_t /2) \left\| \beta_t - z_t \right\|_2^2 $. We also define the distance $r_n = \bar{g}(\beta_t^{n}) - \bar{g}(\beta_t^\ast)$ to the minimum of $\bar{g}$.

\begin{theorem}\label{thm: linear convergence}
Assume that $\hat{\beta}_{t-1}$ and $\hat{\beta}_{t+1}$ are not solutions of   \eqref{block descent}, 
that $\beta_t^\ast \notin \{ \hat{\beta}_{t-1},\hat{\beta}_{t+1}\} $, 
and that the sequence $(\beta_t^{n})_{n \ge 0 }$ generated by 
\eqref{fp operator}-\eqref{fp iteration} has its first term satisfying 
$ \bar{g} (\beta_{t}^{0} )   < \min  (\bar{g} (\hat{\beta}_{t-1} )  , \bar{g} (\hat{\beta}_{t+1} )  ) $. Then 
  the distance $(r_n)_{n\ge 0}$ vanishes exponentially and 
   $(\beta_t^{n})_{n \ge 0 }$ 
 converges linearly to $\beta_t^\ast $, 
that is, there exist constants $C>0$ and $\lambda \in [0,1)$ such that 
$$ \big\|  \beta_{t}^{n} - \beta_t^\ast \big\|_2 \le C \lambda^n . $$ 
\end{theorem}

The first two assumptions of  Theorem \ref{thm: linear convergence} ensure 
that use of the iterative soft-thresholding \eqref{fp operator}-\eqref{fp iteration} is 
warranted, in other words, that \eqref{block descent} and \eqref{prox g} do  not have simple solutions. 
The condition on the starting point $\beta_t^0$ guarantees that the sequence 
$(\beta_t^n)$ does not get stuck in $\hat{\beta}_{t-1}$ or $\hat{\beta}_{t+1}$. 
It is standard for this type of problem, see e.g. \cite{Kuhn1973}. 
In practice this condition is virtually always met by taking the current FISTA iterate as starting point.

\paragraph{Extension to fusion chains.} 
When considering problem \eqref{optimization single fusion chain} over a fusion chain 
$C=\{ t: a \le t \le b\}$, the objective decomposes as
\begin{equation*}
\left\{ 
\begin{array}{l} \medskip
f(\beta_t)  = \displaystyle \frac{1}{2} \sum_{s=a}^{b} \left\| y_s - X_s \beta_{t} \right\|_2^2 , \\ 
g(\beta_{t}) =\displaystyle \lambda_1 n_C    \| \beta_{t} \|_1  +
\lambda_2 \left( w_{a-1} \|  \beta_{t} - \hat{\beta}_{a-1} \|_2 + w_{b} \|  \beta_{t} - \hat{\beta}_{b+1} \|_2 \right)  ,
\end{array}
\right.
\end{equation*}
where $n_C = b-a+1$. 
The conditions for $\hat{\beta}_{a-1}$ or $\hat{\beta}_{b+1}$ to be simple solutions of \eqref{optimization single fusion chain} 
are as follows. 
If $\hat{\beta}_{a-1} = \hat{\beta}_{b+1} $, this vector solves \eqref{optimization single fusion chain} if and only if 
\begin{equation}\label{fusion simple solution 1}
 \big\| \phi\big( \sum_{s=a}^{b} X_s' (X_s \hat{\beta}_{a-1} - y_s)  , \hat{\beta}_{a-1} , \lambda_1  \big) \big\|_2 \le \lambda_2 (w_{a-1} + w_{b} ).
\end{equation}
If $\hat{\beta}_{a-1} \ne \hat{\beta}_{b+1} $, 
$\hat{\beta}_{a-1}$ solves \eqref{optimization single fusion chain} if and only if 
\begin{equation}\label{fusion simple solution 2}
\hspace*{-3mm} \bigg\| \phi\bigg( \sum_{s=a}^b X_s' (X_s \hat{\beta}_{a-1} - y_s) 
 +  \lambda_2 w_{b} \frac{\hat{\beta}_{a-1} - \hat{\beta}_{b+1}} { \| \hat{\beta}_{a-1} - \hat{\beta}_{b+1} \|_2 } 
  , \hat{\beta}_{a-1} , \lambda_1  \bigg) \bigg\|_2 \le \lambda_2 w_{a-1} 
\end{equation}
and $\hat{\beta}_{b+1}$ solves \eqref{optimization single fusion chain} if and only if 
\begin{equation}\label{fusion simple solution 3}
\hspace*{-4mm} \bigg\| \phi\bigg( \sum_{s=a}^b X_s' (X_s \hat{\beta}_{b+1} - y_s) 
 +  \lambda_2 w_{a-1} \frac{\hat{\beta}_{b+1} - \hat{\beta}_{a-1}} { \| \hat{\beta}_{b+1} - \hat{\beta}_{a-1} \|_2 }  , 
 \hat{\beta}_{b+1} , \lambda_1  \bigg) \bigg\|_2 \le \lambda_2  w_{b} .
\end{equation}

If there are no simple solutions to \eqref{optimization single fusion chain},  
we apply Algorithm \ref{alg:FISTA constant step size} to $f+g$.  
The gradient step is given by $\nabla f (\beta_t) = \sum_{s=a}^{b} X_s'\left(   X_s \beta_{t}- y_s\right) $
and its Lipschitz constant $L_{C} = \| \sum_{s=a}^{b} X_s'   X_s \|_2$. 
For a given $z_t \in \mathbb{R}^p$, 
the proximal operator $\mathrm{prox}_{g/L_{C}}(z_t)$ is calculated 
by iteratively applying the soft-thresholding operator 
\begin{equation}\label{fp operator chain}
 \mathcal{T}_{C}(\beta_t ) = \frac{\displaystyle S\left(  L_{C} z_t + 
 \frac{ \lambda_2 w_{a-1} \hat{\beta}_{a-1}}{\big\| \beta_t - \hat{\beta}_{a-1} \big\|_2} + 
  \frac{\lambda_2  w_b \hat{\beta}_{b+1}} {\big\| \beta_t - \hat{\beta}_{b+1} \big\|_2 } \, 
, \ \lambda_1  n_C \right) }
{  \displaystyle L_{C}  + \frac{  \lambda_2 w_{a-1} }{\big\| \beta_t - \hat{\beta}_{a-1} \big\|_2} + 
 \frac{  \lambda_2 w_{b} } {\big\| \beta_t - \hat{\beta}_{b+1} \big\|_2 }
} \, .
  \end{equation}

\subsection{Optimization over all fusion chains}
\label{sec: FISTA all chains}

The optimization \eqref{fixed chains} is carried out 
by applying  Algorithm \ref{alg:FISTA backtracking} (FISTA with backtracking)  
to $\min_{\beta \in \mathbb{R}^{pK}} (f+g)(\beta) $ 
where 
 \begin{equation*}
 \left\{ 
 \begin{array}{l }
 f(\beta)  = \displaystyle
 \frac{1}{2} \sum_{k=1}^{K} \sum_{t = T_{k} }^{T_{k+1}-1} \left\| y_t - X_t \beta_{k}  \right\|_2^2
+  \lambda_2 \sum_{k=1}^{K-1} w_{T_{k+1}-1}  \big\| \beta_{k+1} - \beta_{k}  \big\|_2 \, , \\
 g(\beta)  \displaystyle = \lambda_1 \sum_{k=1}^K n_k    \left\| \beta_{k}  \right\|_1 .
\end{array}\right.
\end{equation*}
For notational convenience, we have relabeled the vectors $\beta_{T_1},\ldots, \beta_{T_{K}}$ of \eqref{fixed chains} 
 as $\beta_{1},\ldots, \beta_{K}$. 
Observe that $f$ is nondifferentiable at points $\beta = (\beta_{1},\ldots, \beta_{K})$ such that $\beta_{k} = \beta_{k+1}$ for some $k$, which  violates the   smoothness  requirements of  Section \ref{sec:FISTA}. We can nonetheless apply FISTA until the algorithm either converges to a minimizer of $f+g$ or to a point of nondifferentiability for $f$. 
In the latter case, we merge the fusion chains $C_k$ and $C_{k+1}$ associated with the equality $\beta_k = \beta_{k+1}$ 
and restart FISTA with the reduced set of chains.  

To fully specify the FISTA implementation, it remains to characterize the gradient of $f$ and proximal operator of $g$.  
The former, wherever it exists, is given by ($1\le k \le K$)
\begin{equation}\label{gradient multichain}
\begin{split}
\frac{\partial f }{\partial \beta_k} (\beta)  & =  
 \sum_{t = T_{k} }^{T_{k+1}-1} X_t' (X_t \beta_{k} - y_t) 
 + \lambda_2  w_{T_{k}-1}  \frac{\beta_{k} - \beta_{k-1}}{\| \beta_{k} - \beta_{k-1} \|_2}\\
&\qquad + \lambda_2  w_{T_{k+1}-1}  \frac{\beta_{k} - \beta_{k+1}}{\| \beta_{k} - \beta_{k+1} \|_2} \, .
\end{split}
\end{equation}
The proximal operator of $g$ performs soft-thresholding by block ($1\le k \le K$): 
\begin{equation}
\big(\mathrm{prox}_{g/L}(\beta)\big)_k = S\left(\beta_k, \lambda_1  n_k\right) . 
\end{equation}

\subsection{Gradient projection method} 
\label{sec: gradient projection}

Here we describe the method of Section \ref{sec: checking global optimality}
to check the optimality of a solution $\hat{\beta}$. For simplicity, we move the regularization parameters 
$\lambda_1,\lambda_2$ and diagonal weight matrix $W$ from the objective in  \eqref{global optimality minimization}  
to the constraint sets $\mathcal{C}_1$ and $\mathcal{C}_2$. This is done with a simple change of variables. 
 
 \paragraph{Gradient step.} Writing the objective as 
$f(U,V) = \frac{1}{2} \left\| Z + U +  V D' \right\|_F^2$, the gradient of $f$ is given by 
\begin{equation}\label{gradient Frobenius}
\frac{\partial f}{\partial U}(U,V) = U +   VD' +  Z  ,\quad   \frac{\partial f}{\partial V}(U,V) =  
 U D + V D'D +  ZD\,.
\end{equation}
Therefore a Lipschitz constant $L$ of $\nabla f (U,V)$  
can be found by evaluating the spectral norm of the $(2T-1)\times (2T-1)$ matrix
$$\left( \begin{matrix} I_{T} & D \\ D' & D'D \end{matrix}\right) . $$ 
Standard calculations show that this matrix has
 spectral norm $1 + \| D'D \|_2$ and that the eigenvalues of $D'D$ are 
$\big\{ 2\big(1-\cos\big(\frac{(2t-1)\pi}{2T}\big)\big), \,1\le t \le T\big\} $. Combining these results, 
one can take $L=5$. 

\paragraph{Projection step.} 
The orthogonal projection $P_{\mathcal{C}_1}(U) $ of $U\in \mathbb{R}^{p \times T}$ 
on $\mathcal{C}_1$ is obtained by applying fixed coefficient constraints 
and clamping values to the interval $[ -\lambda_1  , \lambda_1 ]$ where needed.   
Its coefficients ($1\le t \le T, \, 1\le i \le p$) are given by 
\begin{equation}\label{projection: clamping}
(P_{\mathcal{C}_1}(U))_{it} = 
\begin{cases} 
 \lambda_1  &\textrm{if } (\hat{\beta}_t)_i > 0, \\
 -  \lambda_1  & \textrm{if } (\hat{\beta}_t)_i < 0, \\
 \min( \max((u_t)_i ,- \lambda_1 ) , \lambda_1 ) &\textrm{if } (\hat{\beta}_t)_i = 0 .
\end{cases}
\end{equation}
The orthogonal projection $P_{\mathcal{C}_2}(V) $ 
 of $V \in \mathbb{R}^{p \times (T-1)}$ on $\mathcal{C}_2$ 
 is obtained by rescaling the columns of $V$ ($1\le t < T$) as necessary: 
 \begin{equation}\label{projection: rescaling}
(P_{\mathcal{C}_2}(V))_t = \min\left( \frac{ \lambda_2 w_t}{\| v_t\|_2 } , \, 1\right) v_t  .
\end{equation}

Writing $I_{\mathcal{C}}$ for the indicator function of a set $\mathcal{C}$ 
($I_{\mathcal{C}}(x) = 0$ if $x\in\mathcal{C}$ and $I_{\mathcal{C}}(x)=+\infty$ otherwise) 
and $g(U,V)= I_{\mathcal{C}_1}  (U) + I_{\mathcal{C}_2}  (V)$, 
the constrained problem \eqref{global optimality minimization} reformulates as $\min  (f+g)$.
We can now apply FISTA ( Algorithm \ref{alg:FISTA constant step size})
 to solve this problem with the gradient step given by \eqref{gradient Frobenius} 
 and the Lipschitz constant $L=5$ and the proximal step 
 $\mathrm{prox}_{g/L}(U,V) = P_{\mathcal{C}_1}(U) + P_{\mathcal{C}_2}(V)$ given by 
 \eqref{projection: clamping}-\eqref{projection: rescaling}.

\subsection{Main algorithm} 
\label{sec: main algorithm}

Having presented all the components of the hybrid algorithm, we now collect them in Algorithm \ref{alg:main}, the main algorithm of the paper. The global convergence of Algorithm \ref{alg:main} is stated hereafter while its proof is deferred to Appendix \ref{Appendix B}. 

\begin{algorithm}[ht!]
\caption{Sparse Group Fused Lasso} 
\label{alg:main} 
\begin{algorithmic}
\REQUIRE Starting point $\beta^{0} \in \mathbb{R}^{pT}$, regularization parameters $\lambda_1 \ge 0 , \lambda_2 \ge 0$, 
tolerance $\epsilon > 0$ 
\ENSURE $\beta^{n}$
\STATE $ progressDescent \leftarrow $ \TRUE, $progressFusion \leftarrow $ \TRUE
\STATE $n \leftarrow 0$ 

\REPEAT
\WHILE{$ progressDescent = $ \TRUE}
\STATE $n \leftarrow n+1$
\STATE Apply  Algorithm \ref{alg:BCD} to $\beta^{n-1}$ and output $\beta^{n}$
\COMMENT {Block descent}
\IF {$F(\beta^{n}) \ge (1-\epsilon)F(\beta^{n-1}) $}
\STATE $ progressDescent \leftarrow $ \FALSE
\ENDIF
\ENDWHILE

\WHILE{$progressFusion =$ \TRUE}
\STATE $n \leftarrow n+1$
\STATE Apply  Algorithm \ref{alg:BCD fusion} to $\beta^{n-1}$ and output $\beta^{n}$ 
\COMMENT {Fusion: single chains}
\IF {$F(\beta^{n}) < (1-\epsilon)F(\beta^{n-1}) $}
\STATE $ progressDescent \leftarrow $ \TRUE , $ progressFusion \leftarrow $ \TRUE
\ELSE 
\STATE $ progressFusion \leftarrow $ \FALSE
\ENDIF
\ENDWHILE

\IF{$progressDescent =$ \FALSE \ \AND $progressFusion =$ \FALSE}
\STATE $n \leftarrow n+1$
\STATE Apply  Algorithm \ref{alg:FISTA backtracking} to  $\beta^{n-1}$  as in Section \ref{sec: FISTA all chains} 
and output  $ \beta^{n}$
\COMMENT {Fusion: all chains}
\STATE Apply  Algorithm \ref{alg:FISTA constant step size} to $ \beta^{n}$ 
as in Section \ref{sec: gradient projection} 
and output subgradient $g \in \mathbb{R}^{pT}$

\IF{ $ g \ne 0_{pT} $} 
\STATE $n\leftarrow n+1$
\STATE  $\alpha^{\ast} \leftarrow \mathrm{argmin}_{\alpha > 0} F(\beta^{n-1} - \alpha g )$
\STATE $\beta^{n} \leftarrow \beta^{n-1} - \alpha^{\ast} g$ 
\COMMENT {Subgradient step}
\ENDIF 
\ENDIF
\UNTIL{$F(\beta^{n}) \ge (1-\epsilon)F(\beta^{n-1}) $} 
\end{algorithmic}
\end{algorithm}

\begin{theorem}\label{thm: convergence}
For any starting point $\beta^0 \in \mathbb{R}^{pT}$,  
the sequence $(\beta^n)_{n\ge 0} $ generated by Algorithm \ref{alg:main} 
converges to a global minimizer 
of $F$ and the objective sequence $(F(\beta^n))_{n\ge 0 }$ decreases to 
the minimum of $F$ as the tolerance $\epsilon $ decreases to  0.
\end{theorem}


\section{Numerical experiments}
\label{sec:experiments}

\subsection{Simulations: optimization performance}
\label{sec: simulations optimization}

A simulation study was carried out to compare the proposed hybrid approach to SGFL with state-of-the-art optimization methods. The main focus here is on computational speed. Indeed, high-accuracy solutions are not needed in typical applications of SGFL; it is sufficient to correctly identify the optimal model segmentation and the sparsity structure of the minimizer of \eqref{objective}. 
Two sweeping patterns were examined for the hybrid approach: cyclical (HYB-C) and simple random sampling without replacement (HYB-R).

 \paragraph{Benchmark methods.} 

We provide a brief overview of the optimization methods used as benchmarks for the hybrid method. 
We refer the reader to the articles mentioned below for full details. 

\medskip

\begin{itemize}

\item \emph{Smoothing proximal gradient} (SPG) \cite{Chen2012b,Hadj-Selem2018,Kim2012}. 
This method deals with structured penalized regression problems 
where the penalty term admits a simple dual formulation, for example, 
group lasso and fused lasso.
The idea of SPG is to add quadratic regularization to the dual expression of the penalty 
and to solve the smooth approximate problem by FISTA \cite{Beck2009}.  In the context of SGFL, the objective \eqref{objective} is approximated by  $\min_{\beta} \max_{\alpha} \break \{ \frac{1}{2} \sum_t \| X_t\beta_t - y_t\|_2^2 + \lambda_1 \sum_t \| \beta_t\|_1 +\lambda_2  \sum_t ( \alpha_t ' (\beta_{t+1} -\beta_t) - \mu \|\alpha_t \|_2^2 )\}$ where $\beta \in\mathbb{R}^{pT} $, $\alpha\in\mathbb{R}^{p(T-1)}$,  $\| \alpha_t\|_2 \le w_t$ for all $t$, and $\mu>0$ is a regularization parameter.

\smallskip

\item \emph{Primal-dual method} (PD) \cite{Condat2013,Vu2013}. 
This method pertains to the general convex optimization problem $\min_{x} f(x) + g(x) + (h\circ L) (x)$ where $f$ is a smooth function, $g$ and $h$ are proximable functions, and $L$ is a linear operator. In SGFL, $f$ is taken to be the squared loss, $g$ the lasso penalty, $h$ the mixed $\ell_{2,1}$ norm, and $L$ the first-order differencing operator. At each iteration, the algorithm essentially requires a few matrix-vector multiplications and two easy evaluations of proximal operators: soft-thresholding and projection on $\ell_2$ balls.

\smallskip

\item \emph{Alternative direction of multipliers method} (ADMM). 
This widespread optimization method (see e.g. \cite{Boyd2011,Combettes2011}) is suitable  for convex programs of the form $\min_{x,z} f(x) + g(z)$ subject to  linear constraints $Ax+Bz+c = 0$. 
The SGFL problem \eqref{objective} can be expressed in this form by setting $x=\beta$,  $f$ equal to the squared loss plus lasso penalty, and $g(z) = \lambda_2 \sum_t w_t \| z_t\|_2$ where $z_t = \beta_{t+1}-\beta_t$ for $1\le t <T$. 
ADMM works by forming an augmented Lagrangian function 
$L_{\rho}(x,z,u) = f(x) + g(z) + \frac{\rho}{2} \sum_t \| u_t + z_t - (\beta_{t+1}-\beta_t)\|_2^2$ and optimizing it alternatively with respect to $x$ (lasso problem) and to $z$ (projection on $\ell_2 $ balls), along with closed-form updates of the dual variable $u$. The regularization parameter $\rho > 0$ must be selected by the user. 
 \cite{Cao2018}
 
 \smallskip

\item \emph{Linearized ADMM} (LADMM) \cite{Li2014}. 
This technique is used in instances where one or both of the $x$- and $z$- updates in ADMM are computationally expensive. When applying ADMM to \eqref{objective}, one may linearize the squared loss and regularization term $ \frac{\rho}{2} \sum_t \| u_t + z_t - (\beta_{t+1}-\beta_t)\|_2^2$ in the augmented Lagrangian $L_{\rho}$. 
This replaces the burdensome lasso problem ($x$-update) by a simple soft-thresholding operation. 

\end{itemize}

 \paragraph{Selection of tuning parameters.} 

All the above methods have tuning parameters whose selection is nontrivial. In addition, the numerical performances of these methods are highly sensitive to their tuning parameters. We adopt the following strategies in the simulations. 
\medskip

\begin{itemize}
\item SPG. The parameter $\mu$ sets an upper bound on the gap between the minima of the original objective and  its smooth approximation. However, suitably small values of $\mu$ yield unacceptably slow convergence. 
For this reason, we employ SPG with restarts, starting from a relatively large $\mu$ and decreasing it along a logarithmic scale when the algorithm fails to reduce the objective for 100 successive iterations. 

\item  PD. Two proximal parameters $\tau, \sigma$ and a relaxation parameter $\rho$ must be specified. 
Following the recommendations of the author of \cite{Condat2013} (personal communication), we set $\rho=1.9$, 
$\sigma = 0.25 (1/\tau - \max_{t} \| X_t'X_t\|_2)$, 
and select $\tau$ from the grid $\{ 10^{-6}, 10^{-5}, \ldots, 10^{6}\}$ by trial and error. Specifically, we run 100 iterations of the PD algorithm with $\tau = 10^{-6}$,  $\tau=10^{-5}$, and so on so forth until the best performance over 100 iterations decreases. (The best performance first increases with $\tau$ and then decreases).

\item ADMM and LADMM. The regularization parameter $\rho$ is selected by trial and error as above (best performance over 100 iterations) but going from large to small values: $\rho=10^{4}, 10^{3}, \ldots$ 
 \end{itemize}

  \paragraph{Simulation setup.}
 
 We consider the piecewise multivariate linear regression model $y_t = X_t \beta_t + \varepsilon_t $ where $\beta_t$ ($1\le t \le T$) is constant on each  segment $C_k =\{ t: T_{k} \le t <  T_{k+1}\}$ ($1\le k \le K$) with $T_{k} = \frac{(k-1) T}{K }  + 1$ and  $K=10$. Two combinations of data dimensions were used: 
 $(d,p,T)=(100,500,200)$ for a problem of moderate size  ($10^4$ optimization variables)
 and $(d,p,T)=(100,1000,1000)$ for a larger problem ($10^6$ variables). 
 Different correlation levels $\rho_{X}$ in the predictor variables 
 and noise levels $\sigma_{\varepsilon}$ 
 were examined. The predictors $X_t$ were sampled from a multivariate normal distribution 
with mean zero, unit variance, and exchangeable correlation structure: 
 $\mathrm{Cor}((X_{s})_i, (X_{t})_{j}  ) = \rho_{X}$ if $(s,i)\ne (t,j)$ for $1\le s,t\le T$ and $1\le i,j\le p$. 
 Note that correlation occurs both across components and across time. 
 The regression vectors $\beta_{T_k}$ were first obtained as independent realizations of $N(0_p,I_p)$, after which a fraction $s=0.9$ of each vector was selected randomly and set to zero. As a result each $\beta_t$ has sparsity level 0.9. The response vectors $y_t$ were obtained by adding   white noise $\varepsilon_t \sim N(0,\sigma_{\varepsilon}^2 I )$ to $X_t\beta_t$. 
The regularization parameters $\lambda_1$ and $\lambda_2$ were taken so that the SGFL solution $\hat{\beta}$ has the same change points and  sparsity level as the true $\beta$. For each  setup $(d,p,T,\rho_{X},\sigma_{\varepsilon})$, the simulation (data generation + optimization) was replicated 100 times if $(d,p,T)=(100,500,200)$ and 10 times if $(d,p,T)=(100,1000,1000)$.

 The simulations were realized in the R programming environment \cite{R2019} 
 on  a 24 Intel Xeon Gold processor with 46GB RAM (Ubuntu 18.04.4 LTS). 
The  SPG, PD, ADMM, and LADMM methods were written in C++ 
using the Armadillo library \cite{Sanderson2016} and wrapped in R with \texttt{RcppArmadillo}. 
 The proposed hybrid approach uses a mix of C++ and R; 
 it is implemented in the R package \texttt{sparseGFL}.
 The package and simulation scripts are available at \url{https://github.com/ddegras/sparseGFL}. 
 Each simulation was run on a single CPU core without parallelizing the execution of optimization methods.

The SPG, PD, ADMM, and LADMM methods were executed without stopping criterion for a number of iterations sufficient to reach convergence (3000-5000). The SPG used restarts as described above for $10^4$ iterations at most. For the hybrid approach (HYB-C and HYB-R), the tolerance $\epsilon$ used in the stopping criterion of Algorithm \ref{alg:main} must be specified, as it determines not only the total number of iterations realized but also the type of optimization realized at each iteration (block coordinate descent, fusion cycle, etc.). It was set to $ 10^{-6}$ to reflect the target relative accuracy of the solution to \eqref{objective}.

  \paragraph{Results.}
 
 The main performance measure used in the simulation study  is the CPU runtime needed to reach a sufficiently accurate solution to \eqref{objective}.  We select a target level of $10^{-6}$ for the relative accuracy of a solution $\hat{\beta} \in\mathbb{R}^{pT}$. That is, we deem a  solution $\hat{\beta}$ to be sufficiently accurate if $ F(\hat{\beta}) \le (1+10^{-6}) \min_{\beta} F(\beta)$.  This  level of accuracy is sufficient to guarantee that a solution $\hat{\beta}$ has the same change points and (exactly or very nearly) the same sparsity structure as the minimizer $\beta^\ast$ of $F$. For PD, ADMM, and LADMM, the initial time spent selecting suitable tuning parameters is included in the CPU runtime. (This initial time represents a relatively small fraction of the total runtime.) We point out that it is quite difficult to know good values of the tuning parameters \emph{a priori} and that the performance of these three methods largely depends on their tuning parameters. Badly chosen tuning parameters may lead to excessively slow convergence or, in the other direction, to numerical overflow and divergence. If in a given simulation, an optimization method fails to reach a relative accuracy $10^{-6}$, the total runtime of this method is reported.

\begin{figure}[htbp]
\begin{center}
\includegraphics[scale=.65]{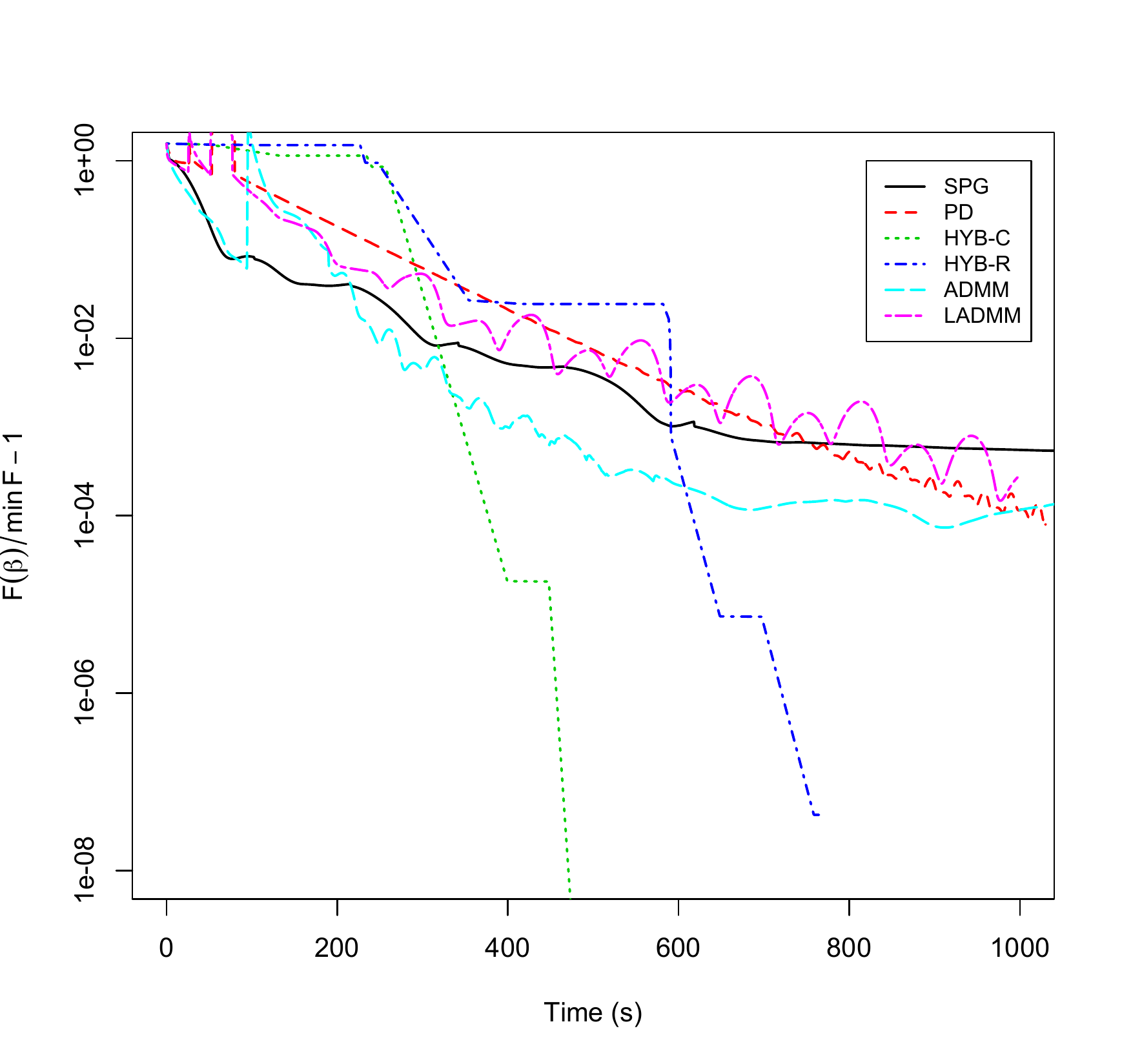}
\caption{Simulation study: relative accuracy of solution versus CPU runtime in the high-dimensional setup $(d,p,T,\sigma_{\varepsilon},\rho_{X}) =(100,1000,1000,0,0.25)$  (typical example).}
\label{fig:optimization history}
\end{center}
\end{figure}

The runtimes of the methods (to reach relative accuracy $10^{-6}$) are summarized in Table \ref{tab:runtime}. 
SPG  is by far the slowest method, taking an order of magnitude more time than all other methods to converge. This method would likely perform better with more sophisticated or more finely tuned restarting rules than the one used here, for example \cite{Hadj-Selem2018}. ADMM is the next slowest method and is  not  competitive for  SGFL because of the necessity to solve a lasso problem at each iteration. PD and LADMM show comparable runtimes, with a very slight advantage for LADMM on problems of moderate size and a more marked advantage for PD on larger problems. Given that these are generic methods, their speed is quite satisfactory in comparison to the proposed hybrid method which is tailored for SGFL. In all setups, either HYB-C or HYB-R shows the best average runtime. 
HYB-R is the fastest method in about  56\% of all simulations, HYB-C in 30\%, LADMM in 9\%, and  PD in 5\%. 
Unsurprisingly, HYB-R has a more variable runtime than HYB-C because of the additional randomization of the sweeping pattern. 
Interestingly, HYB-C performs best in the presence of correlation among predictor variables ($\rho_X \in\{ 0.10 , 0.25\}$)
whereas HYB-R shows superior performance when $\rho_X = 0$. 
The fact that HYB-C and HYB-R improve upon PD (the next best method)  
by respective speedup factors of 40\% and 33\% in the high-dimensional and correlated setup $p=1000, T=1000, \rho_{X}=0.25$ 
is particularly promising for real world applications. See Figure \ref{fig:optimization history} for an illustration.

\begin{table}[ht]
\begin{center}
\scriptsize{
\caption{CPU runtime (in seconds) required to solve the SGFL problem \eqref{objective} with relative accuracy $10^{-6}$.  
 Runtimes are averaged across 100 replications if $(d,p,T)=(100,500,200)$ and 10 replications if  $(d,p,T)=(100,1000,1000)$.
 Best results are indicated in bold.}
 \label{tab:runtime}
 \hspace*{-4mm}
\begin{tabular}{ *{11}{c}}
  \hline\noalign{\smallskip}
  $d$ & $p$ & $T$ & $\sigma_{\varepsilon}$ & $\rho_X$ & SPG & PD & HYB-C & HYB-R & ADMM & LADMM \\ 
\noalign{\smallskip}\hline\noalign{\smallskip}
100 & 500 & 200 & 0.00 & 0.00 & 225.7  & 31.7  & 26.7  & \textbf{22.3}  & 46.7  & 28.1 \\
100 & 500 & 200 & 0.25 & 0.00 & 212.0  & 31.8  & 26.3  & \textbf{22.3}  & 46.7  & 28.4 \\
100 & 500 & 200 & 2.50 & 0.00 & 174.8  & 30.8  & 27.9  & \textbf{23.1}  & 45.2  & 27.4 \\
100 & 500 & 200 & 5.00 & 0.00 & 210.8  & 31.5  & \textbf{16.9}  & 17.1  & 50.8  & 28.3 \\
100 & 500 & 200 & 0.00 & 0.10 & 173.4  & 35.7  & \textbf{30.8}  & 45.5  & 51.4  & 40.7 \\
100 & 500 & 200 & 0.00 & 0.25 & 176.1  & 50.4  & \textbf{38.0} & 48.2  & 76.8  & 52.1 \\
100 & 500 & 200 & 0.25 & 0.25 & 177.6  & 48.6  & \textbf{37.1}  & 48.4  & 69.4  & 47.1 \\
100 & 1000 & 1000 & 0.00 & 0.00 & 6120.3  & 454.0  & 698.0  & \textbf{373.6}  & 2261.5  & 971.8 \\
100 & 1000 & 1000 & 0.00 & 0.25 & 7435.5  & 758.2  & \textbf{458.1}  & 507.7  & 2384.6  & 879.0 \\
100 & 1000 & 1000 & 0.25 & 0.25 &  6753.9  &657.8  &     \textbf{394.8}   & 454.4  & 2376.3   &  893.8  \\           
\noalign{\smallskip}\hline
\end{tabular}}
\end{center}
\end{table}

 We now turn to the accuracy of the methods, keeping in mind that the target accuracy is $ F(\hat{\beta}) \le (1+10^{-6}) \min_{\beta} F(\beta)$. 
 Table \ref{tab:accuracy} displays the worst-case accuracy of each method in each simulation setup. 
For a given method and setup, the worst-case accuracy is calculated as the quantile of level 99\% of $ F(\hat{\beta}) /\min_{\beta} F(\beta) - 1$ across all simulations. Therefore, values inferior to $10^{-6}$ in the table indicate that the target accuracy is virtually always met. It is important to remember that for HYB-C and HYB-R, the stopping tolerance $\epsilon = 10^{-6}$ is set to achieve the target accuracy level $10^{-6}$, not to produce highly accurate solutions. Despite this fact, the worst-case accuracy of HYB-C is well below $10^{-6}$ in all setups 
 and so is that of HYB-R (except for the high-noise setup $\sigma_{\varepsilon} = 5$). No other method achieves the target accuracy so consistently, although they run for a much longer time. Globally, HYB-C meets the target accuracy $10^{-6}$ in 100\% of the simulations, HYB-R in 99.4\%, ADMM in 96.2\%, PD in 95.8\%, LADMM in 95.1\%, and SPG  in 91.4\%.

\begin{table}[ht]
\begin{center}
\caption{Relative accuracy: worst-case performance. For each method and each setup, 
the quantile of level 0.99 of $(F(\hat{\beta}) / \min_{\beta}F(\beta) ) - 1$ across all replications is displayed, 
where $\hat{\beta}$ is the final estimate produced by the method. For HYB-C and HYB-R, 
the optimization is stopped whenever the relative decrease in $F$ between two successive iterations 
is less than $10^{-6}$, whereas the other methods run for many iterations without stopping criterion. 
The numbers in the table must be compared to the target accuracy level $10^{-6}$. }
   \label{tab:accuracy}
\hspace*{-8mm}
\begin{tabular}{*{11}{c}}
  \hline\noalign{\smallskip}
  $d$ & $p$ & $T$ & $\sigma_{\varepsilon}$ & $\rho_X$ & SPG & PD & HYB-C & HYB-R & ADMM & LADMM \\ 
\noalign{\smallskip}\hline\noalign{\smallskip}
 100 & 500 & 200 & 0.00 & 0.00 & 5.7e-06 & 6.0e-15 & 1.9e-09 & 2.0e-09 & 1.3e-14 & 3.2e-12 \\ 
 100 & 500 & 200 & 0.20 & 0.00 & 8.8e-06 & 6.0e-15 & 1.7e-09 & 1.9e-09 & 9.1e-15 & 2.5e-10 \\ 
 100 & 500 & 200 & 2.50 & 0.00 & 4.1e-07 & 5.1e-15 & 2.6e-07 & 2.3e-07 & 6.2e-08 & 5.8e-09 \\ 
 100 & 500 & 200 & 5.00 & 0.00 & 6.6e-07 & 2.6e-10 & 7.5e-07 & 6.3e-04 & 1.8e-06 & 5.7e-07 \\ 
 100 & 500 & 200 & 0.00 & 0.10 & 6.8e-06 & 4.4e-07 & 2.7e-08 & 4.4e-08 & 2.7e-10 & 6.2e-10 \\ 
 100 & 500 & 200 & 0.00 & 0.25 & 2.0e-06 & 3.6e-04 & 3.1e-07 & 6.5e-07 & 2.3e-06 & 2.0e-05 \\ 
 100 & 500 & 200 & 0.25 & 0.25 & 5.6e-06 & 3.9e-04 & 2.2e-07 & 2.3e-07 & 1.2e-06 & 3.5e-05 \\ 
 100 & 1000 & 1000 & 0.00 & 0.00 & 8.1e-06 & 0.0e+00 & 1.8e-10 & 2.9e-10 & 1.3e-05 & 1.9e-06 \\ 
 100 & 1000 & 1000 & 0.00 & 0.25 & 5.9e-06 & 2.8e-04 & 9.6e-08 & 8.7e-07 & 3.3e-05 & 4.2e-04 \\ 
 100 & 1000 & 1000 & 0.25 & 0.25   & 5.3e-06 &  8.7e-04 & 1.7e-07 & 6.3e-08 & 2.3e-05 & 8.3e-05 \\
\noalign{\smallskip}\hline
\end{tabular}
   \end{center}
\end{table}

\subsection{Simulations: statistical accuracy}
\label{sec: stats simulations}

In this part we assess the statistical accuracy of the SGFL in change point detection and parameter estimation. 
We also  compare the SGFL to relevant methods for change point detection in high-dimensional linear regression
listed below.

 \paragraph{Benchmark methods.}

\begin{itemize}
\item \emph{Binary segmentation.} The binary segmentation algorithm (BSA) of \cite{Leonardi2016} is designed to detect change points and estimate sparse regression coefficients in high dimensional linear regression. It enjoys theoretical guarantees of consistency and convergence rates for the estimation of change points and regression coefficients. We note that the BSA is one of two estimation methods proposed in \cite{Leonardi2016}. The other one utilizes dynamic programming and produces exact solutions to the statistical problem under study (unlike BSA) at the cost of much more intensive calculations.

\item \emph{Two-step approximation of SGFL.} The difficulty of the SGFL comes from the fact that it tackles time series segmentation (or change point detection) and sparse estimation of regression coefficients \emph{simultaneously}. 
A simpler device would be to accomplish these tasks \emph{sequentially}.
We thus consider an approximate scheme that breaks the SGFL in two steps (2S). 
The first step segments the time series by minimizing the SGFL objective \eqref{objective} without lasso penalty ($\lambda_1 = 0$). 
As noted in \cite{Bleakley2011}, this can be efficiently implemented as a group lasso over the new variables $\theta_t = \beta_t - \beta_{t-1}$. 
In the second step, lasso regression is carried out over each segment obtained in the first step, 
yielding sparse estimates of the regression coefficients $\beta_t$. 
The two-step method is  extremely efficient from a computational perspective, especially in view of fitting entire regularization paths. 
It also simplifies the selection of the regularization parameters $\lambda_1$ and $\lambda_2$, 
which can be chosen separately in each step. 
\end{itemize}

\paragraph{Simulation setup.} 
The  simulation framework of Section \ref{sec: simulations optimization} was reused with different parameters. 
The data dimensions were set to $(d,p,T)=(20,200,100)$ (moderate dimension) or $(d,p,T)=(100,500,200)$ (high dimension). 
The change points were set to $\{0.2T+1,0.5 T+1,0.9T+1\}$ (few change points with one near the boundary) or $\{ 0.2T+1,0.4T+1,0.6T+1,0.8T+1\}$ (equispaced change points with higher density). 
Due to high computational load, only the 3-change points scenario was considered in  high dimension. 
The noise level $\sigma_{\varepsilon}$ was either 0.25 (low noise) or 1 (moderate noise). The sparsity level of the regression vectors $\beta_t$ was set to $s=0.99$. 
The predictors $X_t$ were generated independently ($\rho_X = 0$).  
Qualitatively similar results were obtained with different levels of signal sparsity, noise variance, and predictor correlation. 
They are provided as Supplementary Materials.

The BSA has several tuning parameters: the maximum number $k_{max}$ of segments allowed (set to 10), the minimum distance $\delta$ between change points (set to 0.1 on a unit interval), and a regularization parameter $\lambda$ for lasso regression.  The two-step method and the SGFL require two regularization parameters: $\lambda_1$ (lasso penalty) and $\lambda_2$ (total variation penalty). 
To our knowledge no selection method with theoretical guarantees has yet been developed  for our particular context (high-dimensional regression with multiple change points). We also note that (i) applying cross-validation techniques in this context is   methodologically and computationally difficult, and (ii)  in high dimension the usual  Akaike Information Criterion (AIC) and Bayesian Information Criterion (BIC) tend to  not penalize enough model complexity and lead to overfitting the data (e.g. \cite{Chen2008}; also confirmed in simulations). For these reasons we utilize a high-dimensional version of BIC  to select $\lambda ,\lambda_1,\lambda_2$ (HBIC, \cite{Wang2011}). 	
Ignoring irrelevant additive constants, this criterion can be formulated as 
\begin{equation}\label{HBIC}
\begin{split}
\mathrm{HBIC}(\lambda)  & = 
dT \log \Big( \sum_{t=1}^{T} \big\| y_t - X_t \hat{\beta}_t(\lambda) \big\|^2 \Big) \\ 
& \qquad + 2\gamma \log(p) \ (\# \textrm{ free parameters in } \hat{\beta}(\lambda))
\end{split}
\end{equation}
where $\lambda$ denote all regularization parameters and $\gamma \ge 1$ must be specified. The number of free parameters is the total number of nonzero coefficients in the vectors $ \hat{\beta}_t$ associated with change points, or equivalently the number of nonzero coefficients in the increments $\theta_t = \beta_{t} - \beta_{t-1}$ with $\beta_0 = 0_p$. 
The choice of a value $\gamma$ that produces good estimates $\hat{\beta}$ is in itself a difficult problem; 
in practice, both  $\lambda$ and $\gamma$ should  be considered as regularization parameters to  select. 
Values $\gamma \in \{ 1,2,\ldots, 10\}$ were used in the simulations to assess the effect of this parameter.

For each combination of change points and noise level, the simulation was replicated 100 times. 
In each replication, the BSA, two-step method, and SGFL method were fitted to the data 
over a fine grid of regularization parameters. 
(Between 10 and 100 values were considered for each grid, depending on the sensitivity of the parameter $\lambda$, $ \lambda_1,$
or $\lambda_2$ and on the computational requirements of the estimation method. The largest grid value was obtained by theoretical considerations to make the estimate uniformly zero or time-invariant; the smallest grid value was taken as a sufficiently small ratio of the largest, between 0.1 and 0.001. Grid values were equispaced on a linear or logarithmic scale as required.)
For each estimation method 
and value $\gamma$, the estimate $\hat{\beta}(\lambda)$ minimizing \eqref{HBIC} as a function of $\lambda$ 
was determined.  
The best estimate $\hat{\beta}(\lambda)$ with regards to change point detection was also determined, more precisely the one minimizing the Hausdorff distance between the estimated and the true change points. For reference, the Hausdorff distance between two non empty subsets $X$ and $Y$ of a metric space $(M,d)$ is $$d_{H}(X,Y) = \max \left( \sup_{x\in X} \inf_{y\in Y} d(x,y) , \sup_{y\in Y} \inf_{x\in X} d(x,y)\right).$$

We note that the two methods used to select the regularization parameters $\lambda$ or $(\lambda_1,\lambda_2)$ are infeasible in practice as they exploit information on the true regression coefficients $\beta$. These methods are meant to describe the estimation performance when a good value $\gamma$ is used in \eqref{HBIC} (for the first method) and when the change points are estimated with high accuracy (for the second). In a controlled simulation environment, this makes it possible to ignore the uncertainty associated with selecting regularization parameters and to compare estimation methods at their best performance level.

\paragraph{Results.}

Table \ref{table: stat sim} reports simulation results for all configurations of model dimension, change points,  noise level and for all estimation methods (BSA, 2S, SGFL). In each simulation setup, the results are averaged across 100 replications. The performance measures in each column are the number of estimated change points (NCP), the Hausdorff distance between estimated and true change points ($d_H$), the true positive rate and positive predictive value in detecting nonzero regression coefficients (TPR,PPV), the sparsity level of the estimate ($\widehat{s}$),  and a pseudo $R^2$ coefficient. 

In the column NCP, perfect detection would yield $\mathrm{NCP}=3$ in the case where the change points are $\{0.2T+1,0.5 T+1,0.9T+1\}$  and  $\mathrm{NCP}=4$ in the case  $\{ 0.2T+1,0.4T+1,0.6T+1,0.8T+1\}$. In the column $d_H$, the best possible value is zero and the worst is $T$ (when no change point id detected). The true positive rate TPR (also called sensitivity) is the proportion of nonzero coefficients in $\beta$ detected by a method. (There were $(1-s) pT $ nonzero coefficients in $\beta$.) The predictive positive value PPV (also called specificity) is the proportion of nonzero coefficients in $\hat{\beta}$ that are truly nonzero in $\beta$. The sparsity level  $\widehat{s}$ of an estimate $\hat{\beta}$ should be close to the sparsity  $s=0.99$ of the true $\beta$. The pseudo $R^2$ measures  goodness of fit  and is calculated as $R^2 = 1 -( \sum \| y_t - X_t \hat{\beta}_t\|^2 /  \sum \| y_t - \overline{y} \|^2 )$. 

The column $\lambda$ of  Table \ref{table: stat sim} refers to the selection method for $\lambda$. The method based on the Hausdorff distance ($d_H$) has been discussed before. Estimation performance with the HBIC selection method corresponds to the ``best" possible choice of $\gamma \in \{ 1 , 2, \ldots, 10\}$ defined as follows. 
For each estimation method, replication, and value $\gamma $, the estimate $\hat{\beta}(\lambda)$ minimizing the HBIC \eqref{HBIC} was determined and the corresponding measures NCP, ..., $R^2$ in Table \ref{table: stat sim}  were calculated.
 A composite score giving equal weight to each measure was constructed to reflect best overall performance. Specifically, each measure was linearly transformed to produce the value zero as best performance (e.g. NCP equal to the true number of change points, $d_H = 0$, $\mathrm{TPR}=1$, $\widehat{s} = s$) and the value one as worst performance. 
The results displayed in the table are for the value $\gamma $ with minimum average composite score across replications.

Looking at Table \ref{table: stat sim}, it appears that the BSA performs extremely well in low noise situations ($\sigma_{\varepsilon}=0.25$) both in the  3- and 4- change point scenarios (see HBIC rows). The method  accurately detects all change points while simultaneously recovering nearly all nonzero coefficients of $\beta$, attaining a sparsity level very close to the target $0.99$, 
and achieving a very high $R^2$ coefficient. However this high performance seems to break down in moderate noise situations ($\sigma_{\varepsilon} = 1.0$). 
There, the method largely fails to estimate the number of change points and their  locations; 
the sparsity of its estimates falls to about 0.5-0.6 and the PPV falls correspondingly to 0.03.  
The rows BSA/$d_H$ in the table reveal that the BSA can in fact accurately locate the change points for a suitable choice of $\lambda$. But even for this choice of $\lambda$, the estimate  $\hat{\beta}$ keeps a low sparsity (0.63 or 0.65) and extremely  low PPV (0.04). 
It is worth noting that in less sparse regression models ($s=0.9$ instead of $s=0.99$), 
the BSA shows good statistical performance at all noise levels (see Supplementary Materials for details).

The two-step method (2S) performs remarkably well with respect to all performance measures in all simulation setups. Its performance degrades in the moderate noise setup with 4 change points but remains at an acceptable level. Concerning the estimation of sparsity patterns in $\beta$, this method consistently achieves a high TPR and high PPV (at least in low noise: between 0.90 and 0.93). 
The SGFL performs very well in change point detection but less so in parameter estimation (its TPR and PPV stay between 0.73 and 0.89 in all setups) and goodness of fit. Its performance is very stable in all change point configurations and noise levels; in particular, its sparsity level stays at the target level 0.99 throughout. Additional simulations showed that this method can still accurately detect change points under high noise levels ($\sigma_{\varepsilon}=2.5$).


\begin{table}[!ht]
\centering
\caption{Simulations: statistical accuracy. Columns indicate the simulation setup, estimation method, selection method for  regularization parameters, number of estimated change points, Hausdorff distance between estimated and true change points, true positive rate and positive predictive value in detecting  nonzero regression coefficients, sparsity level (target level $s=0.99$), and pseudo-$R^2$. Results are averaged over 100 replications (standard deviation in brackets). For the HBIC selection method, the 
best results are in bold.}
\label{table: stat sim}
\bigskip
\hspace*{-10mm}
\footnotesize{
\begin{tabular}{c r c  c c c c c  c}
  \hline
Setup & Method & $\lambda$ & NCP & $d_H$ & TPR & PPV & $\widehat{s}$ & $R^2$ \\ 
\hline
\multirow{5}{12mm}{\vspace{-3.5mm}
\begin{equation*}
\begin{array}{r  l}
d&=20\\ 
p&=200\\
T&=100\\
\sigma_{\varepsilon} &= 0.25 \\
\noalign{$\{21, 51, 91\}$}
\end{array}
\end{equation*}
} 
 & BSA & HBIC & 3.1 (0.3) & \textbf{0.5} (2.2)& {\bf 0.97} (0.06)& 0.77 (0.32)& 0.98 (0.03) & \textbf{0.95} (0.03)\\ 
& 2S & HBIC & 3.8  (0.8)& 1.9  (6.2)& 0.93  (0.10)&  \textbf{0.93}  (0.06) &  \textbf{0.99} (0.00) & 0.95 (0.03) \\ 
  & SGFL & HBIC & \textbf{3.0} (0.2) & 1.1 (6.3) & 0.89  (0.12)& 0.83 (0.11) & 0.99  (0.00) & 0.90 (0.03)\\  
  & BSA & $d_H$ &   3.0 (0.2) &0.3 (1.7) &0.98 (0.05) &0.77 (0.32) &0.98 (0.04) &0.95 (0.03)\\
& 2S & $d_H$ & 4.0 (1.0) &0.9 (1.0) &0.95 (0.08) &0.97 (0.04) &0.99 (0.00) &0.91 (0.06)\\
 & SGFL & $d_H$ & 3.0 (0.0) &0.0 (0.0) &0.91 (0.11) &0.92 (0.15) &0.99 (0.01) &0.82 (0.10)\\
\hline
\multirow{5}{12mm}{\vspace{-3.5mm}
\begin{equation*}
\begin{array}{r  l}
d&=20\\ 
p&=200\\
T&=100\\
\sigma_{\varepsilon} &= 1.00 \\
\noalign{$\{21, 51, 91\}$}
\end{array}
\end{equation*}
} 
& BSA & HBIC  & 1.5 (1.4)& 51.5 (43.4) & {\bf 0.94}  (0.09)& 0.03 (0.03)  & 0.58 (0.2)& 0.50  (0.25) \\  
& 2S & HBIC   & 5.5  (3.8) & 26.2 (20.2) & 0.72 (0.19)  & 0.71 (0.14)& {\bf 0.99}  (0.00)& {\bf 0.51}  (0.19) \\  
& SGFL & HBIC   & {\bf 2.7}  (0.6)& \textbf{11.4}  (20.2) & 0.77 (0.16) & {\bf 0.73} (0.19)& 0.99  (0.01)& 0.48 (0.17) \\  
& BSA & $d_H$ & 3.3 (0.6) &2.9 (4.2) &0.96 (0.07) &0.04 (0.03) &0.63 (0.17) &0.68 (0.09) \\   
& 2S & $d_H$ & 10.6 (5.0) &12.0 (5.8) &0.73 (0.16) &0.81 (0.13) &0.99 (0.00) &0.48 (0.16) \\
& SGFL & $d_H$ & 3.2 (0.6) &0.8 (2.2) &0.8 (0.16) &0.70 (0.38) &0.92 (0.20) &0.43 (0.16) \\
\hline
\multirow{5}{12mm}{\vspace{-3.5mm}
\begin{equation*}
\begin{array}{r  l}
d&=20\\ 
p&=200\\
T&=100\\
\sigma_{\varepsilon} &= 0.25 \\
\noalign{\hspace*{-1.5mm}$\{21, 41,61, 81\}$}
\end{array}
\end{equation*}
}   
 & BSA & HBIC & {\bf 4.0} (0.2)  & {\bf 0.1} (1.0)& {\bf 0.96} (0.06)& 0.80  (0.26)  & 0.98 (0.02) & {\bf 0.96} (0.02)\\  
 & 2S & HBIC & 5.6  (1.3)& 2.2  (3.3) & 0.92 (0.09)& {\bf 0.93}  (0.04)& {\bf 0.99} (0.00)  & 0.95  (0.03) \\  
   & SGFL & HBIC & 4.0  (0.2)& 0.8  (3.9)& 0.88  (0.11)& 0.74  (0.14)& 0.99  (0.00) & 0.90  (0.04) \\  
   & BSA & $d_H$ & 4.0 (0.2) &0.1 (1.0) &0.96 (0.06) &0.80 (0.26) &0.98 (0.02) &0.96 (0.02)\\
   & 2S & $d_H$ & 5.7 (1.4) &1.9 (2.3) &0.94 (0.08) &0.96 (0.05) &0.99 (0.00) &0.92 (0.06)\\
   & SGFL & $d_H$ & 4.0 (0.3) &0.2 (1.2) &0.87 (0.10) &0.95 (0.15) &0.99 (0.01) &0.81 (0.08)\\
\hline
\multirow{5}{12mm}{\vspace{-3.5mm}
\begin{equation*}
\begin{array}{r  l}
d&=20\\ 
p&=200\\
T&=100\\
\sigma_{\varepsilon} &= 1.0 \\
\noalign{\hspace*{-1.5mm}$\{21, 41,61, 81\}$}
\end{array}
\end{equation*}
}   
& BSA & HBIC & 1.7 (2.1) & 62.3 (41.8)& {\bf 0.94}  (0.07) & 0.03 (0.04) & 0.54 (0.21)& 0.42  (0.28)\\  
  & 2S & HBIC & 7.5 (2.9)  & 11.1 (10.4)& 0.69  (0.18)& {\bf 0.76}  (0.10)& {\bf 0.99}  (0.00)& {\bf 0.55} (0.16) \\  
   & SGFL & HBIC & {\bf 3.6} (0.6)& {\bf 8.1}  (10.2)& 0.74  (0.15) & 0.76 (0.21) & 0.99 (0.01) & 0.47 (0.16) \\  
   & BSA & $d_H$ & 5.7 (2.1) &4.8 (4.7) &0.95 (0.06) &0.04 (0.04) &0.65 (0.17) &0.74 (0.06)\\
   & 2S & $d_H$ & 10 (3.3) &6.1 (3.0) &0.75 (0.14) &0.82 (0.12) &0.99 (0.00) &0.52 (0.14)\\
   & SGFL & $d_H$ &4.1 (0.5) &0.8 (1.8) &0.84 (0.13) &0.55 (0.42) &0.83 (0.29) &0.46 (0.14)\\
\hline
\multirow{5}{12mm}{\vspace{-3.5mm}
\begin{equation*}
\begin{array}{r  l}
d&=100\\ 
p&=500\\
T&=200\\
\sigma_{\varepsilon} &= 0.25 \\
\noalign{\hspace*{-1mm}$\{ 41,101, 181\}$}
\end{array}
\end{equation*}
}   
& BSA & HBIC & {\bf 3.0} (0.0) &{\bf 0.0} (0.0) & {\bf 0.98} (0.03) &{\bf 1.00} (0.02) &0.99 (0.00) &{\bf 0.99} (0.00) \\ 
  & 2S & HBIC & 3.0 (0.2) &0.0 (0.2) &0.98 (0.03) &0.99 (0.01) &{\bf 0.99} (0.00) &0.99 (0.00)\\ 
   & SGFL & HBIC & 3.0 (0.1) &0.6 (6.0) &0.96 (0.05) &0.52 (0.04) &0.98 (0.00) &0.96 (0.00) \\ 
   & BSA & $d_H$ & 3.0 (0.0) &0.0 (0.0) &0.99 (0.02) &1.00 (0.02) &0.99 (0.00) &0.99 (0.00)\\
   & 2S & $d_H$ & 3.0 (0.2) &0.0 (0.2) &0.98 (0.03) &1.00 (0.01) &0.99 (0.00) &0.98 (0.01)\\
   & SGFL & $d_H$ & 3.0 (0.0) &0.0 (0.0) &0.90 (0.08) &0.94 (0.04) &0.99 (0.00) &0.89 (0.05)\\
\hline
\multirow{5}{12mm}{\vspace{-3.5mm}
\begin{equation*}
\begin{array}{r  l}
d&=100\\ 
p&=500\\
T&=200\\
\sigma_{\varepsilon} &= 1.0 \\
\noalign{\hspace*{-1.5mm}$\{ 41,101, 181\}$}
\end{array}
\end{equation*}
}   
& BSA & HBIC & 3.2 (0.4) &4.0 (8.0) &{\bf 0.99} (0.02) &0.08 (0.06) &0.84 (0.08) & {\bf 0.83} (0.04)\\  
  & 2S & HBIC & 3.1 (0.3) &0.1 (0.3) &0.94 (0.06) &{\bf 0.97} (0.02) & {\bf 0.99} (0.00) &0.82 (0.04)\\  
   & SGFL & HBIC & {\bf 3.0} (0.0) & {\bf 0.0} (0.1) & 0.92 (0.07) & 0.75 (0.08) &0.99 (0.00) &0.79 (0.04)\\ 
   & BSA & $d_H$ & 3.2 (0.4) &4.0 (8.0) &0.99 (0.02) &0.08 (0.06) &0.84 (0.08) &0.83 (0.04)\\
   & 2S & $d_H$ & 3.1 (0.3) &0.1 (0.3) &0.95 (0.05) &0.99 (0.02) &0.99 (0.00) &0.81 (0.04)\\
   & SGFL & $d_H$ & 3.0 (0.0) &0.0 (0.0) &0.89 (0.07) &0.94 (0.05) &0.99 (0.00) &0.74 (0.05)\\
\hline
\end{tabular}
}
\end{table}

In comparison to the moderate dimension setup  $(d,p,T)=(20,200,100)$, the statistical performance of methods   
 does not seem to degrade in the high dimension setup $(d,p,T)=(100,500,200)$. 
 One possible explanation is that while the number of change points remain essentially in the two setups, i.e. 3, 
 the associated time segments where the regression vectors are constant are twice as long when $T=200$, 
 which makes the detection of change points easier.

We conclude this section with remarks on the HBIC and on computations. 
Regarding HBIC,  the statistical performance all estimation methods is not very sensitive to $\gamma$ in low noise but it is very much so in moderate-to-high noise. In the latter case, the HBIC may be too stringent as the best performance of estimation methods is attained at the boundary value $\gamma=1$. 
Concerning speed and scalability, the 2S method was an order of magnitude faster than the BSA and SGFL in the simulations. 
In the moderate dimension setting, it took on average between 15\,s and 20\,s to calculate $\hat{\beta}(\lambda_1,\lambda_2)$ over a dense grid (100 values for $\lambda_1$, 20 for $\lambda_2$) versus 290-370\,s for SGFL (30 values for $\lambda_1$, 20 for $\lambda_2$) and 510-570\,s for BSA (20 values for $\lambda$). In the high dimension setting,  2S  ran in a few minutes,  SGFL  in 2-3 hours, and  BSA in about 5-6 hours. Detailed runtime information can be found in the Supplementary Materials.

\subsection{Air quality data}

We illustrate the SGFL with an application to air quality monitoring. 
The dataset used in this example was analyzed in \cite{DeVito2008,DeVito2009} 
and is available on the UCI Machine Learning Repository 
(\url{https://archive.ics.uci.edu/ml//datasets/Air+quality}). 
It contains 9357 instances of hourly averaged responses from an array of 5 metal oxide chemical sensors embedded in an Air Quality Chemical Multisensor Device. Data were recorded from March 2004 to February 2005 on a 
device  located in a significantly polluted area, at road level, within an Italian city.
Ground truth hourly averaged concentrations for carbon monoxide (CO), Non Metanic Hydrocarbons (NMHC), Benzene (C6H6), Total Nitrogen Oxides (NOx) and Nitrogen Dioxide (NO2) were provided by a co-located reference certified analyzer. 
As described in \cite{DeVito2008}, the data show evidence of cross-sensitivity as well as of concept and sensor drift, which ultimately affects the sensors' capability to estimate pollutant concentration. 

\begin{figure*}[ht]
\begin{center}
\includegraphics[width=.485 \linewidth]{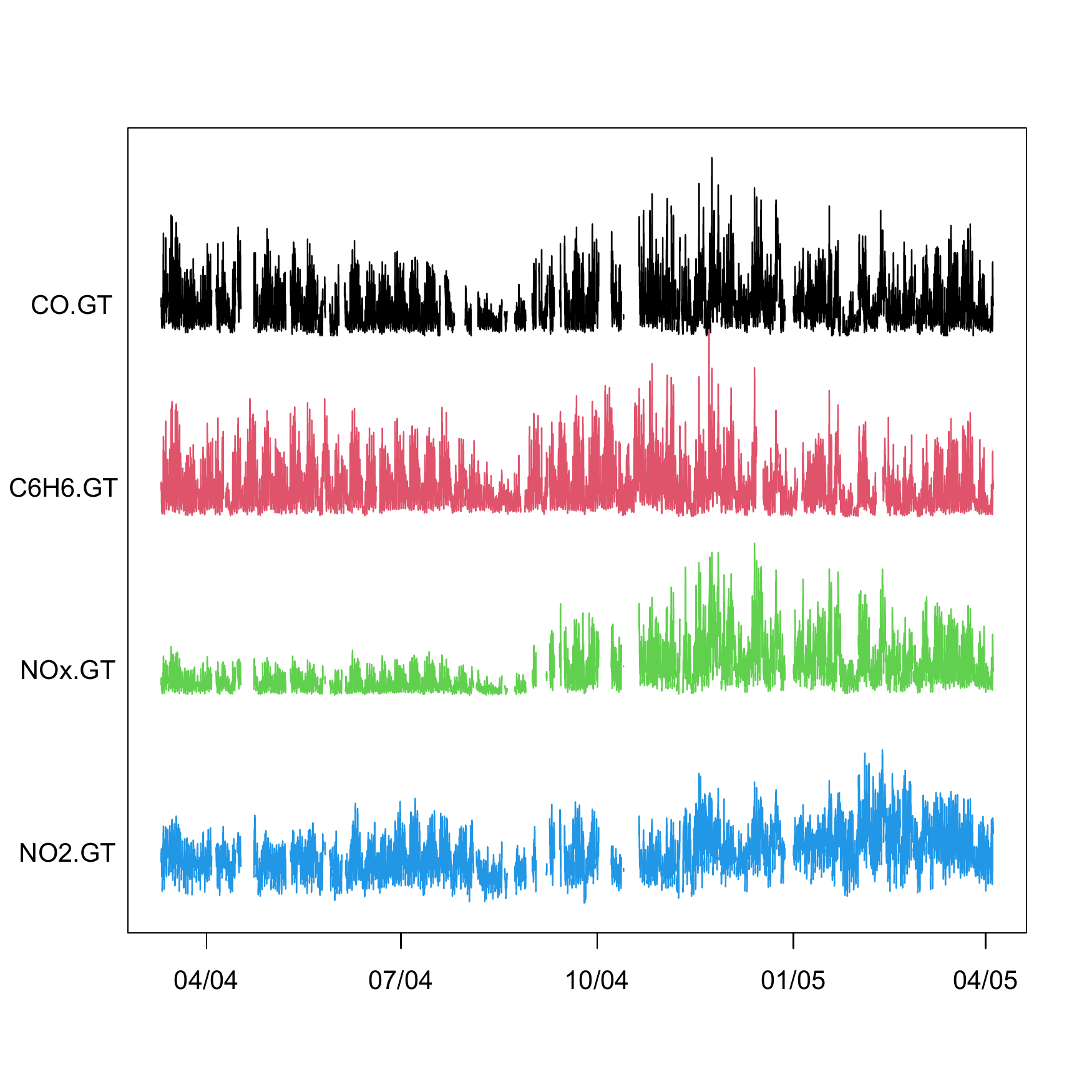} \hfill 
\includegraphics[width=.485 \linewidth]{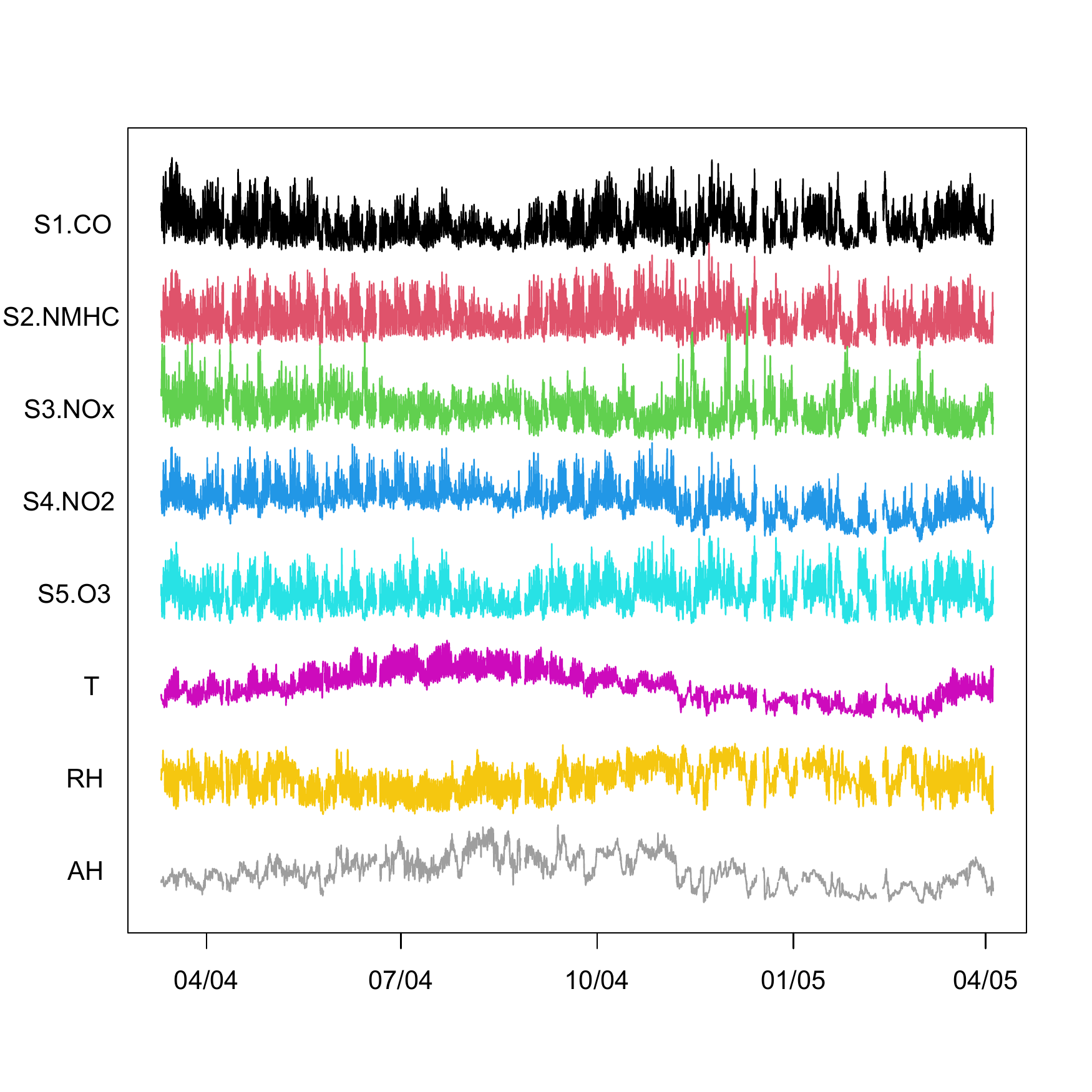} 
\caption{Air quality data. Left: pollutant levels (ground truth). Right:   
sensor measurements and meteorological variables. After each sensor number S1, S2, ... is the pollutant 
nominally targeted by this sensor.} 
\label{fig: air quality}
\end{center}
\end{figure*}

\begin{figure*}[ht]
\begin{center}
\includegraphics[width=.485 \linewidth]{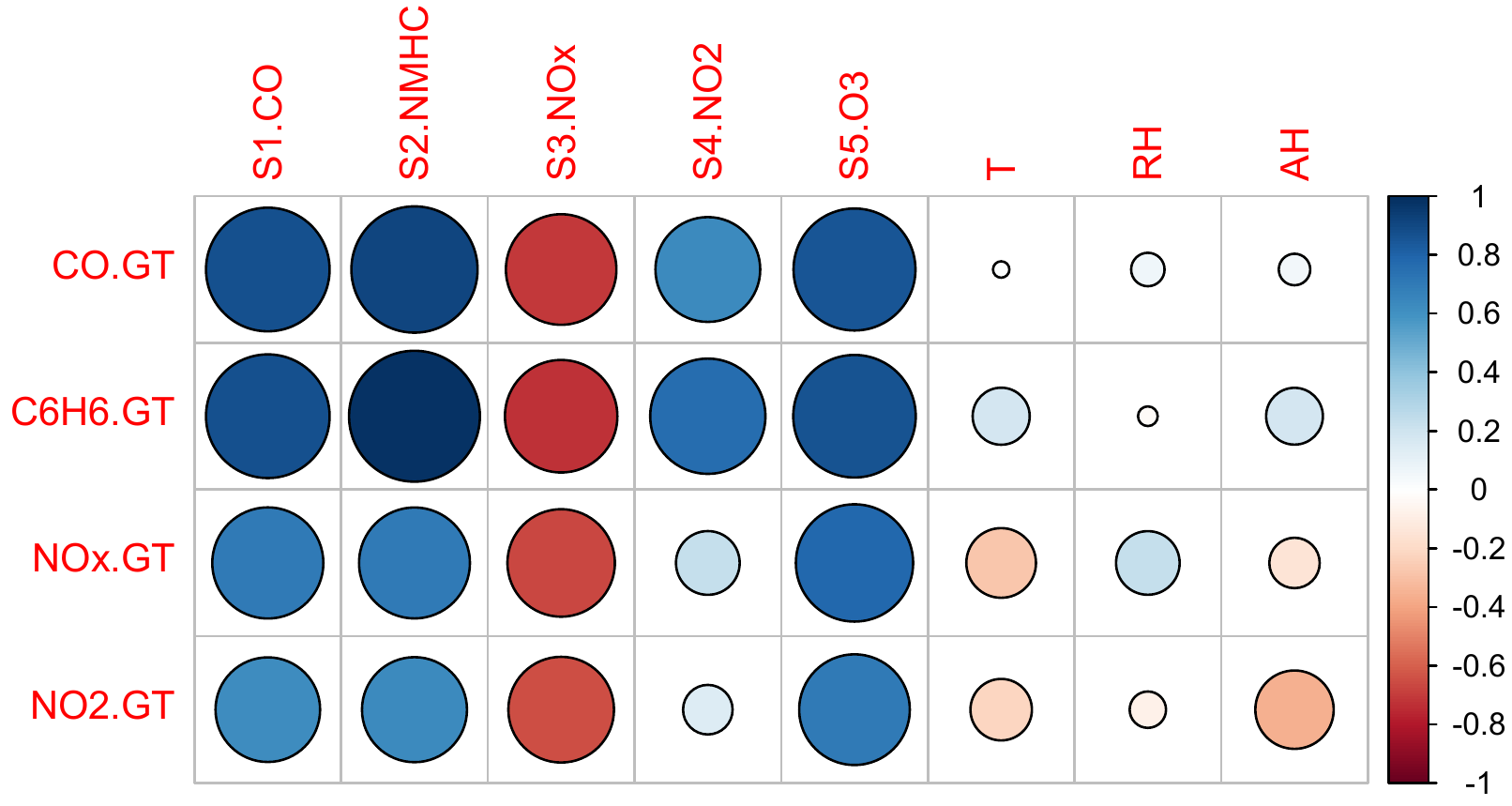} \hfill 
\includegraphics[width=.485 \linewidth]{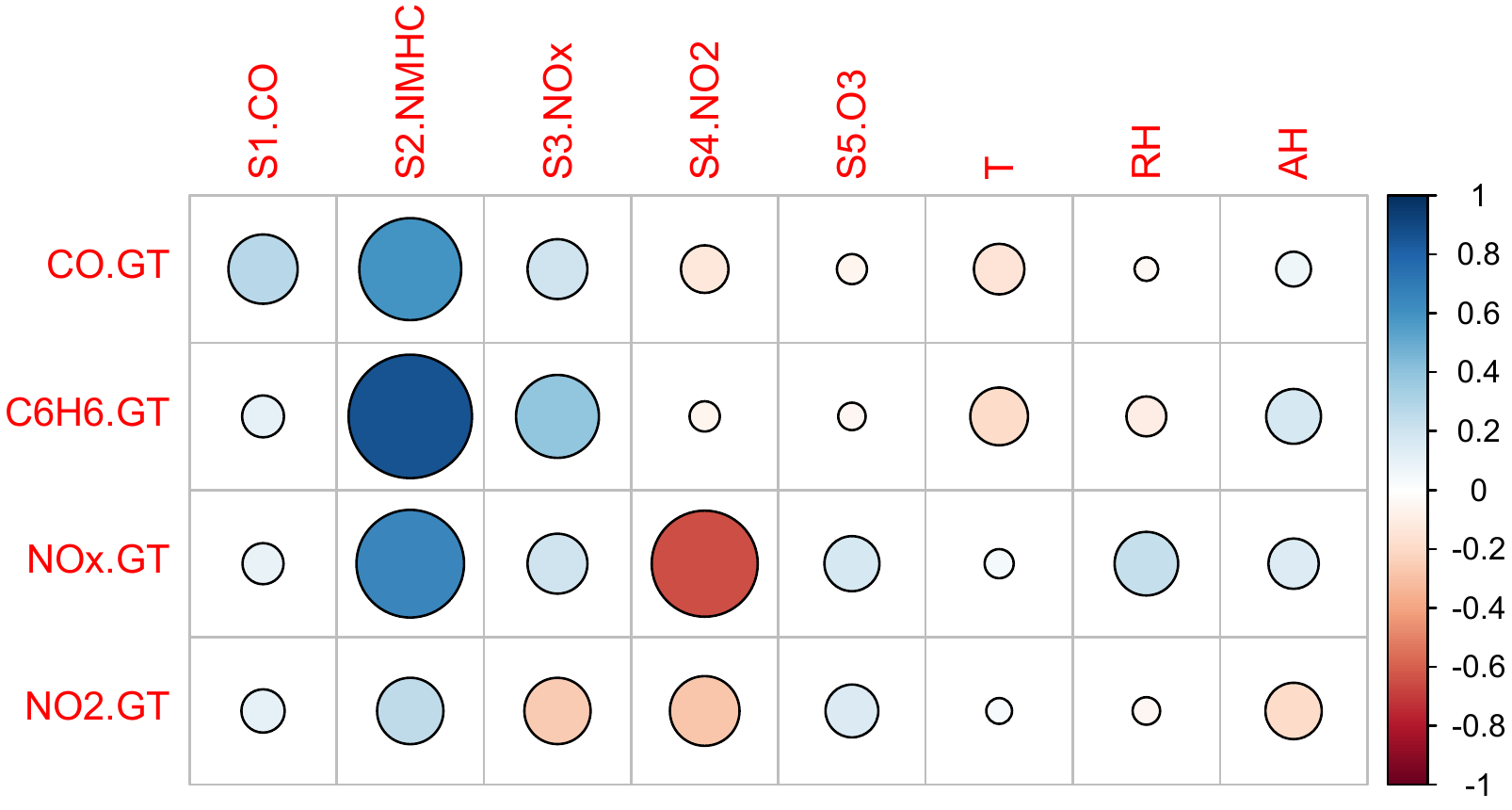} 
\caption{Air quality data:  correlation between pollutant levels and predictors. 
Left: full correlation. Almost all sensors are strongly correlated to all true pollutant concentrations. 
This correlation is positive as expected for all sensors except for S3.NOx, which is surprising. 
Right: partial correlation. S2.NMHC is by far the strongest predictor of all pollutant levels 
(in equality with S4.NO2 for the target NOx.  Interestingly none of the sensors is the best predictor 
for the pollutant it nominally targets. 
} 
\label{fig: correlation}
\end{center}
\end{figure*}

The hourly averaged measurements of the 4 target pollutants, 5 chemical sensors, and 3 meteorological variables (temperature, relative humidity, and absolute humidity)  are displayed in Figure \ref{fig: air quality} after shifting and scaling to facilitate visualization. 
In the statistical analysis, all variables have been centered and scaled. Time points with missing values have been ignored, which reduced the time series length to $T=6930$. Because of this, the quantitative results of this analysis should be interpreted with some caution. (A follow-up analysis, wherein missing values were replaced by the last observed value carried forward, produced roughly the same number of segments as in the present analysis with slight differences in change point location. The regression coefficients were very similar but the goodness of fit was much lower on the imputed dataset.) Correlation patterns between variables are depicted in Figure \ref{fig: correlation}.

The main goals in this application are to: (i) calibrate the sensors so that they accurately estimate the true pollutant concentrations, 
and (ii) determine how often  the sensors must be recalibrated in order to maintain a high accuracy.
Here we use SGFL in an exploratory way to determine which sensors and weather variables 
are predictive of the true pollutant levels, and how the regression relationship evolves over time. 
The relationship between the study variables is conveniently expressed as 
\begin{equation}\label{model yax}
y_t = A_t x_t + \varepsilon_t
\end{equation}
where $y_t \in \mathbb{R}^{d}$ represents the true pollutant concentrations at time $t$, 
 $x_t \in \mathbb{R}^{m}$ the sensor measurements and weather variables, 
 and $A_t \in \mathbb{R}^{d \times m}$ the unknown regression coefficients with $d=4$ and $m=8$ or $m=9$ if the model contains an intercept. This model can be recast in the form $y_t = X_t\beta_t + \varepsilon_t$ considered throughout the paper by setting $\beta_t = \mathrm{vec}(A_t)$ (concatenate the columns of $A_t$) and $X_t = (x_t')\otimes I_{d}$ (Kronecker product). However with this formulation the matrix $X_t \in \mathbb{R}^{d\times dm}$ becomes large and sparse, which tends to slow down calculations.  
 For computational speed and  user convenience, the R package \texttt{sparseGFL} has dedicated functions for both models \eqref{model yxb} and \eqref{model yax}.

 \paragraph{Model fitting.} 

The main model considered in the data analysis is \eqref{model yax} with $x_t$ containing all sensor measurements, weather variables, plus an intercept ($m=9$, $p=dm=36$). For comparison, we have also examined the corresponding time-invariant model $y_t = A x_t + \varepsilon_t$ as well as a much more complex piecewise regression model containing all sensor measurements, lagged versions thereof, weather variables, and interaction terms. (This model had $p=156$ regression coefficients per time point, that is, about 1 million optimization variables.) The motivation for this model was to investigate whether exploiting sensor measurements from the recent past could enhance estimation accuracy and whether weather conditions did modulate the regression relationship between sensors and targets. Our results were inconclusive with regards to these questions and because the complex model did not decisively improve upon the main-effects-only model, we did not pursue it further. We thus focus on model \eqref{model yax} with $m=9$ predictors and on the time-invariant model. 

The time-invariant regression model was fitted to the data by ordinary least squares (OLS). 
 The SGFL was fitted with Algorithm \ref{alg:main} over a lattice of regularization parameters $(\lambda_1,\lambda_2)$ 
spanning several orders of magnitudes: $[10^{-4}, 1]$ for $\lambda_1$ and $[5,200]$ for $\lambda_2$. The total variation weights $w_t$ in \eqref{objective} were set to 1 and a small ridge regression penalty was added to 
the lasso penalty to stabilize the estimation (mixing coefficient $\alpha=0.9$ in \eqref{SGFEN}). 
For each $(\lambda_1,\lambda_2)$, after calculating the SGFL solution $\tilde{A}=\tilde{A}(\lambda_1,\lambda_2) 
 \in (\mathbb{R}^{d\times m})^{T}$, segments of length less than 72 time points (3 days) were fused with contiguous longer segments. For each segment $C_k= \{t: T_k \le t < T_{k+1}\}  $, the matrix $A_{T_k}$ 
 was re-estimated by OLS while preserving the zero coefficients of $\tilde{A}_{T_k}$: 
$\min_{A_{T_k} } \frac{1}{2} \sum_{t\in C_k} \| A_{T_k} x_t - y_t \|_2^2 $ subject to $(A_{T_k})_{ij} = 0$ if $(\tilde{A}_{T_k})_{ij} = 0$.
This re-estimation step is common in penalized regression and reduces the bias induced by the penalty.   
We denote by SGFL-OLS this two-stage estimation procedure and by $\hat{A} = \hat{A}(\lambda_1,\lambda_2)$ 
the associated estimator. 

The best SGFL-OLS solution $ \hat{A} (\lambda_1,\lambda_2)$ was taken as the one minimizing the BIC score (additive constants omitted)
\begin{equation}\label{BIC}
\begin{split}
\mathrm{BIC}(\lambda_1,\lambda_2) & = dT \log \Big( \sum_{t=1}^{T} \big\| y_t -  \hat{A}_t(\lambda_1,\lambda_2)x_t \big\|^2 \Big) \\
& \qquad +  \log(dT) \ (\# \textrm{ free parameters in } \hat{A}(\lambda_1,\lambda_2)) 
\end{split}
\end{equation}
with the number of free parameters calculated as in Section \ref{sec: stats simulations}. BIC (or AIC) are appropriate  selection methods here  because of the low-dimensional setting $p \ll T $.

 \paragraph{Results.}

Due to the close connection between partial correlation and multiple regression, 
the time-invariant regression estimate $\hat{A} = (YX') (XX')^{-1} \in \mathbb{R}^{4\times 9}$ 
is qualitatively comparable to the partial correlation matrix of Figure \ref{fig: correlation}.
In model  \eqref{model yax} with $m=9$ predictors ( sensors, weather variables, intercept), 
the optimal SGFL-OLS solution $\hat{A}(\lambda_1,\lambda_2)$ for the BIC \eqref{BIC}
is obtained for $(\lambda_1, \lambda_2)=(0.0060, 50)$. This solution has a sparsity level of 8.6\% and overall $R^2$ of  94.3\%. 
It  segments the time range $\{1,\ldots,T\}$ into $K=23$ segments. (The AIC solution corresponds to $(\lambda_1, \lambda_2)=(0.0060,40)$. It is similar to the BIC solution but slightly less sparse and with more segments: 35.)

Table \ref{tab:air quality R2} reports the $R^2$ coefficient of each fitted model for each pollutant. For the time-invariant model, this measure varies from 0.756 for NO2 to 0.974 for C6H6. 
Although the piecewise model \eqref{model yax} improves upon the time-invariant model by 7.8\% overall in terms of $R^2$, more sophisticated methods may be required to capture the nonlinear component of the relationship between target pollutants and sensors, e.g. neural network architectures as in \cite{DeVito2009}.

\begin{table}[ht]
\caption{Air quality data: goodness of fit ($R^2$) of multivariate linear regression models.   
The time-invariant model is fitted by ordinary least squares. 
The piecewise model \eqref{model yax} is estimated by solving the SGFL \eqref{objective} 
for a range of values $(\lambda_1, \lambda_2)$ and selecting the solution that 
minimizes the BIC score \eqref{BIC}.}
\label{tab:air quality R2}       
\begin{center}
\begin{tabular}{lllll}
\hline
 Regression model & CO & C6H6 & NOx & NO2  \\
\hline
 Time-invariant & 0.886 & 0.974 & 0.842 & 0.758 \\
 Time-varying & 0.941  & 0.992 & 0.943 & 0.897 \\
 \hline
\end{tabular}
\end{center}
\end{table}

\begin{figure*}[h]
\begin{center}
\includegraphics[scale=.65]{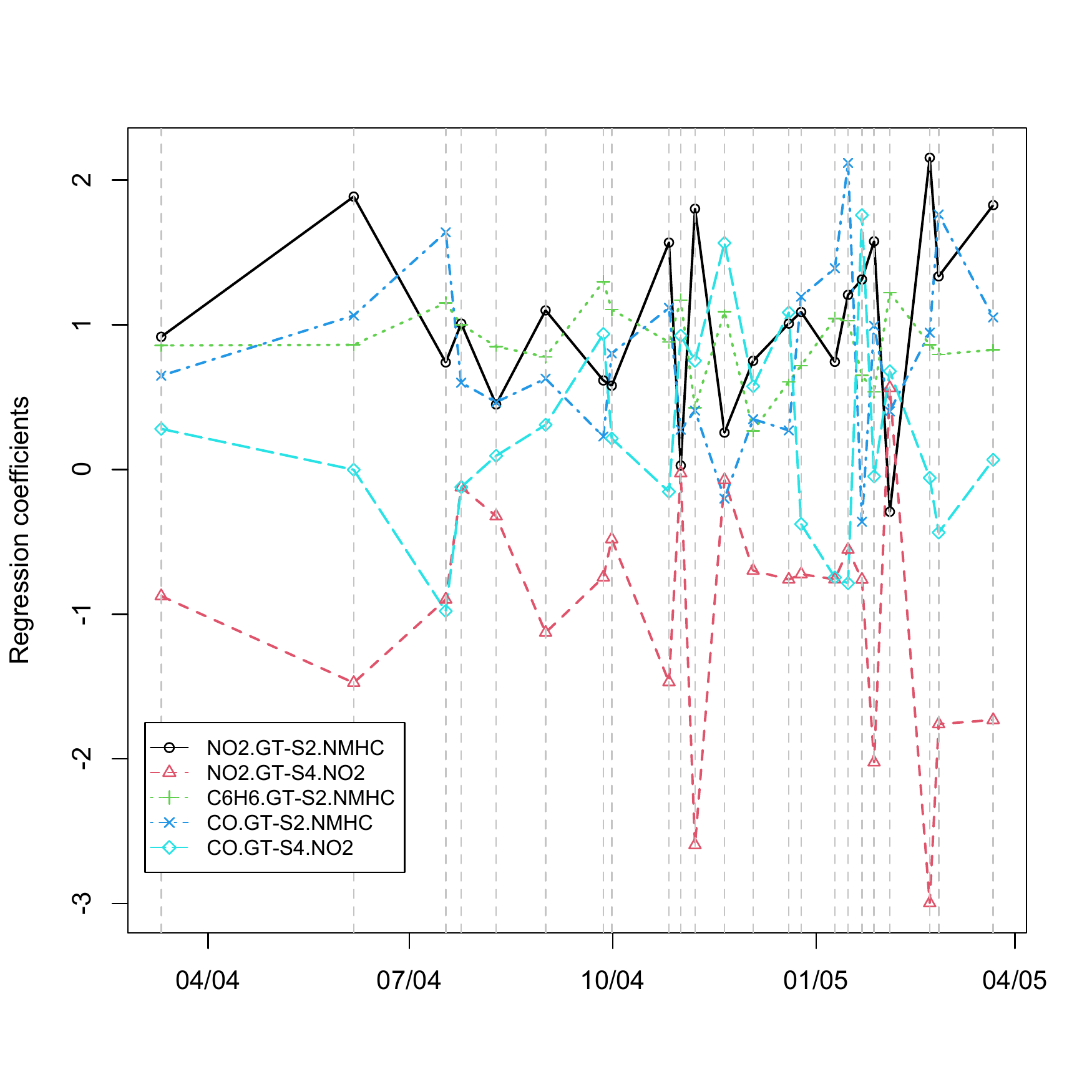}
\caption{Air quality data: time-varying regression coefficients and model segmentation obtained by SGFL. 
For clarity, the piecewise-constant coefficients are represented with continuous lines rather than step functions.} 
\label{fig: reg}
\end{center}
\end{figure*}

Figure \ref{fig: reg} shows the five largest regression coefficients (by average magnitude) of the SGFL-OLS solution over time. 
Only sensor-related coefficients are displayed there, but temperature and absolute humidity are also strong predictors of the target pollutant levels. The associated pairs of sensors and targets are also strongly related in the time-invariant regression model (see Figure \ref{fig: correlation}).  The figure reveals  that this regression relationship varies considerably over time, which justifies the need for regularly recalibrating the sensors. Although the changes in regression are mostly smooth over time, they can sometimes be very abrupt and large in magnitude. These observations must however be tempered by the fact that other SGFL-OLS estimates $\hat{A}(\lambda_1,\lambda_2)$ can fit the data nearly as well as the BIC solution with far fewer segments. For example, a highly parsimonious SGFL-OLS estimate can attain an overall $R^2=0.910$ while producing only $K=4$ segments of average length 72 days.

In conclusion of this example, the SGFL-OLS methodology combined with BIC selection suggests that the linear relationship between pollutant levels and sensor measurements stays homogenous over segments of about one week. This scale is consistent with the observation in \cite{DeVito2009} that the optimal training period for neural network-based predictive models is 10 days, after which no improvement in performance in observed. Also, to give perspective on the accuracy of SGFL-OLS in this example, we note that its mean absolute error in estimating the NO2 level is $11.2\,\mathrm{\mu} g/m^3$ (NO2  is the most difficult of the 4 pollutants to predict as Table \ref{tab:air quality R2} shows). The best mean absolute error for NO2 with a neural network in  \cite{DeVito2009} is $19.8\,\mathrm{\mu} g/m^3$. These results are encouraging although the two figures are not directly comparable. Indeed our approach continuously adjusts the regression model using both sensor and pollutant data whereas \cite{DeVito2009} makes predictions for most of the observation period using only sensor readings.


\section{Discussion}  
 \label{sec:discussion}

This paper has introduced the sparse group fused lasso (SGFL) as a statistical paradigm for segmenting regression models in the context of high-dimensional time series. The objective function used in SGFL favors sparsity in regression coefficients via a lasso penalty and parsimony in the number of segments or change points via an $\ell_2$ total variation penalty. This combination of penalties is essential for  model segmentation in high dimension and has very rarely been studied in the literature. 
More common variants of fused lasso based on $\ell_1$ total variation penalties are not appropriate for this purpose. 

A hybrid algorithm has been developed to optimize the SGFL objective function, which is a highly nontrivial problem of nonsmooth and nonseparable convex optimization. This algorithm exploits the problem's structure by operating at different levels 
(block, single fusion chain, all fusion chains, all blocks) with optimization techniques such as block coordinate descent and subgradient methods. 
With its ability to perform local or global, aggregating or splitting iterations, the hybrid algorithm can flexibly explore the search space but also ``lock in" a given model segmentation and make extremely fast progress. 
In addition to its computational speed, the hybrid algorithm bears the advantage of not requiring any complicated selection of tuning parameters, which renders it more accessible to non-expert users. 

In our simulations, the proposed algorithm has realized significant speed gains in comparison to state-of-the-art techniques like ADMM and primal-dual methods. The speedup was particularly important in high-dimensional situations with millions of optimization variables (30\%-40\% speedup), which is particularly encouraging in view of large scale applications. 
With regards to statistical accuracy, the SGFL has shown reasonable performance in recovering nonzero regression coefficients although competing methods were more accurate. It performed extremely well in change point detection, even at high noise levels, which the other methods could not achieve. A simplified two-step version of the SGFL has shown excellent computational speed and accuracy in the recovery of sparsity patterns as well as very good performance in change point detection (at least in low to moderate noise). This simplified approach emerges as an interesting alternative to SGFL and is certainly worthy of further investigation. 
The practical effectiveness of the SGFL has been illustrated with the analysis of air quality data, which has given insights on the frequency of changes in the relationship between sensor readings and target pollutant levels and on the attainable prediction performance.  
A challenge in our simulations and data analyses using SGFL was the selection of regularization parameters ($\lambda_1,\lambda_2$). 
While the standard AIC/BIC lead to overfitting the data in multiple instances (i.e. too many change points and/or too many nonzero regression coefficients), the alternative selection criterion we found, HBIC, seemed overly conservative in moderate to high noise. In addition, the HBIC itself requires selecting an extra regularization parameter ($\gamma$). 
Theoretical investigations of the asymptotics of SGFL and more numerical experimentation will be needed to find suitable selection methods.

The hybrid method for the SGFL is implemented in the R package \texttt{sparseGFL} available at \url{https://github.com/ddegras/sparseGFL}.

 \subsection*{Extensions}

 \paragraph{Elastic net penalty.} 

The sparse group fused lasso can be extended to encompass an \emph{elastic net} penalty (\cite{Zou2005}):
 \begin{equation}\label{SGFEN}
 \begin{split}
\min_{\beta \in \mathbb{R}^{pT}} & \left\{ 
\frac{1}{2} \sum_{t=1}^{T}  \|  y_{t} -  X_{t} \beta_{t}    \|_2^2 
+ \lambda_1\sum_{t=1}^{T} \left( \alpha  \, \| \beta_{t}  \|_1 
+ \displaystyle \frac{ (1-\alpha) }{2} \,  \| \beta_{t}  \|_2^2 \right) \right. \\
& \left. \quad + \lambda_2 \sum_{t=1}^{T-1} w_{t}   \, \| \beta_{t+1} - \beta_{t}  \|_2 \right\} .
\end{split}
\end{equation}
 The elastic net penalty combines the lasso penalty and the (squared $\ell_2$) ridge regression penalty thanks to a mixing coefficient $\alpha \in [0,1]$ ($\alpha = 1 $ corresponds to pure lasso, $\alpha = 0$  to pure ridge). This penalty seeks the ``best of both worlds", namely the sparsity-inducing effect of the lasso and the stabilization effect of ridge regression. In particular the ridge penalty can  mitigate the adverse effects of high correlation among predictors in lasso. Also, from a theoretical perspective, when $\alpha < 1$, the objective function $F$ is strictly convex and thus admits a unique minimizer.

 \paragraph{Loss function.} 
For simplicity of exposition, we have developed the SGFL and hybrid optimization method in the context of  linear regression. However, they are by no means restricted to linear regression: the squared loss in the objective function \eqref{objective} can be replaced by any differentiable loss function. For example, our methodology can be used for classification problems using exponential,  logistic, or generalized smooth hinge loss functions.

\subsection*{Future lines of research}

\paragraph{Pathwise implementation of SGFL.} 

In practice, one rarely solves the SGFL for a single pair of regularization parameters $(\lambda_1,\lambda_2)$ but rather along a path of values. %
The selection of such path is in itself a nontrivial problem for two main reasons: first the interplay between $\lambda_1$ and $\lambda_2$ in the lasso and total variation penalty and second, the computation time required to fit the SGFL. 
To elaborate on the first point, for a given $\lambda_1$, varying $\lambda_2$ may not only affect the number of segments produced by the SGFL solution $\hat{\beta}=\hat{\beta}(\lambda_1,\lambda_2)$ but also the sparsity of this solution. Conversely, for fixed $\lambda_2$, varying $\lambda_1$ may not affect only the sparsity of the solution but also the model segmentation.  
  
One strategy may be to calculate regularization paths over a product grid $G=G_1\times G_2$.
 The problem then is to determine ranges and sizes for $G_1$ and $G_2$ that yield a 
sufficiently fine-grained path of solutions $\hat{\beta}(\lambda_1,\lambda_2)$ with a  
 good variety of sparsity levels and segmentations while keeping the computational load reasonable. 
The largest value in $G_1$, say  $\lambda_{1\mathrm{max}}$, can be determined analytically, 
for example as  $\lambda_{1\mathrm{max}} = \frac{1}{T} \| \sum_{t=1}^T X_t'y_t \|_{\infty}$, 
so that for $\lambda_2 = \infty$ (time-invariant solution)
$\hat{\beta}(\lambda_1,\infty) \equiv 0$ if and only if  $\lambda_1 \ge \lambda_{1\mathrm{max}}$. 
One can then take $G_1$ as an equispaced grid on a linear or logarithmic scale going from $\lambda_{1\mathrm{max}}$ to a small fraction of $\lambda_{1\mathrm{max}}$, say 0.001 or 0.0001. 
The question remains to select a suitable grid $G_2$. More precisely, for a given $\lambda_1 \in G_1$, it would be useful to know the smallest $\lambda_2=\lambda(\lambda_1)$ such that $\hat{\beta}(\lambda_1,\lambda_2) = \hat{\beta}(\lambda_1,\infty)$, 
say $\lambda_{2 \mathrm{max}}(\lambda_1)$. With this information, one can avoid wasting time solving \eqref{objective} for $(\lambda_1,\lambda_2)$ with $\lambda_2 \ge \lambda_{2 \mathrm{max}}(\lambda_1)$, 
which by definition results in the solution $\hat{\beta}(\lambda_1,\infty)$. The problem of finding $\lambda_{2 \mathrm{max}}(\lambda_1)$ can be formulated as $\min_{\lambda_2,U,V} \lambda_2$ subject to the  optimality conditions \eqref{global optimality}-\eqref{constraints u} where $U=(u_1,\ldots,u_T)$ and $V=(v_1,\ldots,v_{T-1})$.
By eliminating the $v_t$ from \eqref{global optimality}, 
this problem can be formulated as the min-max problem  
\begin{equation}
 \min_{U} \max_{1\le t \le T} \frac{1}{w_t} \Big\| \sum_{s=1}^t z_s + \lambda_1 \sum_{s=1}^t u_s \Big\|_2 
\end{equation}
subject to the constraints \eqref{constraints u} and  $\sum_{t=1}^T z_s + \lambda_1 \sum_{t=1}^T u_t = 0_p$. 
We leave this nontrivial problem as a direction for future research.

 \paragraph{Screening rules.}

In lasso and fused lasso regression, there exist powerful screening rules for identifying zero coefficients  or fused coefficients in solutions before computing these solutions \cite{Tibshirani2012b,Wang2015,Wang2015b}.  Such screening rules often greatly reduce the number of variables to optimize in the objective function and considerably speed up calculations. It would be interesting to see if existing rules can be adapted to the more difficult problem of SGFL or if novel screening rules can be devised for it. 
A difficulty of SGFL is that it does not possess the monotonic inclusion property of lasso and fused lasso. 
For example, the fact that two successive vectors $\hat{\beta}_t$ and $\hat{\beta}_{t+1}$ are fused in a SGFL solution $\hat{\beta}(\lambda_1,\lambda_2)$ does \emph{not} imply that they stay fused in a solution $\hat{\beta}(\lambda_1,\lambda_2')$ with $\lambda_2' \ge \lambda_2$.


\appendix

\section{Proof of Theorem \ref{thm: linear convergence}}
\label{sec: linear convergence}

Recall the notations of Section \ref{sec:IST}: 
\begin{equation*}
\begin{split}
g(\beta_t)& = \lambda_1  \| \beta_t \|_1 + \lambda_2 w_{t-1}\| \beta_t - \hat{\beta}_{t-1} \|_2 + \lambda_2 w_t \| \beta_t - \hat{\beta}_{t+1} \|_2,  \\ 
g_1(\beta_t)& = \lambda_1   \| \beta_t \|_1 , \\
g_2(\beta_t) &= \lambda_2 w_{t-1}  \| \beta_t - \hat{\beta}_{t-1}  \|_2  +\lambda_2w_t \| \beta_t- \hat{\beta}_{t+1}   \|_2  + (L_t/2) \|  \beta_t -z_t \|_2^2 ,\\
 \bar{g} (\beta_t)& = g(\beta_t) + (L_t/2) \left\| \beta_t - z_t \right\|_2^2 = g_1  (\beta_t)+   g_2 (\beta_t) , \\ 
 \quad \gamma_n &= \big(  L_t + (\lambda_2 w_{t-1} / \| \beta_t^{n} - \hat{\beta}_{t-1} \|_2 ) +
(\lambda_2 w_t / \| \beta_{t}^{n} - \hat{\beta}_{t+1}  \|_2 ) \big)^{-1},  \\
\beta_t^\ast & = \mathrm{argmin}\, \bar{g} = \mathrm{prox}_{g/L_t}(z_t) , \quad r_n =  \bar{g} (\beta_t^n) -  \bar{g} (\beta_t^\ast).
\end{split}
\end{equation*}

In view of  Remark \ref{fixed point as forward-backward} framing the iterative soft-thresholding scheme \eqref{fp operator}-\eqref{fp iteration} as a proximal gradient method, we can establish the linear convergence of this scheme to $\mathrm{prox}_{g/L_t}(z_t)$ by adapting the results of \cite{Bredies2008} to a nonsmooth setting.  
Essentially, the proof of linear convergence in \cite{Bredies2008} works by first establishing a lower bound on $\bar{g}(\beta_t^{n})-\bar{g}(\beta_t^n)$, the decrease in the objective function between successive iterations of the proximal gradient method (Lemma 1). This general result shows in particular that when using sufficiently small step sizes, the proximal gradient is a descent method. After that, under the additional assumptions that $g_2$ is convex and that $\| \beta_t^n - \beta_t^\ast  \|_2^2 \le c r_n$ for some $c>0$, 
the lower bound of Lemma 1 is exploited to show the exponential decay of $(r_n)$ and the linear convergence of $(\beta_t^n)$ (Proposition 2). In a third movement, the lower bound of Lemma 1 is decomposed as a Bregman-like distance term involving $g_1$ plus a Taylor remainder term involving $g_2$.  The specific nature of $g_1$ ($\ell_1$ norm) and possible additional regularity conditions on $g_2$ (typically, strong convexity) are then used to establish the linear convergence result (Theorem 2). For brevity, we refer the reader to \cite{Bredies2008} for the exact statement of these results. 

Lemma 1 and Proposition 2 of \cite{Bredies2008} posit, among other things, that the ``smooth" part of the objective, $g_2$ in our notations, is  differentiable everywhere and has a Lipschitz-continuous gradient. In the present case, $g_2$ is not differentiable at $\hat{\beta}_{t-1}$ and  $\hat{\beta}_{t+1}$; however it is differentiable everywhere else and its gradient is Lipschitz-continuous in a local sense. The main effort required for us is to show that Lemma 1 still holds if the points of nondifferentiability of $g_2$ are not on segments joining the iterates $\beta_t^n , n\ge 0$. Put differently, the iterative soft-thresholding scheme should not cross $\hat{\beta}_{t-1}$ and  $\hat{\beta}_{t+1}$ on its path. This is where the requirement that $\bar{g}(\beta_t^0) < \min( \bar{g}(\hat{\beta}_{t-1}), \bar{g}(\hat{\beta}_{t+1}))$ in  Theorem \ref{thm: linear convergence} plays a crucial part. We now proceed to adapt Lemma 1, after which we will establish the premises of Theorem 2 of  \cite{Bredies2008}.

\paragraph{Adaptation of Lemma 1 of  \cite{Bredies2008}.}

The main result we need prove is that 
\begin{equation}\label{clear path}
\forall n\in \mathbb{N}, \quad \big\{ \hat{\beta}_{t-1} , \hat{\beta}_{t+1} \big\} \bigcap^{\,}
 \left\{ \alpha \beta_{t}^{n} + (1-\alpha)  \beta_{t}^{n+1} : 0 \le \alpha  \le 1 \right\}  = \emptyset \,.
\end{equation}
Once this is established, we may follow the proof of Proposition 2 without modification. 
In particular, we will be in position to state that 
\begin{equation}\label{eq: local Lipschitz}
 \left\| \nabla g_2(\beta_t^n + \alpha (\beta_t^{n+1} - \beta_t^n )
 ) - \nabla g_2(\beta_t^{n})\right\|_2 \le \alpha \tilde{L}_n \left\| \beta_t^{n+1} - \beta_t^{n} \right\|_2 
\end{equation}
for all $n\in \mathbb{N}$ and $ \alpha \in[0,1]$, where 
$$ \tilde{L}_n =   L_t + \frac{2\lambda_2 w_{t-1} }{ \| \beta_t^n - \hat{\beta}_{t-1} \|_2 } +\frac{2\lambda_2 w_t }{ \| \beta_t^n - \hat{\beta}_{t+1} \|_2 } \, . $$ 
Note that the left-hand side in \eqref{eq: local Lipschitz} is not well defined if \eqref{clear path} does not hold. 
Combining the local Lipschitz property \eqref{eq: local Lipschitz} 
with the step size condition $\gamma_{n} < 2 / \tilde{L}_n$, 
we may go on to establish the descent property (3.5) of \cite{Bredies2008}: 
\begin{equation}\label{descent property}
\bar{g}(\beta_t^{n+1}) \le \bar{g}(\beta_t^{n})  - \delta D_{\gamma_n}(\beta_t^{n})
\end{equation}
where 
\begin{equation*}
D_{\gamma_n}(\beta_t^{n}) = g_1(\beta_t^{n}) - g_1(\beta_t^{n+1}) +  \nabla g_2(\beta_t^{n})'(\beta_t^{n}-\beta_t^{n+1}) \quad \textrm{ and } \quad \delta = 1 - \frac{\max_{n} \gamma_n \tilde{L}_n }{ 2} \, .
\end{equation*}
Lemma 1 shows that $D_{\gamma_n}(\beta_t^{n}) \ge \| \beta_t^{n}-\beta_t^{n+1}\|_2^2 / \gamma_n  \ge 0$. 
To show the positivity of $\delta$, note that 
\begin{equation*}
\gamma_n \tilde{L}_n = 2 - L_t \left( \displaystyle L_t + \frac{\lambda_2 w_{t-1} }{ \| \beta_t^{n} - \hat{\beta}_{t-1}\|_2} + \frac{\lambda_2 w_{t}}{  \| \beta_t^{n} - \hat{\beta}_{t+1}\|_2 } \right)^{-1} \,.
\end{equation*}
Given  the descent property of $(\beta_n)$ for $\bar{g}$, the assumption 
$\bar{g}(\beta_t^0) < \min(\bar{g}(\hat{\beta}_{t-1}),\bar{g}(\hat{\beta}_{t+1})) $, and the convexity of the sublevel sets of $\bar{g}$, 
it holds that $  
 \| \beta_t^{n} - \hat{\beta}_{t-1}\|_2 \ge d(\hat{\beta}_{t-1} , \{ \beta_t : \bar{g}(\beta_t)\le \bar{g}(\beta_t^0)  \}) $ 
 for all $n\in\mathbb{N}$; an analog inequality holds for $\hat{\beta}_{t+1}$. 
 Denoting these positive lower bounds by $m_{t-1} $ and $m_{t+1}$, we have 
\begin{equation}\label{eq: delta positive}
0 < \frac{1}{2} \left( \displaystyle 1 + \frac{\lambda_2 w_{t-1} }{L_t m_{t-1}} + \frac{\lambda_2 w_{t}}{ L_t m_{t+1} } \right)^{-1} \le 
\delta \le \frac{1}{2} \, .
\end{equation}
Together, the step size condition $\gamma_{n} < 2 / \tilde{L}_n$, descent property \eqref{descent property}, 
and lower bound \eqref{eq: delta positive} finish to establish Lemma 1 and the precondition of Proposition 2 of \cite{Bredies2008}.

It remains to prove \eqref{clear path}. We will show a weaker form of \eqref{descent property}, namely that $\bar{g}(\beta_t^{n+1}) \le \bar{g}(\beta_t^{n}) $ for all $n$. This inequality, combined with the convexity of $\bar{g}$ and the assumption $\bar{g}(\beta_t^0) < \min(\bar{g}(\hat{\beta}_{t-1}),\bar{g}(\hat{\beta}_{t+1})) $, implies that $\hat{\beta}_{t-1}$ and $\hat{\beta}_{t+1}$ cannot be on a segment joining $\beta_t^{n}$ and $\beta_t^{n+1}$. Otherwise, the convexity of $\bar{g}$ would imply that, say, $\bar{g}(\hat{\beta}_{t-1}) \le \max(  \bar{g}(\beta_t^{n}), \bar{g}(\beta_t^{n+1})) \le \bar{g}(\beta_t^{n})\le \cdots \le \bar{g}(\beta_t^{0}) < \bar{g}(\hat{\beta}_{t-1})$, a contradiction. 

To prove the simple descent property, we start with an easy lemma stated without proof. 
\begin{lemma}
For all $x,y \in\mathbb{R}^p$ such that $y \ne 0_p$, 
\begin{equation*}
\| x \|_2 \le \| y \|_2 + \frac{ y'(x-y) }{\| y \|_2 } + \frac{\| x - y\|_2^2 }{2\| y\|_2} \,.
\end{equation*}
\end{lemma}
Applying this lemma to $x= \beta_t - \hat{\beta}_{t \pm 1}$ and $y  = \beta_t^{n} - \hat{\beta}_{t \pm 1}$, 
we deduce that for all $\beta_t \in \mathbb{R}^p $, 
\begin{align}
\label{g2 tm1}
 \|  \beta_t - \hat{\beta}_{t - 1}\|_2 &  \le \| \beta_t^{n} - \hat{\beta}_{t - 1}\|_2
 +   \frac{ ( \beta_t - \hat{\beta}_{t - 1})'(\beta_t - \beta_t^n) }{\|  \beta_t^n - \hat{\beta}_{t - 1} \|_2 } + \frac{\| \beta_t - \beta_t^n \|_2^2 }{2\|  \beta_t^n - \hat{\beta}_{t - 1} \|_2} \, , \\
 \label{g2 tp1}
 \|  \beta_t - \hat{\beta}_{t + 1}\|_2 &  \le \| \beta_t^{n} - \hat{\beta}_{t + 1}\|_2
 +   \frac{ ( \beta_t - \hat{\beta}_{t + 1})'(\beta_t - \beta_t^n) }{\|  \beta_t^n - \hat{\beta}_{t +1} \|_2 } 
 + \frac{\| \beta_t - \beta_t^n \|_2^2 }{ 2 \|  \beta_t^n - \hat{\beta}_{t + 1} \|_2} \, . \\
\noalign{\noindent In addition, it is immediate that} \nonumber \\
\label{g2 quad}
\| \beta_t - z_t \|_2^2 & = \| \beta_t^n - z_t \|_2^2  - 2 (\beta_n - z_t)'(\beta_t - \beta_t^n ) +   \| \beta_t - \beta_t^n \|_2^2 \, .
\end{align}
Multiplying \eqref{g2 tm1} by $\lambda_2 w_{t-1}$, \eqref{g2 tp1} by $\lambda_2 w_{t} $, \eqref{g2 quad} by $L_t/2$, 
summing these relations, and adding  $g_1(\beta_t)$ on each side, 
we obtain 
\begin{equation}\label{gbar majorant}
(g_1 + g_2)(\beta_t) \le g_1(\beta_t) + g_2(\beta_t^n) + \nabla g_2(\beta_t^n) ' (\beta_t - \beta_t^n) + \frac{1}{2\gamma_n} \| \beta_t - \beta_t^n \|_2^2 . 
\end{equation}
 The left-hand side  of \eqref{gbar majorant} is simply $\bar{g}(\beta_t)$. Also, in view of Remark \ref{fixed point as forward-backward}, 
the minimizer of the right-hand side of \eqref{gbar majorant} is $\mathcal{T}(\beta_t^n) = \beta_t^{n+1}$. 
Evaluating \eqref{gbar majorant} at $ \beta_t^{n+1}$ and exploiting this minimizing property, 
it follows that 
\begin{equation}
\begin{split}
 \bar{g}(\beta_t^{n+1} )& \le g_1(\beta_t^{n+1} ) + g_2(\beta_t^n) + \nabla g_2(\beta_t^n) ' (\beta_t^{n+1} - \beta_t^n) + \frac{1}{2\gamma_n} \| \beta_t^{n+1}  - \beta_t^n \|_2^2  \\
 & \le g_1(\beta_t^{n} ) + g_2(\beta_t^n) + \nabla g_2(\beta_t^n) ' (\beta_t^n - \beta_t^n) + \frac{1}{2\gamma_n} \| \beta_t^{n}  - \beta_t^n \|_2^2  \\
 & = \bar{g}(\beta_t^n). 
 \end{split}
\end{equation}
This establishes the desired descent property. 

\paragraph{Prerequisites of Theorem 2 of  \cite{Bredies2008}.}

The distance $r_n = \bar{g}(\beta_t^n) - \bar{g}(\beta_t^\ast)$ to the  minimum of the objective 
can be usefully decomposed as 
\begin{equation}
\begin{split}
r_n & = R(\beta_t^n) + T(\beta_t^n)  \\
R(\beta_t) & =   \nabla g_2(\beta_t^\ast) '( \beta_t - \beta_t^\ast )  + g_1(\beta_t) - g_1(\beta_t^\ast) \\
T(\beta_t) & = g_2(\beta_t) - g_2(\beta_t^\ast) -    \nabla g_2(\beta_t^\ast)'( \beta_t - \beta_t^\ast )
\end{split}
\end{equation}
where $R(\beta_t)$ is a Bregman-like distance and 
$T(\beta_t)$ is the remainder of the Taylor expansion of $g_2$ at $\beta_t^\ast$.  

To obtain the linear convergence of $(\beta_t^n)$ to $\beta_t^\ast $ 
and the exponential decay of $(r_n)$ to 0 with Theorem 2 of  \cite{Bredies2008}, it suffices to show that 
\begin{equation}\label{RT bound}
 \| \beta_t - \beta_t^\ast\|_2^2 \le c \left(R(\beta_t) + T(\beta_t)\right) 
\end{equation}
for some constant $c>0$ and for all $\beta_t\in\mathbb{R}^p$.

Invoking the convexity of $\| \cdot \|_2$ and strong convexity of $\| \cdot \|_2^2$, one sees that 
$ T(\beta_t) \ge (L_t/2) \| \beta_t - \beta_t^\ast\|_2^2$ for all $\beta_t$. 
Also,  $R(\beta_t) \ge 0$ for all $\beta_t$ (Lemma 2 
of \cite{Bredies2008}) so that $c$ can be taken as $2/L_t$ in \eqref{RT bound}. $\square$


\section{Proof of Theorem \ref{thm: convergence}}
\label{Appendix B}

We first observe that by design, each of the four steps or components of Algorithm \ref{alg:main} is nonincreasing in the objective function $F$. Indeed the first three steps (optimization with respect to single blocks, single chains, and descent over fixed chains) are  all based on FISTA (Algorithms \ref{alg:FISTA constant step size} and \ref{alg:FISTA backtracking}) which is globally convergent  (\cite[Theorem 4.4]{Beck2009}). As each of these components minimizes $F$ under certain constraints (namely, some blocks or fusion chains are fixed), the objective value of their output, say $F(\beta^{n+1})$, cannot be lower than that of their input, $F(\beta^n)$. The fourth step, subgradient descent is also nonincreasing because the subgradient of minimum norm - if it is not zero - provides a direction of (steepest) descent. The line search for the step size in the subgradient step then guarantees that the objective does not decrease after this step. As a result,  Algorithm \ref{alg:main} as a whole is nonincreasing in $F$.

Let us denote a generic segmentation of the set $\{ 1, \ldots, T\}$ by $C=(C_1,\ldots, C_K)$ where $C_k =\{ T_k ,\ldots, T_{k+1} - 1\} $ and  $1= T_1 \le \cdots \le T_K < T_{K+1}=T+1$. There are $2^{T-1}$ such segmentations. 
Let $S_C$ be the associated open set for the parameter $\beta$: 
\begin{equation*}
S_C  = \left\{ \beta \in \mathbb{R}^{pT}: \beta_{T_k} = \cdots = \beta_{T_{k+1}-1},\  \beta_{T_{k+1}-1} \ne  \beta_{T_{k+1}} , 1\le k \le K \right\} .
\end{equation*}
To each segmentation $C$ is associated an infimum value of $F$: $\inf_{\beta \in S_C}F(\beta)$. 
Let $(\beta^n)_{n\ge 0}$ be the sequence of iterates generated by Algorithm \ref{alg:main} and let $C^{n}$ the associated segmentations of $\{1,\ldots,T\}$.  
By setting the tolerance $\epsilon$ of Algorithm \ref{alg:main} sufficiently small, each time the third optimization component (descent over fixed chains) is applied, say with $\beta^n$ as input and $\beta^{n+1}$ as output, the objective $F(\beta)$ can be made arbitrarily close to $\min_{ \beta \in S_{C^n}}F(\beta)$ or even become inferior to this value if a fusion of chains occurs during this optimization.  
If the segmentation $C^n$ is optimal, i.e. $ \min_{  S_{C^n} } F (\beta) = \min_{\beta \in \mathbb{R}^{pT}}F (\beta)$, then Algorithm \ref{alg:main} has converged: for all subsequent iterates $m\ge n$, 
$F(\beta^m)$ and $\beta^m$ will stay arbitrarily close to the minimum of $F$ and to the set of minimizers, respectively, 
 because of the nonincreasing property of Algorithm \ref{alg:main}. 
If the segmentation $C^n$ is suboptimal, i.e. $ \min_{\beta \in S_{C^n} } F(\beta) > \min_{\beta \in \mathbb{R}^{pT}}F (\beta)$, 
provided that $\epsilon$ is sufficiently small, the (fourth) subgradient step of Algorithm \ref{alg:main} 
will eventually produce an iterate $\beta^m$ ($m \ge n$) such that  $F(\beta^m) <  \min_{ \beta \in S_{C^n} }F(\beta) $. 
This is because each subgradient step brings the iterates closer to the set of global minimizers of $F$. 
Once this has happened, the nonincreasing property of the algorithm guarantees that the segmentation $C^n$ will not be visited again. 
Because the segmentations of $\{ 1,\ldots,T\}$ are in finite number, Algorithm \ref{alg:main} eventually finds an optimal segmentation $C$ such that $\min_{\beta \in S_C} F(\beta) = \min_{\beta \in\mathbb{R}^{pT}} F(\beta)$. Then, through its third level of optimization (descent over fixed chains), it reaches the global minimum of $F$. 
We note that the first and second components of Algorithm \ref{alg:main} (block coordinate descent over single blocks and single chains) are not necessary to ensure global convergence; they only serve for computational speed. 
$\square $

\newpage

\bibliographystyle{apalike}      
\bibliography{biblio}



\clearpage

\begin{center}
\LARGE{Supplementary Materials for\\ ``Sparse Group Fused Lasso for Model Segmentation"}
\end{center}

\vspace*{2cm}

This document contains additional simulation results in complement of section 4.2 (Simulations: statistical accuracy)  of the main manuscript. These results were obtained in the same simulation setup as section 4.2 but with different values of the simulation parameters $s$ (sparsity level of regression vectors), $\rho_{X}$ (correlation level among predictors), and/or $\sigma_\varepsilon$ (noise level).

\section*{Statistical accuracy} 

This section provides additional results on statistical accuracy. As in section 4.2, the following tables have columns that indicate the  estimation method, selection method for  regularization parameters, number of change points, Hausdorff distance between estimated and true change points, true positive rate and positive predictive value in detecting  nonzero regression coefficients, sparsity level, and pseudo-$R^2$. Results are averaged over all replications of a setup, with standard deviations in brackets. The best results for the HBIC  selection method are in bold.

\clearpage

Table \ref{table: s99 more} shows results under the following scenario: sparsity level $s=0.99$, 
model dimensions $(d,p,T)=(20,200,100)$ or $(d,p,T)=(100,500,200)$, no correlation between predictors ($\rho_X=0)$. 
In comparison to Table 3 of the main article, the noise levels $\sigma_{\varepsilon} = 0 $ (no noise) and $\sigma_\varepsilon = 2.5$ (high noise) are examined. The excellent performances of the BSA and 2S methods in a noiseless situation are confirmed as well as their breakdown in high noise. Also confirmed is the lesser performance of SGFL in recovering nonzero coefficients in low noise ($\sigma_\varepsilon = 0$) and its super performance in high noise  ($\sigma_\varepsilon = 2.5$). 
The high-dimensional setup does not seem to affect statistical performance although, 
as noted in the main article, the increase in $d$ (20 to 100) and $p$ (100 to 500) may be counterbalanced by the increase in $T$ (100 to 200), which makes the model segments longer and the associated change points (possibly) easier to detect.


\begin{table}[H]
\centering
\caption{Simulation study of statistical accuracy. True sparsity level $s=0.99$, correlation $\rho_X =0$, number of replications: 100.}
\label{table: s99 more}
\bigskip
\hspace*{-8mm}
\footnotesize{
\begin{tabular}{c r c  c c c c c  c}
  \hline
Setup & Method & $\lambda$ & NCP & $d_H$ & TPR & PPV & $\widehat{s}$ & $R^2$ \\ 
\hline
\multirow{5}{15mm}{\vspace{-4mm}
\begin{equation*}
\begin{array}{r  l}
d&=20\\ 
p&=200\\
T&=100\\
\sigma_{\varepsilon} &= 0 \\
\noalign{$\{21, 51, 91\}$}
\end{array}
\end{equation*}
} 
&BSA & HBIC & \textbf{3.0} (0.0) & \textbf{0.0} (0.0) & \textbf{1.00} (0.00) & \textbf{1.00} (0.00) & \textbf{0.99} (0.00) & \textbf{1.00} (0.00) \\ 
 & 2S & HBIC & 3.4 (0.6) & 0.4 (0.5) & 1.00 (.02)&  1.00 (0.02) &  0.99 (0.00) & 1.00 (0.00) \\ 
  & SGFL & HBIC & 3.0  (0.1) & 1.1 (5.6) & 0.93 (.09)& 0.72 (0.11)& 0.99 (0.00) & 0.96 (0.01) \\  
 & BSA & $d_H$ & 3.0 (0.0) & 0.0 (0.0)& 1.00 (.01)& 1.00 (0.00)& 0.99 (0.00)& 1.00 (0.00)\\ 
 & 2S & $d_H$ & 3.4 (0.5) & 0.4 (0.5)& 1.00 (.02) & 1.00 (0.01) & 0.99 (0.00)& 0.97 (0.05)\\ 
  & SGFL & $d_H$ & 3.0 (0.0) &0.0 (0.0) &0.91 (.10) &0.93 (0.13) &0.99 (0.02) &0.86 (0.10)\\  
\hline
\multirow{5}{12mm}{\vspace{-3.5mm}
\begin{equation*}
\begin{array}{r  l}
d&=100\\ 
p&=500\\
T&=200\\
\sigma_{\varepsilon} &= 0 \\
\noalign{\hspace*{-1.5mm}$\{ 41,101, 81\}$}
\end{array}
\end{equation*}
}   
& BSA & HBIC & {\bf 3.0} (0.0) & {\bf 0.0} (0.0) & {\bf 1.00} (0.01) &{\bf 1.00} (0.00) & 0.99 (0.99) &{\bf1.00} (0.00)\\  
  & 2S & HBIC &  {\bf 3.0} (0.0) & {\bf 0.0} (0.0) & {\bf 1.00} (0.01) &{\bf 1.00} (0.00) & 0.99 (0.99) &{\bf1.00} (0.00)\\ 
   & SGFL & HBIC & {\bf 3.0} (0.0) &{\bf 0.0} (0.0) &0.97 (0.04) &0.51 (0.05) &0.98 (0.00) &0.97 (0.00)\\   
   & BSA & $d_H$ & 3.0 (0.0) &0.0 (0.0) &1.00 (0.01) &1.00 (0.00) &0.99 (0.00) &1.00 (0.00)\\
   & 2S & $d_H$ & 3.0 (0.0) &0.0 (0.0) &1.00 (0.01) &1.00 (0.00) &0.99 (0.00) &0.99 (0.02)\\
   & SGFL & $d_H$ &3.0 (0.0) &0.0 (0.0) &0.89 (0.06) &0.94 (0.05) &0.99 (0.00) &0.90 (0.06)\\
\hline
\multirow{5}{15mm}{\vspace{-3.5mm}
\begin{equation*}
\begin{array}{r  l}
d&=100\\ 
p&=500\\
T&=200\\
\sigma_{\varepsilon} &= 2.5 \\
\noalign{\hspace*{-3mm}$\{ 41,101, 181\}$}
\end{array}
\end{equation*}
}   
& BSA & HBIC & 1.6 (1.2) & 95.0 (77.1) &{\bf 0.98} (0.03) &0.02 (0.00) &0.45 (0.10) &0.35 (0.17)\\
  & 2S & HBIC & 4.4 (1.8) &18.8 (32.2) & 0.84 (0.14) & 0.82 (0.10) &0.99 (0.00) &{\bf 0.40} (0.10) \\ 
   & SGFL & HBIC & {\bf 3.0} (0.2) & {\bf 2.3} (12.7) &0.81 (0.12) & {\bf 0.95} (0.05) &{\bf 0.99} (0.00) &0.37 (0.09) \\ 
   & BSA & $d_H$ & 3.0 (0.0) &0.0 (0.0) &0.99 (0.02) &1.00 (0.02) &0.99 (0.00) &0.99 (0.00)\\
   & 2S & $d_H$ & 3.0 (0.2) &0.0 (0.2) &0.98 (0.03) &1.00 (0.01) &0.99 (0.00) &0.98 (0.01)\\
   & SGFL & $d_H$ & 3.0 (0.0) &0.0 (0.0) &0.90 (0.08) &0.94 (0.04) &0.99 (0.00) &0.89 (0.05)\\
\hline
\end{tabular}}
\end{table}


\clearpage

Table \ref{table: s90 hd} displays results for the setup $s=0.90$, $(d,p,T)=(100,500,200)$, and $\rho_X=0$. 
Only 10 replications were performed in this case due to the large computational demand of the high model dimensions.  
The table shows a better performance of BSA and 2S, in comparison to the situation where $s=0.99$. 
This can be interpreted as follows: the lower sparsity $s=0.90$ produces more marked changes between regression vectors $\beta_t$ at the change points $t = T_k$. More precisely, the jump size $\| \beta_{T_{k+1}} - \beta_{T_k} \|$ becomes larger (in expected value). 
As a result, change points become easier to detect, which improves performance.

\begin{table}[!ht]
\centering
\caption{Simulation study of statistical accuracy. True sparsity level $s=0.90$, correlation $\rho_X =0$, number of replications: 10.}
\label{table: s90 hd}
\bigskip
\hspace*{-5mm}
\footnotesize{
\begin{tabular}{c r c  c c c c c  c}
  \hline
Setup & Method & $\lambda$ & NCP & $d_H$ & TPR & PPV & $\widehat{s}$ & $R^2$ \\ 
\hline
\hspace*{-6mm}
\multirow{5}{15mm}{\vspace{-4mm}
\begin{equation*}
\begin{array}{r  l}
d&=100\\ 
p&=500\\
T&=200\\
\sigma_{\varepsilon} &= 0 \\
\noalign{$\{41, 101, 181\}$}
\end{array}
\end{equation*}
} 
&BSA & HBIC & \textbf{3.0} (0.0) & \textbf{0.0} (0.0) & 1.00 (0.00) & \textbf{1.00} (0.00) & 0.90 (0.00) & 1.00 (0.00) \\ 
 & 2S & HBIC & {\bf 3.0} (0.0) &{\bf 0.0} (0.0) &{\bf 1.00} (0.00) &{\bf 1.00} (0.00) & {\bf 0.90} (0.00) &{\bf 1.00} (0.00) 
 \\ 
  & SGFL & HBIC & {\bf 3.0} (0.0) & {\bf 0.0} (0.0) &  0.90 (0.03) &0.86 (0.03) &0.90 (0.01) &0.94 (0.01)\\  
 & BSA & $d_H$ & 3.0 (0.0) &0.0 (0.0) &1.00 (0.00) &1.00 (0.00) &0.90 (0.00) &1.00 (0.00)\\
 & 2S & $d_H$ & 3.0 (0.0) &0.0 (0.0) &1.00 (0.00) &1.00 (0.00) &0.90 (0.00) &1.00 (0.00)\\
  & SGFL & $d_H$ &3.0 (0.0) &0.0 (0.0) &0.87 (0.03) &0.92 (0.03) &0.91 (0.01) &0.93 (0.01)\\
\hline
\hspace*{-10mm}
\multirow{5}{12mm}{\vspace{-3.5mm}
\begin{equation*}
\begin{array}{r  l}
d&=100\\ 
p&=500\\
T&=200\\
\sigma_{\varepsilon} &= 0.25 \\
\noalign{$\{41, 101, 181\}$}
\end{array}
\end{equation*}
}   
& BSA & HBIC & {\bf 3.0} (0.0) &{\bf 0.0} (0.0) & {\bf 0.99} (0.01) &{\bf 0.99} (0.01) &{\bf 0.90} (0.0) &{\bf 1.00} (0.00)\\ 
  & 2S & HBIC &  {\bf 3.0} (0.0) &{\bf 0.0} (0.0) & 0.99 (0.01) &0.98 (0.01) &{\bf 0.90} (0.0) &{\bf 1.00} (0.00)\\ 
   & SGFL & HBIC &{\bf 3.0} (0.0) & {\bf 0.0} (0.0) &0.89 (0.04) &0.89 (0.03) &0.90 (0.01) &0.92 (0.01)\\  
   & BSA & $d_H$ & 3.0 (0.0) &0.0 (0.0) &0.99 (0.01) &1.00 (0.00) &0.90 (0.00) &1.00 (0.00)\\ 
   & 2S & $d_H$ &3.0 (0.0) &0.0 (0.0) &0.99 (0.01) &1.00 (0.00) &0.90 (0.00) &1.00 (0.00)\\ 
      & SGFL & $d_H$ & 3.0 (0.0) &0.0 (0.0) &0.87 (0.04) &0.93 (0.02) &0.91 (0.00) &0.92 (0.01)\\ 
\hline
\hspace*{-10mm}
\multirow{5}{15mm}{\vspace{-3.5mm}
\begin{equation*}
\begin{array}{r  l}
d&=100\\ 
p&=500\\
T&=200\\
\sigma_{\varepsilon} &= 1.0 \\
\noalign{$\{41, 101, 181\}$}
\end{array}
\end{equation*}
} 
& BSA & HBIC & {\bf 3.0} (0.0) &{\bf 0.0} (0.0) & {\bf 0.97} (0.01) &0.88 (0.07) &0.89 (0.01) &{\bf 0.98} (0.00)\\ 
  & 2S & HBIC &  {\bf 3.0} (0.0) & {\bf 0.0} (0.0) &0.96 (0.01) &{\bf 0.98} (0.00) &  0.90 (0.00) & 0.98 (0.00)\\ 
   & SGFL & HBIC &  {\bf 3.0} (0.0) &{\bf 0.0} (0.0) &0.89 (0.04) &0.89 (0.03) & {\bf 0.90} (0.01) &0.92 (0.01)\\
   & BSA & $d_H$ & 3.0 (0.0) &0.0 (0.0) &0.97 (0.01) &0.88 (0.07) &0.89 (0.01) &0.98 (0.00)\\
   & 2S & $d_H$ &  3.0 (0.0) &0.0 (0.0) &0.95 (0.01) &0.99 (0.01) &0.90 (0.00) &0.98 (0.00)\\
   & SGFL & $d_H$ & 3.0 (0.0) &0.0 (0.0) &0.88 (0.04) &0.93 (0.02) &0.91 (0.00) &0.90 (0.02)\\
\hline
\hspace*{-10mm}
\multirow{5}{15mm}{\vspace{-3.5mm}
\begin{equation*}
\begin{array}{r  l}
d&=100\\ 
p&=500\\
T&=200\\
\sigma_{\varepsilon} &= 2.5 \\
\noalign{$\{41, 101, 181\}$}
\end{array}
\end{equation*}
} 
& BSA & HBIC & {\bf 3.0} (0.0) &{\bf 0.0} (0.0) & {\bf 0.97} (0.01) &0.28 (0.03) &0.65 (0.04) &{\bf 0.89} (0.01)\\
  & 2S & HBIC &  {\bf 3.0} (0.0) & {\bf 0.0} (0.0) &0.89 (0.02) & {\bf 0.97} (0.01) & {\bf 0.91} (0.00) &0.87 (0.01)\\ 
   & SGFL & HBIC &   {\bf 3.0} (0.0) & {\bf 0.0} (0.0) &0.84 (0.03) &0.92 (0.02) &0.91 (0.00) &0.81 (0.01)\\
   & BSA & $d_H$ & 3.0 (0.0) &0.0 (0.0) &0.97 (0.01) &0.28 (0.03) &0.65 (0.04) &0.89 (0.01)\\
   & 2S & $d_H$ & 3.0 (0.0) &0.0 (0.0) &0.90 (0.03) &0.97 (0.01) &0.91 (0.00) &0.87 (0.01)\\
   & SGFL & $d_H$ & 3.0 (0.0) &0.0 (0.0) &0.83 (0.02) &0.95 (0.02) &0.91 (0.00) &0.80 (0.02)\\
\hline
\end{tabular}}
\end{table}

\newpage


Table \ref{table: s90 md} displays results for the setup $s=0.90$, $(d,p,T)=(20,200,100)$, and $\rho_X=0$. 
The results of this table are qualitatively very similar to those of  \ref{table: s90 hd}. 
In comparison to Table 3 of the main article, they show that the increase in ``signal amplitude" associated 
to the decrease in sparsity of the regression vectors $\beta_t$ (from $s=0.99$ to $s=0.90$) 
offsets the increase in noise so that even when $\sigma_\varepsilon = 1.0 $ or $\sigma_\varepsilon = 2.5 $, the statistical performance of these methods remains good.

\begin{table}[!ht]
\centering
\caption{Simulation study of statistical accuracy. True sparsity level $s=0.90$, correlation $\rho_X =0$,  number of replications: 100.}
\label{table: s90 md}
\bigskip
\hspace*{-5mm}
\footnotesize{
\begin{tabular}{c r c  c c c c c  c}
  \hline
Setup & Method & $\lambda$ & NCP & $d_H$ & TPR & PPV & $\widehat{s}$ & $R^2$ \\ 
\hline
\hspace*{-7mm}
\multirow{5}{15mm}{\vspace{-4mm}
\begin{equation*}
\begin{array}{r  l}
d&=20\\ 
p&=200\\
T&=100\\
\sigma_{\varepsilon} &= 0 \\
\noalign{$\{21, 51, 91\}$}
\end{array}
\end{equation*}
} 
&BSA & HBIC & {\bf 3.0} (0.0) & {\bf 0.0} (0.0) &1.00 (0.00) &{\bf 1.00} (0.00) &0.90 (0.00) &{\bf 1.00} (0.00)\\ 
 & 2S & HBIC &3.3 (0.5) &0.3 (0.5) & {\bf 1.00} (0.01) &0.99 (0.01) &{\bf 0.90} (0.00) &1.00 (0.00)\\ 
   & SGFL & HBIC & {\bf 3.0} (0.0) & {\bf 0.0} (0.0) &0.84 (0.05) &0.85 (0.04) &0.90 (0.01) &0.91 (0.02)\\ 
 & BSA & $d_H$ & 3.0 (0.0) &0.0 (0.0) &1.00 (0.00) &1.00 (0.00) &0.90 (0.00) &1.00 (0.00)\\
 & 2S & $d_H$ & 3.3 (0.5) &0.3 (0.5) &1.00 (0.01) &0.99 (0.01) &0.90 (0.00) &1.00 (0.00)\\
  & SGFL & $d_H$ &3.0 (0.0) &0.0 (0.0) &0.83 (0.05) &0.89 (0.04) &0.91 (0.01) &0.89 (0.03)\\

  \hline
\hspace*{-7mm}
\multirow{5}{15mm}{\vspace{-4mm}
\begin{equation*}
\begin{array}{r  l}
d&=20\\ 
p&=200\\
T&=100\\
\sigma_{\varepsilon} &= 0.25 \\
\noalign{$\{21, 51, 91\}$}
\end{array}
\end{equation*}
} 
&BSA & HBIC & {\bf 3.0} (0.0) & {\bf 0.0} (0.0) & {\bf 0.96} (0.02) & {\bf 0.98} (0.01) &0.90 (0.00) & {\bf 0.99} (0.00)\\ 
 & 2S & HBIC & 3.4 (0.5) &0.4 (0.5) &0.95 (0.03) &0.96 (0.01) &0.90 (0.00) &0.99 (0.00)\\ 
   & SGFL & HBIC & {\bf 3.0} (0.0) & {\bf 0.0} (0.0) &0.83 (0.05) &0.84 (0.04) & {\bf 0.90} (0.01) &0.91 (0.02)\\ 
 & BSA & $d_H$ & 3.0 (0.0) &0.0 (0.0) &0.96 (0.02) &0.99 (0.01) &0.90 (0.00) &0.99 (0.00)\\
 & 2S & $d_H$ & 3.4 (0.5) &0.4 (0.5) &0.95 (0.02) &0.98 (0.01) &0.90 (0.00) &0.99 (0.00)\\
  & SGFL & $d_H$ &3.0 (0.0) &0.0 (0.0) &0.81 (0.05) &0.89 (0.05) &0.91 (0.01) &0.89 (0.03)\\
\hline
\hspace*{-7mm}
\multirow{5}{15mm}{\vspace{-4mm}
\begin{equation*}
\begin{array}{r  l}
d&=20\\ 
p&=200\\
T&=100\\
\sigma_{\varepsilon} &= 1 \\
\noalign{$\{21, 51, 91\}$}
\end{array}
\end{equation*}
} 
&BSA & HBIC & {\bf 3.0} (0.1) &{\bf 0.1} (1.0) & {\bf 0.93} (0.03) &0.61 (0.13) &0.84 (0.04) &{\bf 0.95} (0.01)\\ 
 & 2S & HBIC &3.5 (0.6) &0.5 (0.5) &0.87 (0.05) & {\bf 0.91} (0.03) & {\bf 0.90} (0.01) & 0.93 (0.01)\\ 
   & SGFL & HBIC &3.0 (0.3) &1.1 (10.0) &0.77 (0.11) &0.84 (0.11) &0.91 (0.02) &0.82 (0.11)\\ 
    & BSA & $d_H$ & 3.0 (0.1) &0.1 (1.0) &0.93 (0.03) &0.61 (0.13) &0.84 (0.04) &0.95 (0.01)\\
 & 2S & $d_H$ & 3.5 (0.6) &0.5 (0.5) &0.87 (0.05) &0.95 (0.03) &0.91 (0.01) &0.92 (0.01)\\
  & SGFL & $d_H$ &3.0 (0.0) &0.0 (0.0) &0.77 (0.06) &0.86 (0.08) &0.91 (0.02) &0.8 (0.06)\\
  \hline
\hspace*{-7mm}
\multirow{5}{15mm}{\vspace{-4mm}
\begin{equation*}
\begin{array}{r  l}
d&=20\\ 
p&=200\\
T&=100\\
\sigma_{\varepsilon} &= 2.5 \\
\noalign{$\{21, 51, 91\}$}
\end{array}
\end{equation*}
} 
&BSA & HBIC & {\bf 2.7} (0.9) &11.1 (29.4) & {\bf 0.91} (0.04) &0.21 (0.04) &0.56 (0.10) &{\bf 0.75} (0.16)\\ 
 & 2S & HBIC &4.1 (1.4) &13.3 (17.5) &0.69 (0.08) & {\bf 0.82} (0.06) &{\bf 0.92} (0.01) &0.67 (0.09)\\ 
   & SGFL & HBIC &2.7 (0.4) &{\bf 10.7} (17.3) &0.62 (0.12) &0.79 (0.11) &0.92 (0.02) &0.50 (0.12)\\ 
 & BSA & $d_H$ &3.0 (0.2) &0.5 (2.1) &0.91 (0.03) &0.22 (0.03) &0.58 (0.06) &0.80 (0.03)\\
  & 2S & $d_H$ & 5.6 (2.0) &3.2 (4.2) &0.71 (0.05) &0.86 (0.07) &0.92 (0.01) &0.69 (0.04)\\
    & SGFL & $d_H$ &3.0 (0.0) &0.0 (0.0) &0.66 (0.08) &0.76 (0.16) &0.90 (0.06) &0.49 (0.09) \\
    \hline
\end{tabular}}
\end{table}


\clearpage

Finally, Table \ref{table: s99 rho} shows results in the setup $s=0.99$, $(d,p,T)=(20,200,100)$, and $\rho_X=0.5$.  
There, the presence of relatively strong correlation among predictors has little to no effect on the performance of the BSA method 
(compare to Table \ref{table: s99 more} and Table 3 of the main article where $\rho_X = 0$). 
 The correlation slightly degrades the performance of the 2S method and more substantially degrades that of the SGFL, at least when the HBIC is used to select regularization parameters. On the other hand, the fact that  
 the best HBIC parameter for SGFL is $\gamma = 1$ (at the boundary of the $\gamma$ range) 
 suggest an overpenalization of model complexity. Indeed, the lines SGFL/$d_H$ show that at least for some choice of $(\lambda_1,\lambda_2)$, the SGFL performance can be satisfactory and on par with the other two methods.

\begin{table}[!ht]
\centering
\caption{Simulation study of statistical accuracy. True sparsity level $s=0.99$, correlation $\rho_X =0.5$,  number of replications: 100.}
\label{table: s99 rho}
\bigskip
\hspace*{-5mm}
\footnotesize{
\begin{tabular}{c r c  c c c c c  c}
  \hline
Setup & Method & $\lambda$ & NCP & $d_H$ & TPR & PPV & $\widehat{s}$ & $R^2$ \\ 
\hline
\hspace*{-7mm}
\multirow{5}{15mm}{\vspace{-4mm}
\begin{equation*}
\begin{array}{r  l}
d&=20\\ 
p&=200\\
T&=100\\
\sigma_{\varepsilon} &= 0 \\
\noalign{$\{21, 51, 91\}$}
\end{array}
\end{equation*}
} 
&BSA & HBIC & {\bf 3.0} (0.2) & {\bf 0.3} (1.8) &{\bf 1.00} (0.01) &{\bf 0.99} (0.09) & 0.99 (0.01) & {\bf 1.00} (0.00)\\ 
 & 2S & HBIC &4.3 (1.8) &5.6 (11.1) &0.99 (0.04) &0.95 (0.11) &{\bf 0.99} (0.00) &0.99 (0.03)\\ 
   & SGFL & HBIC & 2.3 (1.2) &24.2 (38.7) &0.85 (0.19) &0.28 (0.15) &0.95 (0.06) & 0.73 (0.31)\\ 
 & BSA & $d_H$ &3.0 (0.0) & 0.0 (0.0) &1.00 (0.01) &1.00 (0.00) &0.99 (0.00) &1.00 (0.00)\\
 & 2S & $d_H$ & 4.3 (1.7) &5.2 (10.9) &0.98 (0.06) &0.97 (0.06) &0.99 (0.00) &0.96 (0.10)\\
  & SGFL & $d_H$ &3.1 (0.7) &2.0 (7.3) &0.89 (0.14) &0.55 (0.43) &0.79 (0.32) &0.69 (0.24)\\
\hline
\hspace*{-7mm}
\multirow{5}{15mm}{\vspace{-4mm}
\begin{equation*}
\begin{array}{r  l}
d&=20\\ 
p&=200\\
T&=100\\
\sigma_{\varepsilon} &= 0.25 \\
\noalign{$\{21, 51, 91\}$}
\end{array}
\end{equation*}
} 
&BSA & HBIC & {\bf 3.0} (0.4) &{\bf 1.8} (10.3) & {\bf 0.96} (0.07) &0.52 (0.31) &0.96 (0.06) & {\bf 0.94} (0.05) \\ 
& 2S & HBIC &4.7 (2.1) &8.3 (12.2) &0.90 (0.12) &{\bf 0.85} (0.15) & {\bf 0.99} (0.00) &0.92 (0.06)\\  
& SGFL & HBIC & 2.0 (1.3) &32.5 (41.3) &0.80 (0.22) &0.38 (0.22) &0.96 (0.06) &0.62 (0.32)\\ 
& BSA & $d_H$ & 3.0 (0.2) &0.4 (2.0) &0.97 (0.07) &0.52 (0.31) &0.96 (0.04) &0.94 (0.03)\\
& 2S & $d_H$ & 5.1 (2.3) &6.2 (10.0) &0.91 (0.10) &0.94 (0.08) &0.99 (0.00) &0.87 (0.10)\\
& SGFL & $d_H$ & 3.4 (1.3) &2.2 (6.0) &0.91 (0.13) &0.47 (0.45) &0.73 (0.35) &0.62 (0.22)\\
\hline
\end{tabular}
}
\end{table}


\vspace*{-5mm}

\section*{Computation time} 

Table \ref{table: runtime} displays the average runtime per replication
 for each method (BSA, 2S, SGFL) in each simulation setup. 
Clearly, 2S is the fastest method of the three by at least 1 order of magnitude. 
SGFL is the second fastest and BSA last. 
On average, 2S is about 40 times as fast as SGFL, which in turn is 2 times as fast as BSA.

\begin{table}[!ht]
\centering
\caption{Average computation time per simulation replicate (in seconds, standard deviations in brackets).}
\label{table: runtime}
\medskip

\begin{tabular}{*{10}{c}}
  \hline
  $d$ & $p$ & $T$ & $s$ & $\sigma$ & $\rho_X$ & NCP & BSA  & 2S  & SGFL  \\ 
  \hline
 20 & 200 & 100 & 0.99 & 0.25 & 0.00 & 3 & 515 (69) & 16  (2) & 290  (53) \\ 
  20 & 200 & 100 & 0.99 & 1.00 & 0.00 & 3 & 564  (76) & 19  (3) & 353  (63) \\ 
   20 & 200 & 100 & 0.99 & 0.00 & 0.00 & 4 & 486  (57) & 15  (2) & 316  (56) \\ 
   20 & 200 & 100 & 0.99 & 0.25 & 0.00 & 4 & 500  (61) & 16  (3) & 334  (64) \\ 
   20 & 200 & 100 & 0.99 & 1.00 & 0.00 & 4 & 546  (62) & 19  (3) & 370  (78) \\ 
   100 & 500 & 200 & 0.99 & 0.00 & 0.00 & 3 & 20941  (3454) & 114  (19) & 8341 (1294) \\ 
   100 & 500 & 200 & 0.99 & 0.25 & 0.00 & 3 & 20778  (3270) & 116  (22) & 8331  (1204) \\ 
   100 & 500 & 200 & 0.99 & 1.00 & 0.00 & 3 & 20997  (3670) & 135  (28) & 8242  (1307) \\ 
   100 & 500 & 200 & 0.99 & 2.50 & 0.00 & 3 & 21926  (3820) & 217  (42) & 9472  (1485) \\ 
   100 & 500 & 200 & 0.90 & 0.00 & 0.00 & 3 & 18597  (36) & 100  (4) & 7841  (227) \\ 
   100 & 500 & 200 & 0.90 & 0.25 & 0.00 & 3 & 18593  (41) & 98  (6) & 7901  (232) \\ 
   100 & 500 & 200 & 0.90 & 1.00 & 0.00 & 3 & 18746  (105) & 100  (9) & 7848  (277) \\ 
   100 & 500 & 200 & 0.90 & 2.50 & 0.00 & 3 & 19509  (165) & 114  (5) & 8308  (159) \\ 
   20 & 200 & 100 & 0.90 & 0.00 & 0.00 & 3 & 620  (58) & 14  (2) & 354  (60) \\ 
   20 & 200 & 100 & 0.90 & 0.25 & 0.00 & 3 & 635  (56) & 14  (2) & 360  (58) \\ 
   20 & 200 & 100 & 0.90 & 1.00 & 0.00 & 3 & 671  (59) & 15  (2) & 393  (61) \\ 
   20 & 200 & 100 & 0.90 & 2.50 & 0.00 & 3 & 727  (64) & 18  (3) & 518  (83) \\ 
   20 & 200 & 100 & 0.99 & 0 & 0.50 & 3 & 2690 (2155) &  15    (5) & 833 (1292) \\
   20 & 200 & 100 & 0.99 & 0.25 & 0.50 & 3 & 3202 (2988) &  15    (6) & 1046 (1694) \\
   \hline
\end{tabular}
\end{table}

One information not shown in the table is the size of the grid at which the methods 
were evaluated. The grid size largely affects the runtime of the BSA and SGFL methods but almost not that of 2S. 
Indeed, the first step of 2S, group lasso, can be implemented quickly for any single regularization parameter $\lambda_2$ 
and warm starts can be exploited along the regularization path to speed up calculations. The second step of 2S, standard lasso, 
can calculate fits $\hat{\beta} = \hat{\beta}(\lambda_1)$ for any number of regularization parameters $\lambda_1$ essentially for free
thanks to its pathwise properties. 
In contrast, every model fit has a nontrivial cost for  BSA and SGFL. 
For BSA, it appears infeasible to efficiently implement pathwise lasso regression because the segments produced by the binary search are in general different for each $\lambda$ (one cannot keep track of a segmentation for each $\lambda$). 
For SGFL, warm start methods do not reduce by much the computational cost of a grid evaluation.    

Importantly, BSA and SGFL take an excessively long time to compute when the regularization parameter $\lambda$ or $(\lambda_1,\lambda_2)$ is taken too small (this typically lead to a lengthy search for optimal segmentation and a final result with far too many segments/change points). One way to avoid this behavior is through pilot attempts, making sure that the smallest value in a regularization grid is not too small. A related approach is through the use of stopping rules: stop the pathwise regularization  whenever the number of segments or nonzero regression coefficients exceeds a threshold. A third way to avoid spending excessive time fitting the model for small $\lambda$ is to limit the maximum number of iterations of each method. We have used some of these techniques in our simulations to keep the computation time manageable although no attempt at optimizing the computation time was made. 

We finally note that in practice, when implementing any of the three above pathwise regularization methods, 
computations can be efficiently parallelized, 
which may well reduce the computation time to a fraction of the values reported in Table \ref{table: runtime}.

\end{document}